\documentclass[a4paper,12pt]{article}
\usepackage[affil-it]{authblk}
\usepackage{cite}
\usepackage{amsmath}
\usepackage{graphicx}
\usepackage{xcolor}
\usepackage[cp1252]{inputenc}
\usepackage{float}
\usepackage{mathtools}
\usepackage{tabu}
\usepackage{amssymb}
\usepackage{tikz}
\usepackage{tabularx}
\usepackage{enumitem}

\providecommand{\keywords}[1]
{
  \small	
  \textbf{\textit{Keywords:}} #1
}
\usepackage{tikz}
\makeatletter
\DeclareRobustCommand*{\bfseries}{%
   \not@math@alphabet\bfseries\mathbf
   \fontseries\bfdefault\selectfont
   \boldmath
}
\makeatother
\usepackage[square,numbers,sort&compress]{natbib}
\usepackage{xfp}
\usepackage{numerica}
\begin{document}
\title{A Generalized Model for Predicting the Drag Coefficient of Arbitrary Bluff Shaped Bodies at High Reynolds Numbers }
\author[1,2]{Yousef M. F. El Hasadi \footnote{Email address for correspondence: yme0001@auburn.edu,g.damianidis.al.chasanti@tue.nl}}
\author[1]{ Johan T. Padding}
\affil[1]{Process and Energy department, Delft University of Technology,Leeghwaterstraat 39,2628 CB Delft, the Netherlands}
\affil[2]{Eindhoven University of Technology, Mechanical Engineering Department, The Netherlands}
\date{}
\maketitle
\begin{abstract}

We propose an accurate model for the drag coefficient of arbitrary bluff bodies that is valid for high Reynolds numbers ($Re$). The model is based on the drag coefficient model derived for the case of a sphere:, $C_D = a_1 +{\frac{K a_2}{Re}} +{a_3\log(Re)+ a_4\log^2(Re) + a_5\log^4(Re)}$ (El Hasadi and Padding, Chemical Engineering Science, Vol. 265, 2023). The coefficients $a_2$, $a_3$, $a_4$, and $a_5$ do not depend on the object's shape or its orientation with respect to the flow, and $K$ is the Stokes drag correction factor, which for the case of the sphere, is equal to 1.0. 
The shape and orientation effects are included in the value of $a_1$ for the high Reynolds number flow regime. Interestingly, we found a strong correlation between the value of the $a_1$ coefficient and the frictional drag derived from boundary layer theory. One of the main findings of this investigation is that the rate of change of the drag coefficient with respect to the Reynolds number in the inertial flow regime is independent of the shape of the body or its orientation. Our model successfully predicts, with acceptable accuracy, the historical data of Wieselsberger (Technical Report, 1922) for the case of an infinite cylinder. Additionally, the model predicts the drag coefficient of other bluff body geometries such as oblate and prolate spheroids, spherocylinders, cubes, normal flat plates and irregular non-spherical particles\@. Additionally, we present a power-based model for the drag coefficient: \( C_D = a_{p_1} + \frac{24K}{Re} + \frac{4.119}{\sqrt{Re}} \). In this model, the term \( a_{p_1} \) represents the asymptotic form drag in the subcritical flow regime for different bluff body geometries\@.

\end{abstract}
\\

\section{Introduction}

The drag force, which is the resistance force that occurs when a body moves through a fluid, is a critical parameter in various fields of science and engineering. Organisms from tiny bacteria to massive whales have evolved adaptations to minimize the drag force they experience when swimming through fluids \citep{vogel2020life,cohen2019fluid,el2016simulating,el2017self,el2020self}. In the transportation industry, drag force is a crucial optimization parameter for reducing fuel consumption in vehicles, aeroplanes, and ships \citep{hucho1993aerodynamics,anderson2011ebook}. In the military sector, drag force is vital in designing projectiles, submarines, and unmanned aerial vehicles \citep{eckert2019ludwig,winslow2018basic}. Sustainable energy applications such as wind turbines and fluidized beds also rely on understanding and controlling drag force \citep{hansen2015aerodynamics,mahajan2019fluid}\@. Fluid drag is one of nature's most vital forces, yet we have a limited ability to predict it, as we will describe in the following paragraphs\@.\\

One of the earliest systems of partial differential equations derived were those of Euler \cite{darrigol2005worlds} in 1752, which described the motion of an ideal fluid (inviscid incompressible flow)\@. However, shortly after their introduction, d' Alembert \cite{darrigol2005worlds} found that fluid resistance on an object moving in a static fluid is zero after applying Euler's equations, which contradicts the behaviour of real fluids\@. This statement made by d'Alembert is now widely known as d'Alembert's paradox, which was a great surprise, since Euler's equations are based on a rigorous mathematical framework of  Newton's laws\@.
With the advancement of the molecular theory of matter, mainly that of gases, Navier and Stokes independently derived a new set of equations that better resembled fluid behaviour, known as the Navier-Stokes equations \cite{darrigol2005worlds}\@. These equations introduced a shearing stress term that depends on the fluid's viscosity\@. In the meantime experiments in the nineteenth century showed that fluids with finite but small viscosity that flow over bluff bodies such as spheres and cylinders are unstable, especially at the rear of the body where eddies are formed\@. Also, the experiments  verified that the drag force depends on the {quadratically} on the flow velocity, as Newton suggested \cite{darrigol2005worlds}\@. The instability of the fluid at the rear side of the bluff bodies were very difficult to  explain by either Euler's, or the  Navier-Stokes equations\@. The  non-linearity  of the  
  Navier-Stokes makes it impossible to find a generic analytical  solution\@. This  inability of the theory to match the experiments led to a division between theoreticians and empiricists, especially when it comes to  predicting the drag force of an object\@. This division continued for the whole period of the nineteenth century\@.\\

One of the first triumphs of the Navier-Stokes equations was when Stokes \cite{stokes1851effect} found an analytical expression for the drag force experienced by a sphere in slow (creeping) flow conditions and showed that the drag force scales linearly  with the velocity\@. However, for high-flow velocities, theoreticians attempted to develop new approaches to tackle the discrepancy in the prediction of the drag force\@. Examples of those are Rayleigh's dead water theory and Poncelet and Saint-Venant's eddy resistance theories, but with no success\cite{darrigol2005worlds}\@.\\

A partial resolution of d'Alembert's paradox came at the beginning of the twentieth century \cite{stewartson1981d} by Prandtl \cite{Prandtl1904} using the idea of the boundary layer, which states that the effects of viscosity are confined to a boundary layer near the wall\@. In contrast, beyond the  the boundary layer, the flow is governed by inviscid flow theory\@. Prandtl \cite{Prandtl1904} suggested that suction pressure generated at the rear part of the bluff body contributes significantly to the drag force, and it is formed  due to the separation of the boundary layer from the wall\@.
Prandtl's ideas are based on the boundary layer equations, a simplified  version of the Navier-Stokes equations\@. Prandtl used the non-slip boundary condition to describe the fluid velocity at the wall\@. The  mathematical correctness of this boundary condition  has yet to be proven \cite{day1990no}\@. However, substantial evidence from experiments suggests that the velocity of the fluid attached to a static wall is zero \cite{day1990no}, which supports the concept of the non-slip boundary condition\@. Blasius \cite{blasius1908grenzschichten} solved the boundary layer equations for the case of a thin flat plate placed horizontally to the flow direction and came up with an estimate of the drag force for high-velocity laminar flow conditions\@.
Despite the usefulness of the boundary layer theory, it has its limitations\@. For example, it is not valid for separated flows, such as those that occur for flow over bluff bodies, and it consists of several paradoxes, as mentioned by Birkhoff \cite{birkhoff1960study}\@. However, the boundary layer theory led to significant progress in understanding fluid mechanics in general. Also, it helped significantly to close the gap between experimental observations and theoretical predictions, and above all, it agglomerated the theoretical and experimental societies together\@. \\

Wieselsberger \cite{wieselsberger1922further} performed experiments with cylinders that were oriented at right angles to the fluid flow direction\@. He was one of the first researchers to propose using the drag coefficient to represent the drag force\@. The drag coefficient is defined as $C_D = F_D/\frac{1}{2}\rho v^2_{\infty} A_p$, where $F_D$ is the drag force, $\rho$ is the density of the fluid, $v_{\infty}$ is the undisturbed velocity of the flow, and $A_p$ is the projected area\@. Also, he showed that the drag coefficient $C_D$ is strongly dependent on   the Reynolds number that can be defined as $Re = \dfrac{\rho v_{\infty} d}{\mu}$, where $d$ is a length scale related to the object, for the case of the cylinder is the diameter, and $\mu$ is the viscosity of the fluid\@. Wieselsberger \cite{wieselsberger1922further} found that the cylinder's drag coefficient depends in a complex non-expected way on the Reynolds number  $Re$, and the flow behaviour over a cylinder can take on a range of different morphologies\@. Churchill's \cite{churchill2013viscous} further elucidates this spectrum of flow patterns, which can range from symmetrical and non-separated at low Reynolds numbers to separated and recirculating at moderate Reynolds numbers to shedding vortexes at higher Reynolds numbers. When the Reynolds number is around $3\times 10^5$, the boundary layer on the surface of the cylinder changes from laminar to turbulent\@. The cascade of the different  flow morphologies significantly effects the evolution   of the drag coefficient with the Reynolds number\@. Several other experiments for different bluff body geometries, such as sphere and the cube \cite{achenbach1972experiments,khan2018flow} showed the same dependence of the drag coefficient on the Reynolds number\@.\\

Only a few analytical solutions are available for the drag coefficient, such as for spherical and cylindrical geometries \cite{proudman1957expansions,kaplun1957asymptotic}\@. However, these solutions are only valid for Reynolds numbers $<$ 1.0\@. At higher Reynolds numbers, analytical solutions of the Navier-Stokes equations no longer exist due to their non-linearity\@. In these cases, the drag coefficient is typically evaluated using  boundary layer theory \cite{thom1928boundary, frossling1958evaporation} or by solving the Navier-Stokes equations numerically \cite{dennis1971calculation, cheng2017large, sanjeevi2022accurate,ouchene2020numerical}\@. The drag coefficient results from the numerical simulations are typically expressed in curve-fitted equations, which we will discuss in more detail later on\@.\\

In real-life applications, the correlations used for the drag coefficient are predominantly based on experimental data rather than theory\@. This is because obtaining theoretical results, even from numerical simulations, can be prohibitively computationally  expensive, particularly at high Reynolds numbers\@. Despite several decades of using boundary-layer theory and CFD simulations, our understanding of the drag force experienced by a bluff body has remained the same since Proudman \cite{proudman1960example} expressed his concerns about the field's lack of progress and made the following statement  nearly 60 years ago:  ``\textit{It is a sobering thought that, despite the fifty or so years that the techniques of boundary-layer theory have been at our disposal, so little progress has been made with the problem that even such a bulk characteristic as the dependence of drag coefficient on (large) Reynolds number is still a matter of pure conjecture\@.}"\@.
Several correlations exist for the drag coefficient of arbitrarily shaped particles, but they are only valid for moderate Reynolds numbers of up to 100\@. One of the earliest correlations proposed for multi-shaped particles is that of Haider and Levenspiel \citep{haider1989drag}\@. They introduced the effects of the shape by introducing the drag coefficient dependence on the sphericity\@. The mathematical form of their correlation is similar to that of the Clift \cite{clift1978bubbles} correlation, developed for the case of sphere geometry at high Reynolds numbers\@.\\

Recently, we employed  a physics-informed symbolic regression machine learning algorithm that uses sparse volumes of data and incorporates knowledge such as analytical solutions in the training process \cite{el2019solving}\@. We applied this algorithm to obtain mathematical models that describe the drag coefficient of a prolate spheroid and sphero-cylinder particles for different particle orientations and aspect ratios for $Re$ less than 2000, using only 72 data points for training \cite{el2019solving}\@. For the geometries we considered, the drag coefficient is best described by logarithmic terms of the Reynolds number\@. \\

We used our machine learning algorithm to verify the conjecture made by Proudman and Pearson \cite{proudman1957expansions}, which states that the drag coefficient of a sphere can be expressed as an expansion of logarithmic terms and powers of the Reynolds number\@. Our findings showed that the drag coefficient for a sphere from the creeping flow regime up to the drag crisis ($Re \simeq 10^5$) can indeed be represented by an expansion of logarithmic terms of the Reynolds number \@. We trained our algorithm using data from the Brown and Lawler \cite{brown2003sphere} correlation for Reynolds numbers ranging from 0\@.1 to $10^5$\@. Our logarithmic-based equation was able to extrapolate beyond its training range\@. For example, it converged to the Oseen constant as $Re \to$ 0 and accurately predicted the onset of the drag crisis for high $Re$\@. Furthermore, we demonstrated that our algorithm could obtain analytical solutions from data based on experimental correlations\@. For instance, the equation we derived for the Nusselt number in the case of forced convection over a sphere converged to the analytical solution of Acrivos and Goddard \cite{acrivos1965asymptotic} in the limit as $Pe$ (Peclet number) $\to \infty$\@. \\

We also found a strong connection between the equations we derived for the drag coefficient of both spherical and non-spherical particles \cite{el2022logarithmic, el2019solving}\@. Regardless of the particle shape, they all contained logarithmic terms, and the coefficients of these terms had very close values\@. This implies that a general model governing the variation of the drag coefficient could possess logarithmic terms of the Reynolds number\@. \\

The main focus of this paper is to develop a drag coefficient model for non-spherical geometries  based on the derived drag coefficient for a sphere, incorporating logarithmic terms of the Reynolds number\@. The paper will be divided into several sections. Firstly, we will discuss and analyze various drag coefficient formulas available in the literature, providing readers with an understanding of the historical evolution of these formulas\@. Subsequently, we will delve into the theory underlying our drag coefficient model and extensively validate it against different geometries documented in the literature\@. The last section of the paper will be devoted to a simple general power-based equation for the prediction of the drag coefficient of the bluff bodies\@.

\section{Correlations for the Drag Coefficient}
This section will summarize and describe the main correlations available for the drag coefficient for different non-spherical shapes\@. We will emphasize the general correlations that can be applied to different particle shapes. Also, we will cover correlations that are specific to the geometries, such as that of oblate and prolate spheroids, since for those geometries there are vast numbers of correlations that are based on data from CFD simulations, which will allow us to compare our results directly with the approximate solution of the Navier-Stokes equations\@. 
\subsection{Correlations for arbitrary particle shapes}
There are few equations in the literature for the drag variation of an arbitrary particle shape; usually, those correlations are derived using either experimental or numerical available data\@. {For all the correlations from the different sources from the literature that we will present for different particle shapes, the Reynolds number is defined based on the volume-equivalent sphere diameter, and the frontal area is based on a frontal area of a sphere with the same diameter\@. This type of definition of the Reynolds number and the drag coefficient is widely used in chemical engineering applications\@. It will not be a problem comparing our results with the available correlations, as we will show later\@. For this reason, we did not change the symbols of the drag coefficient and the Reynolds number in the upcoming equations\@.}  \\

As we mentioned in the introduction, Haider and Levenspiel \citep{haider1989drag} were among the first to suggest a general correlation for the drag for non-spherical particles\@. By fitting a comprehensive variety of data from different experiments for different particle geometries, including spheres, cube octahedrons, octahedrons, cubes, tetrahedrons, and disks\@. The parameter they used to describe the non-spherical particle geometry variation is the sphericity  ($\phi$)\@. Sphericity is defined as the ratio between the surface area of the sphere having the same volume as the particle, divided by the actual surface area of the particle\@. By definition, the value of  $\phi$ is  $\leq$ 1.0,  the Haider and Levenspiel \citep{haider1989drag} correlation is given by the following formula: 
\begin{equation}\label{eqpowg1}
C_D = \frac{24}{Re}\left(1+ARe^B\right)+\frac{C}{1+\frac{D}{Re}}
\end{equation}
The constants in Eq.(\ref{eqpowg1}) are given as the follows: 
\begin{equation}\label{eqpowg2}
\begin{aligned}
A=& \exp \left(2.3288-6.4581 \phi+2.4486 \phi^{2}\right) \\
B=& 0.0964+0.5565 \phi \\
C=& \exp \left(4.905-13.8944 \phi+18.4222 \phi^{2}
-10.2599 \phi^{3}\right) \\
D=& \exp \left(1.4681+12.2584 \phi-20.7322 \phi^{2}+15.8855 \phi^{3}\right)
\end{aligned}
\end{equation}
The correlation of Haider and Levenspiel \citep{haider1989drag} shows that the particle geometry influences the drag coefficient  from the  low to high Reynolds numbers since all constants from $A$ to $D$ are functions of $\phi$\@. The most accurate  general correlation for predicting the drag coefficient of non-spherical particles is that of Holzer and Sommerfeld \cite{holzer2008new}, it is based on the same philosophy as that of  Haider and Levenspiel \citep{haider1989drag} by fitting the experimental drag coefficient values for different particle shapes\@. Compared to the Haider and Levenspiel \citep{haider1989drag}, they added crosswise sphericity($\psi$), which is defined  {as the ratio between the cross-sectional area of the volume equivalent sphere and the projected cross-sectional area of the considered particle perpendicular to the flow}\@. Crosswise sphericity accounts for the effects of the orientation of the flow field on the drag coefficient. At the same time, it significantly increases the complexity of the resulted correlation\@. The Holzer and Sommerfeld correlation in its simple form is given by the following :
\begin{equation}\label{eqpowg3}
C_D = \frac{8}{Re}\frac{1}{\sqrt{\psi}}+\frac{16}{Re} \frac{1}{\sqrt{\phi}}+\frac{3}{\sqrt{Re}}\frac{1}{\phi^{\frac{3}{4}}}+0.42\times 10^{0.4\left(-\log(\phi)\right)^{0.2}}\frac{1}{\psi}
\end{equation}

\subsection{Oblate and prolate spheroids}

 A few correlations deal with predicting the drag coefficient for oblate spheroids, especially at  high Reynolds numbers\@. The correlation provided by Ouchene\cite{ouchene2020numerical}  is applicable for a wide range of aspect ratios that spans from 0.2 to 1.0 and Reynolds numbers from the Stokes regime up to 100, { for case that the flow is parallel to the minor axes of the oblate spheroid  the orientation $\alpha$ = $0^{\circ}$, while if the flow is perpendicular to the minor axes of the oblate spheroid  $\alpha$ = ${90^{\circ}}$\@}. The correlation takes the following form for the case of $\alpha = 90^{\circ}$: 
 \begin{equation}\label{eqpowo1}
 C_{D_{90}} = \frac{24}{Re}\left( K_{o_{90}} +0.15p_a^{100.7}Re^{0.687}+0.14\left(1-p_a^{24.75}\right)\right) Re^{0.7143}
\end{equation} 
Where  $K_{o_{90}}$ is the correction of the Stokes drag for the oblate spheroid derived by Happel \cite{happel2012low} for $\alpha$ = $90^{\circ}$ given by the following relation: 

\begin{equation}\label{eq2olog}
K_{o_{90}} = \frac{8}{3} p_a^{-\frac{1}{3}}\left[-\frac{p_a}{1-p_a^2}-\frac{2p_a^2+3}{(1-pa^2)^{\frac{3}{2}}}\sin^{-1}(1-p_a^2)\right]^{-1}
\end{equation}
Eq.(\ref{eqpowo1}) shows that $C_D$ is dependent on the $Re$, and $p_a$\@. Ouchene\cite{ouchene2020numerical} also provided a correlation for the oblate spheroids for $\alpha = 0^{\circ}$\@. However, In order to keep the paper concise and short, we did not include it\@.

 There is  a correlation that is based on experimental data provided by Clift \citep{clift1978bubbles} for oblate spheroids with aspect ratio $p_a$ =0.5, for $\alpha$ = $90^{\circ}$ valid for a wide range of Reynolds numbers {ranging}  between 40  {and}  $10^4$\@.  The mathematical  formula of the  Clift\citep{clift1978bubbles} correlation is the following:
\begin{equation}\label{eqpowo3}
\log_{10}(C_{D_{90}}) = 2.035-1.660w+0.3958w^2-0.0306w^3
\end{equation} 
where $w = \log_{10}(Re_s)$, and $Re_s = Re/(2p^{1/3}_a) $. \\

 {Example of drag coefficient correlation for prolate spheroids is given in our previous work   \cite{el2019solving} based on the numerical  data of Sanjeevi et al.\cite{sanjeevi2018drag} for the case of $p_a$  = 2.5 and $\alpha$ = $0^{\circ}$, it takes the following form:}

\begin{equation}\label{eq1plog}
C_{D_0} = a_1+ \frac{a_2 K_{p{0}} + a_3+ a_4\log^2(Re)}{Re}+a_5\log(Re)+a_6\log^2(Re)
\end{equation}
 {The values of the coefficients of Eq.(\ref{eq1plog}) are listed in Table \ref{table1plog}\@. Since the drag coefficient and the Reynolds number of the Sanjeevi et al.\cite{sanjeevi2018drag} data  are based on the volume equivalent of a sphere, the coefficients of the logarithmic equation (Eq.(\ref{eq1plog})) are also based on the volume equivalent of a sphere as well\@. The orientation angle $\alpha$ for the case of prolate spheroids is the following $\alpha$ = $0^{\circ}$ when the flow direction of the flow is in a parallel direction to the major axis of the spheroid, and  $\alpha$ = $90^{\circ}$, when the flow direction is perpendicular to the major axis of the spheroid\@. }
\begin{table}[h!]
\begin{center}
\begin{tabu} to 0.8\textwidth { | X[c] | X[c] | }
 \hline
 Coefficients  & Eq.(\ref{eq1plog})   \\
 \hline
 $a_1$  & 3.151    \\
 $a_2$  & 25.873   \\
 $a_3$  & -2.258     \\
 $a_4$  &  0.230\\
 $a_5$  &  -0.841 \\
  $a_6$ & 0.057 \\
\hline
\end{tabu}
\end{center} 
\caption{ Coefficients for Eq.(\ref{eq1plog})}
\label{table1plog}
\end{table}
Where  $K_{p_{0}}$ is a Stokes correction factor for the prolate spheroids  given by the following equation:
\begin{equation}\label{eq3plog}
K_{p_0}=\frac{8}{3} p_a^{-\frac{1}{3}}\left(\frac{-2 p_a}{p_a^2+1}+\frac{2 p_a^2-1}{(p_a^2-1)^\frac{2}{3}} \log \frac{p_a+\sqrt{p_a^2-1}}{p_a-\sqrt{p_a^2-1}}\right)^{-1}
\end{equation}

Several power-based drag coefficient correlations exist for prolate spheroids across a wide range of Reynolds numbers and aspect ratios\@. Ouchene et al.\cite{ouchene2016new}  developed a specific correlation for Reynolds numbers in the range {$1\leq Re\leq 240$} and aspect ratios in the range {$1.25\leq p_a \leq 32$}\@. This correlation is given by:
\begin{equation}\label{eqpowp1}
C_{D_{0}} = \frac{24}{Re}[K_p +0.15p^{-0.8}_aRe^{0.687}+\frac{(p_a-1)^{0.63}}{24} Re^{0.41}]
\end{equation}

Frohlich et al. \cite{frohlich2020correlations}  developed correlations for prolate spheroids using Lattice Boltzmann method and a large number of simulations (around 4400 runs) \cite{frohlich2020correlations}\@. The resulting correlation is valid for Reynolds numbers in the range of 1 to 100 and aspect ratio values $p_a$ between 1 {and} 8\@. The correlation is given by: 
\begin{equation}\label{eqpowp2}
C_{D{_0}} = \frac{24 K_p}{Re}f_{d_0}
\end{equation}
Where $f_{d0}$ is given by the following relation:
\begin{equation} \label{eqpowp3}
f_{d_0} = 1+0.15Re^{0.687}+c_{d1}\log(p_a)^{c_{d2}}Re^{c_{d3}+c_{d4}\log(p_a)}
\end{equation}
$c_{d1}$, $c_{d2}$,  $c_{d3}$, and  $c_{d4}$ are constants\@. Sanjeevi et al. \cite{sanjeevi2022accurate} recently developed correlations for prolate spheroids with aspect ratios $p_a$ in the range of 1 to 16 and Reynolds numbers in the range of 0.1 to 2000\@. Their correlations are given by:
\begin{equation} \label{eqpowp4}
C_{D{_0}}=\left(\frac{a_1}{R e}+\frac{a_2}{R e^{a_3}}\right) \mathcal{R}+a_5(1-\mathcal{R})
\end{equation} 
Where $\mathcal{R}$ is given {by}:
\begin{equation}\label{eqpowp5}
\mathcal{R}=e^{-a_4 R e}
\end{equation}
The coefficients in Eq.(\ref{eqpowp5}) are functions of the aspect ratio\@. Sanjeevi et al.'s \cite{sanjeevi2022accurate} correlation will be used as the main data source for accurate comparison due to its applicability for a wide range of Reynolds numbers and aspect ratios for prolate spheroids\@. The above correlations for the prolate spheroids have analogous correlations for the case of $\alpha$ = $90^{\circ}$\@. The interested reader can refer to the original papers, the same implies to the Stokes correction factor $K$\@.   \\

Recently Chen et al.\cite{chen2021drag} developed a correlation for the drag coefficient specifically   {aimed at} the oblate and prolate spheroids\@. The correlation was obtained by fusing the numerical data obtained by the authors for prolate spheroids geometries with drag coefficient data from other { investigations} for prolate and oblate spheroids\@.  {For example, they used the oblate, and prolate, data  of Sanjeevi et al.\citep{sanjeevi2018drag}} that we used to derive  equations for prolate spheroids \cite{el2019solving}, {simultaneously} { they used the data of prolate  spheroids  from Ouchene et al. 
\citep
{ouchene2016new}\@}. The correlation is valid for the case of oblate { spheroids}  for aspect ratio ($p_a$) and $Re$  in the range of   $0.2\leq p_a <1$, $ 1\leq Re\leq 100$, and for prolate spheroids $1\leq p_a\leq 5$, $1\leq Re\leq 2000$\@.  The mathematical form  of the correlations that they provide, {for} the horizontal orientation ($\alpha = 0^\circ$) and for the vertical orientation ($\alpha = 90^\circ$) are the following:
{ 
\begin{equation}\label{eqpowg5}
C_{D_0} = \frac{26.47}{Re}p^{-0.120}_a +\frac{4.576}{Re^{0.343}}p^{-0.525}_a+\frac{0.413}{1+16300Re^{-1.09}}
\end{equation} 
\begin{equation}\label{eqpowg4}
C_{D_{90}} = \frac{26.58}{Re}p^{0.066}_a +\frac{4.607}{Re^{0.343}}p^{-0.166}_a+\frac{0.413}{1+16300Re^{-1.09}}+p_a^{0.352}-1.0
\end{equation} }

The drag coefficient depends on the aspect ratio for high and low Reynolds number regimes regardless of the flow orientation, as shown by both equations Eq.(\ref{eqpowg4}) and Eq.(\ref{eqpowg5}). It will also be interesting to evaluate the generalization behaviour of Eq.(\ref{eqpowg4}) and Eq.(\ref{eqpowg5}) since they are derived by using a substantial amount of data\@. 

The correlations that we presented for the case of prolate spheroids show a significant  {diversity} in their mathematical forms, even though they describe the same geometry. This shows that, in general, we lack a good understanding of the theory that governs the flow over obstacles\@.

\section{Theory} 

Our predictions for the drag coefficient will be made using a single logarithmic equation that we derived for the case of a sphere, {where} the Reynolds number is based on the diameter of the sphere  \cite{el2022logarithmic} for the most of the current paper\@.{The drag coefficient is defined as $C_D = \dfrac{F_D}{\frac{1}{2} \rho v_{\infty}^2 A_p}$, where $A_p$ is the frontal area of the sphere, which is defined as $\dfrac{\pi}{4} d_p^2$, and $d_p$ is the diameter of the sphere\@.} The equation is valid for large Reynolds numbers up to $10^6$. The equation takes the following form:

\begin{equation}\label{eqloggeneral}
C_D = a_1 +{\frac{K a_2}{Re}} +{a_3\log(Re)+ a_4\log^2(Re) + a_5\log^4(Re)} 
\end{equation}

\begin{table}[h!]
\begin{center}
\begin{tabu} to 1.0\textwidth { | X[c]  | X[c]|}
 \hline
 Coefficients  & Eq.(\ref{eqloggeneral})\\
 \hline
 $a_1$  & shape and orientation dependent \\
 $a_2$  &  24.205   \\
 $a_3$  & -0.818   \\
 $a_4$  &  0.064  \\
  $a_5$  &   -0.000107  \\
\hline
\end{tabu}
\end{center} 
\caption{ Coefficients for Eq.(\ref{eqloggeneral})}
\label{tableloggeneral}
\end{table}
\noindent The values of the coefficients are tabulated in Table \ref{tableloggeneral}\@.{ Comparing the prefactors of the logarithmic terms in Equation (\ref{eqloggeneral}) with the corresponding prefactors of the logarithmic terms in Equation (\ref{eq1plog}), we found that they are alike\@. This similarity suggests that the prefactor values are universal and applicable in cases where the drag coefficient is defined based on the volume equivalent sphere\@.} The value of the $a_1$ coefficient  for the case of the sphere is 3.286\@. For Eq.(\ref{eqloggeneral}) to predict the drag coefficient for a geometry other than a sphere, the following assumptions are made:
\begin{itemize}
\item The rate of change of the drag coefficient with respect to $Re$ at high Reynolds number will not depend on the object's geometry or its orientation with respect to the flow\@. 
\item   { The coefficient $K$ and $a_1$ are the only coefficients in Eq.(\ref{eqloggeneral}) that change with the object's geometry and orientation\@ Throughout this paper, if we mention the $a_1$ coefficient, we mean the $a_1$ coefficient related to Eq.(\ref{eqloggeneral}), unless we say otherwise\@.  Several analytical solutions are available to evaluate the correction factor $K$ for different bluff body geometries\@. In cases where we lack sufficient information about the value of $K$, we assign a value of zero to the correction factor $K$ to cancel out the Stokes term from Eq.(\ref{eqloggeneral})\@. The cancellation of the Stokes term will not significantly impact the value of $C_D$ at high Reynolds numbers since the Stokes term does not play an essential role in that flow regime\@.} 
\item The value of the $a_1$ coefficient by the following two way: 

\begin{itemize} [label=\textbullet]
\item {The value of the coefficient $a_1$ can be deduced from a single measurement of the drag coefficient at low or moderate  Reynolds numbers, whether obtained through experimental or numerical means. This requirement for only one drag coefficient reading is rooted in one of the outcomes of the mean value theorem \cite{flett19582742}, which states that if two functions share the same derivative over a specific interval, they differ only by a constant. This constant difference corresponds to the coefficient $a_1$\@. Once the value of coefficient \(a_1\) is determined, the logarithmic Equation (Eq.(\ref{eqloggeneral})) incorporates all definitions of the drag coefficient and Reynolds numbers from the data source used for $a_1$ calculation\@. This unification of frameworks occurs because of the logarithmic Equation (Eq.(\ref{eqloggeneral})) and the data source becomes identical functions\@. } 
\item {In the forthcoming sections of this paper, we will demonstrate that the coefficient $a_1$ is intricately connected to the shear stress at the wall. This connection is established through the prefactor of the $Re^{-1/2}$ term, which governs the skin friction coefficient as derived from the boundary layer theory\@. By adopting this method to ascertain the value of coefficient $a_1$, Equation (\ref{eqloggeneral}) becomes independent from any requirement for numerical or experimental data pertaining to the drag coefficient.} 
\end{itemize} 

\end{itemize}

We will provide extensive comparison with experimental and numerical results to demonstrate the validity of the above two assumptions\@. \\

 In our previous work \cite{el2022logarithmic}, we demonstrated a potential link between the $a_1$ coefficient and the boundary layer separation point\@. {In this work, we aim to provide an additional explanation for the nature of the $a_1$ coefficient by showing that it is strongly correlated with the prefactor of the $Re^{-1/2}$ term in the frictional  drag {(shear stress at the wall)} experienced by the object and derived using the boundary layer theory\@}.{ To illustrate this correlation, we will compare the coefficient values $a_1$ with the available prefactor values of $Re^{-1/2}$ obtained from  solutions for the skin drag coefficient in the literature for various object geometries\@.} \\

{For relatively high $Re$, the Stokes drag $\dfrac{Ka_2}{Re}$ becomes insignificant\@. The drag coefficient for high Reynolds numbers is given by:}   
 
\begin{equation}\label{eqhighRe}
C_{D_{h}} = a_1  +{a_3\log(Re)+ a_4\log^2(Re) + a_5\log^4(Re)} 
\end{equation} 
Where the total drag  at  high Reynolds number is $C_{D_{h}}$\@. The total drag at high Reynolds numbers  is the summation of the friction drag and pressure drag:
\begin{equation}\label{eqfphighRe}
C_{D_{h}} = C_{D_{hf}}+C_{D_{hp}}
\end{equation} 
From the boundary layer theory \cite{schlichting2016boundary}  we can assume that the skin friction drag in the inertial flow regime  can be approximated {by} the  following expression:
\begin{equation}\label{frictiondrag}
C_{D_{hf}} = \frac{\delta}{\sqrt{Re}}
\end{equation}

 We will demonstrate that there is a strong correlation between the value of $a_1$ and the {shear stress at the wall}, as represented by the constant $\delta$\@.{ For the case of the sphere, for example,  $\delta = a_1 = 3.286$\@}. We know that the sphere is bluff body in which the boundary layer will separate and the above approximation for the frictional drag  may not be valid\@. However,  there are  plenty of drag correlations  {in which  the}   $Re^{-1/2}$ term exists, such that of Abraham \cite{abraham1970functional}\@.  For this reason we will keep the current approximation for the {frictional drag Eq.(\ref{frictiondrag})}, and test its validity by comparing our results for the frictional drag  with numerical, and experimental results available in the literature\@. Churchill \cite{churchill2013viscous} integrated further the  results of Frossling \cite{frossling1958evaporation} for the boundary layer theory for the case of  sphere, and found that the frictional drag is given is by $\dfrac{3.464}{\sqrt{Re}}$\@.{ The value of $\delta$ obtained from the logarithmic equation, which is 3.286, is very close to the value of $\delta$ obtained from Churchill's formula \cite{churchill2013viscous}, which is 3.464. The difference between the two values is less than 5\%\@.}\\

 The overall friction  and pressure drag are obtained by adding the contribution from the Stokes drag, as given by the following equations for the case of the sphere: 

\begin{equation}\label{eqfrictiondraag}
C_{D_{f}} = \frac{2}{3} \frac{24}{Re}+ \frac{3.286}{\sqrt{Re}}
\end{equation}
\begin{equation}\label{eqformdrag}
C_{D_{p}} = \frac{1}{3} \frac{24}{Re}+(C_{D_{h}}- \frac{3.286}{\sqrt{Re}})
\end{equation}
\begin{table}[t]
\begin{center}
\begin{tabu} to 1.25\textwidth { | X[c] | X[c] | X[c] |X[c] | X[c] |   }
\hline
$Re$&$C_{D_{f}}$ Ref\cite{dennis1971calculation} &$C_{D_{f}}/2$ Eq.(\ref{eqfrictiondraag})(\%)&\ $C_{D_{p}}$ Ref\cite{dennis1971calculation}&$C_{D_{p}}/2$ Eq.(\ref{eqformdrag})(\%) \\
 \hline
 0.5& 17.23& 18.46 ($\eval*{$-\frac{CD1-CD2}{CD1}100$}[CD1 = 17.23, CD2 = 18.46][2]$\%)&8.62&7.68($\eval*{$-\frac{CD1-CD2}{CD1}100$}[CD1 = 8.62, CD2 = 7.68][2]$\%)\\
1.0& 9.13& 9.71 ($\eval*{$-\frac{CD1-CD2}{CD1}100$}[CD1 = 9.13, CD2 = 9.7][2]$\%)&4.58&4.03($\eval*{$-\frac{CD1-CD2}{CD1}100$}[CD1 = 4.58, CD2 = 4.03][2]$\%)\\ 
5.0& 2.36& 2.34 ($\eval*{$-\frac{CD1-CD2}{CD1}100$}[CD1 = 2.36, CD2 =2.34][2]$\%)&1.23&1.13($\eval*{$-\frac{CD1-CD2}{CD1}100$}[CD1 = 1.23, CD2 = 1.13][2]$\%)\\ 
10.0& 1.42& 1.32 ($\eval*{$-\frac{CD1-CD2}{CD1}100$}[CD1 = 1.42, CD2 =1.32][2]$\%)&0.78&0.75($\eval*{$-\frac{CD1-CD2}{CD1}100$}[CD1 = 0.78, CD2 = 0.75][2]$\%)\\ 
20.0& 0.85& 0.77 ($\eval*{$-\frac{CD1-CD2}{CD1}100$}[CD1 = 0.85, CD2 =0.77][2]$\%)&0.51&0.53($\eval*{$-\frac{CD1-CD2}{CD1}100$}[CD1 = 0.51, CD2 = 0.53][2]$\%)\\ 
40.0& 0.53& 0.46 ($\eval*{$-\frac{CD1-CD2}{CD1}100$}[CD1 = 0.53, CD2 =0.46][2]$\%)&0.36&0.40($\eval*{$-\frac{CD1-CD2}{CD1}100$}[CD1 = 0.36, CD2 = 0.40][2]$\%)\\ 
 \hline
\end{tabu}
\end{center}
\caption{ Comparison of the predictions of the friction drag Eq.(\ref{eqfrictiondraag}), and pressure drag Eq.(\ref{eqformdrag}) with the results of Dennis and Walker \cite{dennis1971calculation}.}
\label{tablefrictionformdrag}
\end{table}
We will compare the predictions of the friction drag and the pressure drag from Eq.(\ref{eqfrictiondraag}), and Eq.(\ref{eqformdrag}) with the numerical results of Dennis and Walker \cite{dennis1971calculation}, the results of the comparison is shown in Table \ref{tablefrictionformdrag}\@. Our predictions are close to those of Dennis and Walker\cite{dennis1971calculation} for low and moderate $Re$, the largest deviation is about 13\%, which is acceptable for an approximation\@. We will further compare our results with the results of Achenbach \cite{achenbach1972experiments} for extremely high Reynolds numbers, as shown in Table \ref{tablefrictiontototaldrag}\@. We also included comparison with the results of Abraham \cite{abraham1970functional} equation:

\begin{equation}\label{eqfaridspher}
C_D = 0.29+\frac{24}{Re}+\frac{5.291}{\sqrt{Re}}
\end{equation} 
We specifically choose the specific semi-analytical equation because, in its core derivation, it uses the boundary layer theory\@. {We will assume that for the Abraham \cite{abraham1970functional} semi-analytical equation the $C_{D_{f}} = \dfrac{5.291}{\sqrt{Re}}$\@}. The comparison with the experimental results shows the great sensitivity of the results on the value of the constant  ($\delta$). Our results show a maximum difference of about 20\% near the drag criss\@.   At the same Reynolds number, the results of Abraham \cite{abraham1970functional} show a deviation of about 200\%, which is substantial as shown in Table \ref{tablefrictiontototaldrag}\@. { From the results of the compression in 
Table \ref{tablefrictiontototaldrag} we conclude that the value of 5.291  from the Abraham \cite{abraham1970functional} correlation does not correspond to a $\delta$ value related to the frictional drag\@.}{The comparison that we made for the frictional drag with different results from the literature is shown in Table \ref{tablefrictionformdrag}, and Table \ref{tablefrictiontototaldrag}, plus that the value of the $a_1$ is close to the prefactor value of the $Re^{-1/2}$ term in Churchill's \cite{churchill2013viscous} frictional drag formula for the sphere solidifies the idea that the coefficient $a_1$ may also have strong links to the {shear stress} around the sphere\@.} We will  show in the coming sections, that we found similar relation for $a_1$ coefficient of the logarithmic equation with frictional drag for the geometries of flat plate and cylinder, and normal flat plate\@. We emphasise that $\dfrac{1}{\sqrt{Re}}$ terms also appeared in the power-based drag coefficient equations that we derived in our previous work \cite{el2022logarithmic}\@. For example, the most accurate power-based equation that we derived (Eq.(8) Ref\cite{el2022logarithmic}) contains $Re^{-1/2}$ term, similar to the frictional drag, and its $\delta$ value is close to the value of the $a_1$  coefficient of the logarithmic equation for the case of  {a} sphere\@. The inverse square root Reynolds number term also appears in the power-based equation that we derived directly \cite{el2022logarithmic} from the experimental data of Brown and Lawler \cite{brown2003sphere}: 
\begin{equation}\label{eqexpspherebrown}
C_D =0.5+\frac{23.22}{Re}+\frac{2.76}{\sqrt{Re}}
\end{equation}   
Here the value of the coefficient of the square root term is 2.76, which is still close to the value of 3.28\@.  The reappearance of the {constant} $a_1$ coefficient value in different drag correlations of various mathematical structures derived from different data sources shows that the value of the $a_1$ coefficient plays an essential role in the evolution of the drag coefficient for the sphere\@. 

\begin{table}[t]
\begin{center}
\begin{tabu} to 1.25\textwidth { | X[c] | X[c] | X[c]| X[c]|}
\hline
$Re$& Ref\cite{achenbach1972experiments}$\dfrac{C_{D_{f}}}{{C_D}}$ (\%) &$\dfrac{C_{D_{f}}}{{C_D}}$ (\%)& Ref \cite{abraham1970functional} $\dfrac{C_{D_{f}}}{{C_D}}$(\%)\\
 \hline
 $8.9\times 10^{4}$& 2.46& 2.36 ($\eval*{$-\frac{CD1-CD2}{CD1}100$}[CD1 = 2.46, CD2 = 2.36][2]$\%)&5.79 ($\eval*{$-\frac{CD1-CD2}{CD1}100$}[CD1 = 2.46, CD2 =5.79][2]$\%)\\
$1.1\times 10^{5}$&1.92&2.08 ($\eval*{$-\frac{CD1-CD2}{CD1}100$}[CD1 = 1.92, CD2 = 2.08][2]$\%)&5.15 ($\eval*{$-\frac{CD1-CD2}{CD1}100$}[CD1 = 1.92, CD2 =5.15][2]$\%)\\
$1.5\times 10^{5}$& 1.52&1.78 ($\eval*{$-\frac{CD1-CD2}{CD1}100$}[CD1 = 1.52, CD2 = 1.78][2]$\%)&4.40 ($\eval*{$-\frac{CD1-CD2}{CD1}100$}[CD1 = 1.52, CD2 =4.4][2]$\%)\\ 
$1.9\times 10^{5}$& 1.34&1.61 ($\eval*{$-\frac{CD1-CD2}{CD1}100$}[CD1 = 1.34, CD2 =1.61][2]$\%)&3.95 ($\eval*{$-\frac{CD1-CD2}{CD1}100$}[CD1 = 1.34, CD2 =3.95][2]$\%)\\   
 \hline
\end{tabu}
\end{center}
\caption{ Comparison of the predictions of the ratio of between the friction drag and the total drag from our predictions, and the predictions of Abraham \cite{abraham1970functional}, with the experimental results of Achenbach\cite{achenbach1972experiments}.}
\label{tablefrictiontototaldrag}
\end{table}

\begin{figure}[h!]
\begin{center}
\includegraphics [scale=0.8, trim = 50 0 0 0,clip]{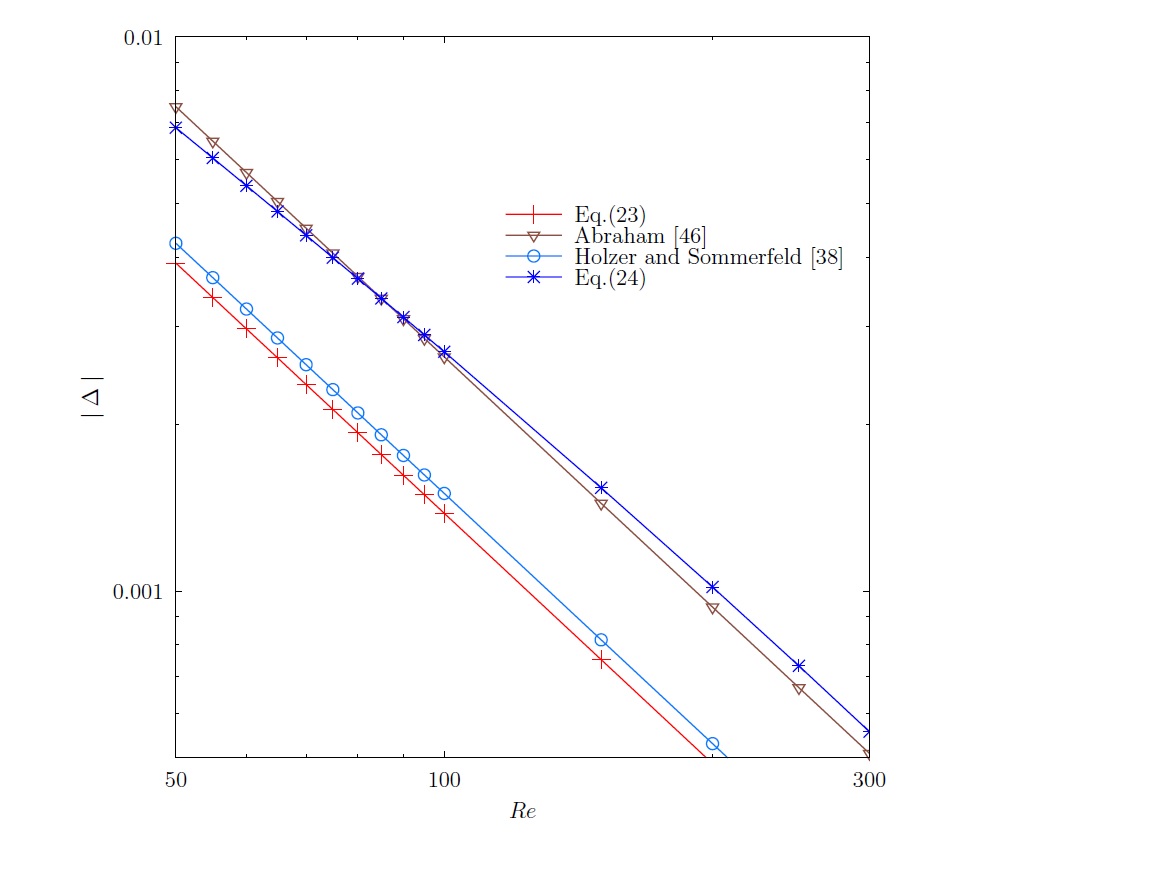}
\end{center}
\caption{Comparison  between the value of the derivative of the drag coefficient {(exclusion the Stokes term $\sim Re^{-1}$)} with respect to the $Re$ for different correlations,  for the case of a sphere.}
\label{Fig1shpherdiv}
\end{figure}

Another interesting property of the evolution of the drag coefficient that we believe that plays an essential role evolution of the drag coefficient is the  partial derivative of drag coefficient with respect to the Reynolds number for high $Re$\@. If we differentiate Eq.(\ref{eqhighRe})  we get the following equation :
\begin{equation}\label{eqhighRe2}
\Delta =\frac{dC_{D_{h}}}{dRe} =\frac{a_3+2a_4\log(Re)+4a_5\log^3(Re)}{Re}
\end{equation} 
Where $C_{D{s}}$ is the Stokes drag\@. We mention that when we will take the derivative of the drag coefficient, we will omit any terms containing the Stokes term ($\sim Re^{-1}$)\@. We will plot the value of $\Delta$   for a number of correlations and models available for a sphere as shown in Figure \ref{Fig1shpherdiv}\@. The values of logarithmic-based Eq.(\ref{eqhighRe2}) are close to the results of Abraham \cite{abadi2016tensorflow}, even though the two equations have different mathematical structures\@.  The trend of the two previous equations shows that the derivative values obtained from Eq.(\ref{eqhighRe2}) do not belong to a specific source of the training data\@. On the contrary, it shows that derivative values obtained from Eq.(\ref{eqhighRe2}) represent the actual general behaviour of the change of the drag coefficient with the Reynolds number for {a} sphere\@. The derivative values obtained from  Eq.(\ref{eqexpspherebrown}), and Holzer and Sommerfeld \cite{holzer2008new} correlation, differ significantly from those obtained by the logarithmic equation or that of Abraham \cite{abraham1970functional}\@. Despite all of these equations having similar accuracy in predicting the drag coefficient {at the specific $Re$ regime that are tested}, the deviation in the $\Delta$ values indicates that the coefficients in the drag equations must be carefully adjusted to produce the $\Delta$ values obtained from Eq.(\ref{eqhighRe2})\@. For example,  {the correlations  by} Abraham \cite{abraham1970functional}, Holzer and Sommerfeld \cite{holzer2008new}, and Eq.(\ref{eqexpspherebrown}) have the same mathematical structure. Still, the difference in the coefficient values produces a significant deviation in $\Delta$ values of the different correlations, and models\@. In the coming sections, we will demonstrate that the values of $\Delta$ obtained from Eq.(\ref{eqhighRe2}) are universal and apply  not only spheres but also to other non-spherical geometries\@. The values of $\Delta$ are indicators of the accuracy of drag coefficient equations, particularly in the high Reynolds number regime\@.

\section{Results}
 
 We will first confirm our assumption about the universality of the drag coefficient's rate of change with the Reynolds number\@. Then, we will demonstrate how our theory can be used to determine the drag coefficient for different shapes, including flat plates and irregularly shaped particles\@.

\subsection{The variation of $\Delta$ for different geometries}
We will demonstrate in the current subsection that at moderate and high Reynolds numbers, regardless of the shape of the object and their orientation to the flow direction, the value of $\Delta$ is the same\@. The value of  $\Delta$ indicates how the drag force changes with the changes in the flow field around the object\@.   {To the best of  our knowledge}, no one investigated a similar parameter such $\Delta$ before\@. We used an automatic differentiation Python-based library to obtain the derivative of different correlations and predictive models\@. In the process of estimating the derivatives of the different correlations, we omitted the Stokes term $Re^{-1}$ when we calculated the derivatives of the different correlations \@. { Only for the case of the experimental results of Wieselsbergerg \cite{wieselsberger1922further} for the case of the cylinder, we could not omit the Stokes drag from the data, because the analytical solution of Lamb's for the case of Stokes flow around the cylinder diverges for $Re > 10$. However, this will not affect the accuracy of the results since the Reynolds numbers considered are at the high end.}   \\

\begin{figure}[h!]
\begin{center}
\includegraphics [scale=0.8, trim = 50 0 0 0,clip]{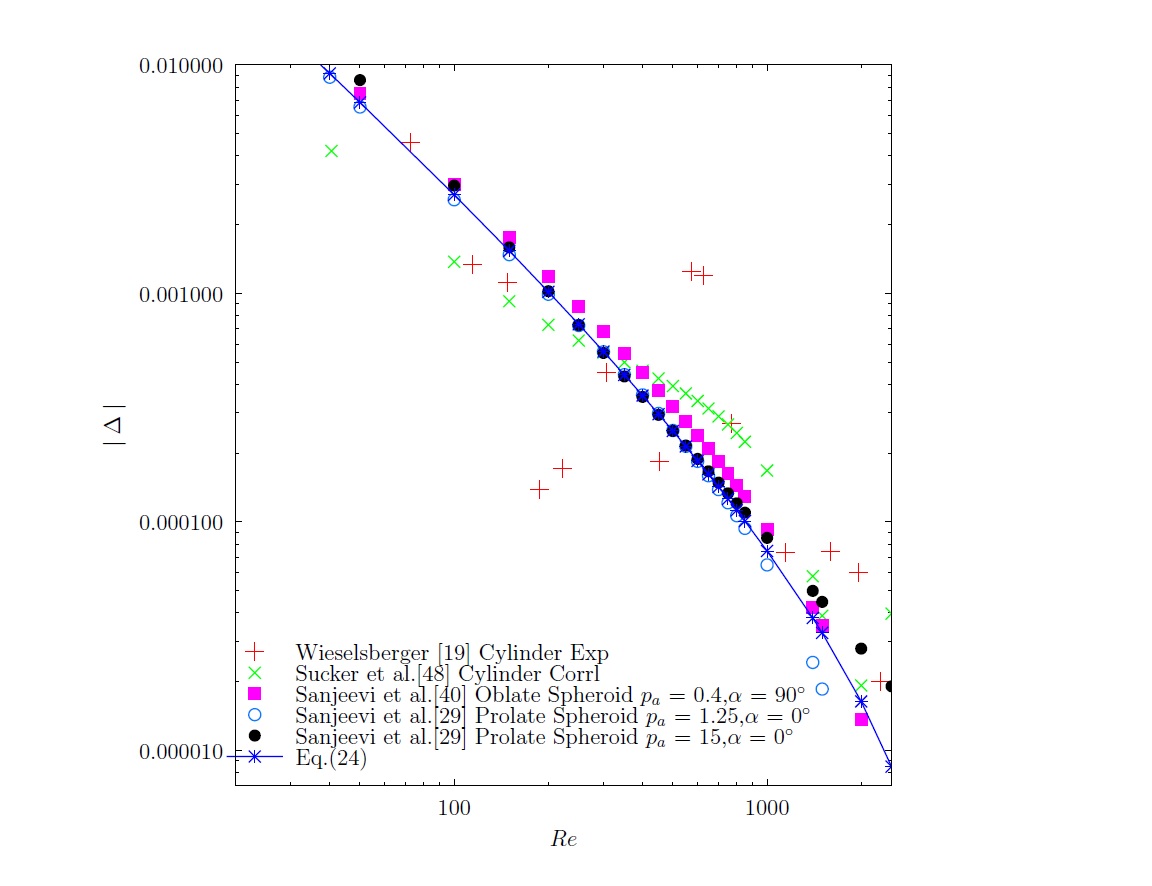}
\end{center}
\caption{Comparison of the $\Delta$ value obtained from Eq.(\ref{eqhighRe2}) with $\Delta$ values  from different sources available in the literature\@.}
\label{Figgeneralderavitive}
\end{figure}

To assess our results of $\Delta$, we will examine the variation of $\Delta$ from Eq.(\ref{eqhighRe2}) with other correlations and experimental data found in the literature 
 Figure \ref{Figgeneralderavitive}\@. Our selection includes the oblate spheroid geometry from Sanjeevi et al.'s numerical results \cite{sanjeevi2018drag}, as well as two prolate spheroid geometries from Sanjeevi et al. \cite{sanjeevi2022accurate}\@. These geometries were chosen due to their capacity to provide information on the $C_D$ variation with Reynolds numbers up to 3000, which would enable us to test the $\Delta$ value variation for a broad Reynolds number range\@. Furthermore, experimental-based results will be utilized to compare our findings\@. We have opted to use Wieselsberger's results \cite{wieselsberger1922further} for the scenario of an infinite cylinder with right angles to the flow,and the  correlation of  Sucker et al.\cite{sucker1975fluiddynamik} for the same geometry\@. \\
From Figure \ref{Figgeneralderavitive}, it appears that the values of $\Delta$ are converging toward the values of $\Delta$ from Eq.(\ref{eqhighRe2}), regardless of the source of the data, the type, size, and orientation of the bluff body\@. This observation is consistent with our theory, and leads us to conclude that a single value for $C_D$ at high $Re$ can be used to determine the value of $a_1$ in Eq.(\ref{eqloggeneral}) for flows beyond that of a sphere\@. The deviations we observe in the $\Delta$ values from Wieselsberger's \cite{wieselsberger1922further} experimental results can be attributed to the fact that the digitally extracted data is not uniformly distributed, resulting in inaccurate gradient calculations. However, overall, the values of $\Delta$ for Wieselsberger's \cite{wieselsberger1922further} experiments are close to the $\Delta$ values calculated by Eq.(\ref{eqhighRe2}), even for high $Re$ where $\Delta$ values are on the order of $10^{-4}$\@. The \(\Delta\) results from Eq.(\ref{eqhighRe}) closely align with those from the correlation by Sucker et al. \cite{sucker1975fluiddynamik}, which is based on Wieselsberger's data \cite{wieselsberger1922further}, supporting our previous claim. The convergence of our \(\Delta\) results with those of Wieselsberger and Sucker et al. \cite{sucker1975fluiddynamik} demonstrates that both 2D and 3D flows respond similarly to changes in the flow field.

\subsection{Drag coefficient}
In this subsection we will deploy Eq.(\ref{eqloggeneral}) to different bluff body geometries to test our theory\@. 

\subsubsection{Horizontal flat plate}
Flat plate geometry plays a vital role in aerodynamics as it serves as a benchmark for testing fluid mechanics theories. The boundary layer theory solution by Blasius and Prandtl\cite{blasius1907boundarylayers, Prandtl1904} applies to a flat plate with flow velocity parallel to its main side\@.  It is worth noting that for flat plate geometry, only the skin friction drag contributes to the total drag force, horizontal flat plates are not considered as bluff body, but a stream line object\@.  we can evaluate the validity of our theory  that the coefficient $a_1$ is linked to the shear stress, since the for the case of the flat plate we have an analytical solution for the frictional   drag\@. We will compare our results to those obtained by Melnik and Chow \cite{melnik1975asymptotic} using triple deck boundary layer theory\@. Specifically, their analytical solution for a single wetted side finite flat plate is the following:
\begin{equation}\label{eqflat1}
C_{D_{f}} = \frac{1.328}{\sqrt{Re}}+\frac{2.67}{Re^{7/8}}
\end{equation} 
While, for the double wetted side finite flat, the drag coefficient is given by the following equation:
\begin{equation}\label{eqflat2}
C_{D_{f}} = \frac{2.656}{\sqrt{Re}}+\frac{5.34}{Re^{7/8}}
\end{equation}

\begin{figure}[H]
\begin{center}
\includegraphics [scale=0.8, trim = 50 0 0 0,clip]{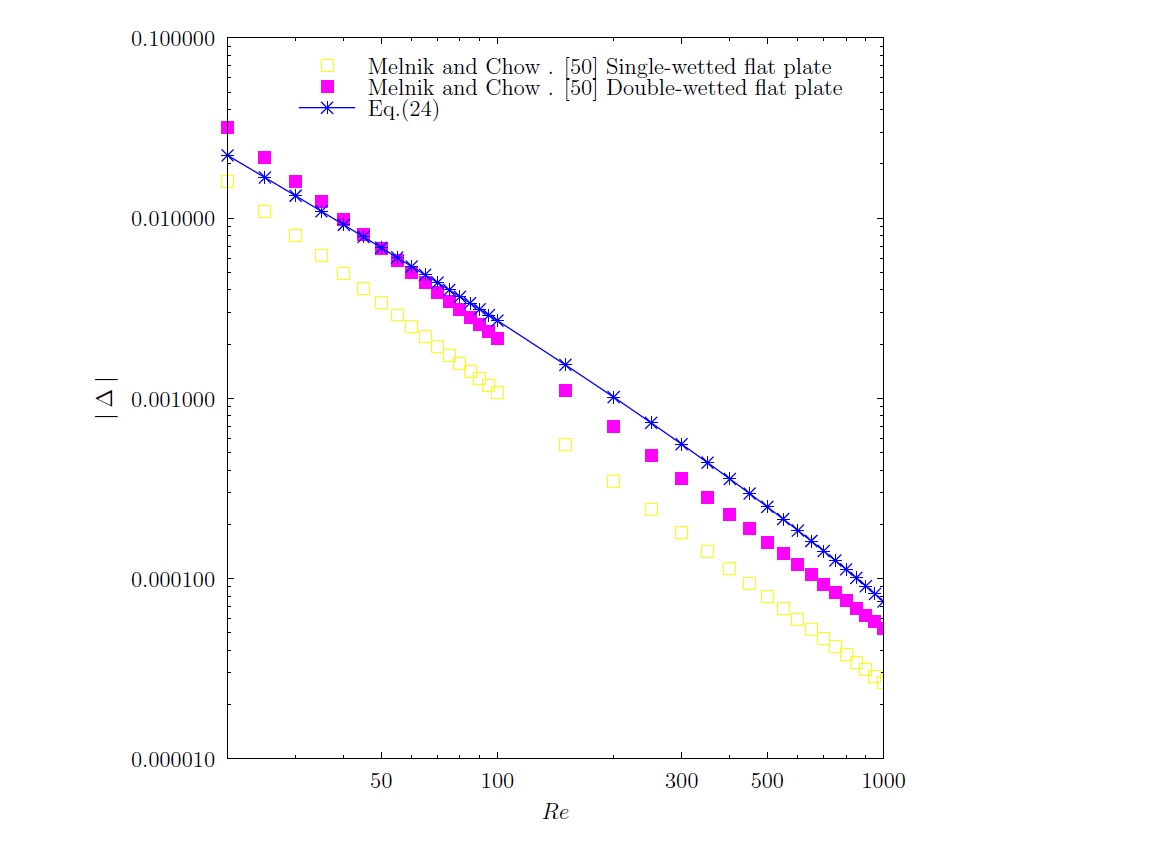}
\end{center}
\caption{Comparison of $\Delta$ values from  Melnik and  Chow \cite{melnik1975asymptotic} analytical solution for the case of the flat plate for the single, and double wetted cases  with the logarithmic based equation Eq.(\ref{eqhighRe2}) for different Reynolds numbers\@. }
\label{Fig1plate}
\end{figure}

{In Melnik and Chow \cite{melnik1975asymptotic} model, the Reynolds number is defined based on the plate length, and the drag coefficient is based on the plate's surface area\@.}  We will compare the $\Delta$ values from the above equations with the logarithmic based equation Eq.(\ref{eqhighRe2}) to gauge the validity of our assumptions, as illustrated in Figure \ref{Fig1plate}\@. A close examination of Figure \ref{Fig1plate} reveals two intriguing insights\@. Firstly, the $\Delta$ values for the single-wetted flat plate deviate significantly from those of the logarithmic-based Eq.(\ref{eqhighRe2})\@. In contrast, the $\Delta$ values for the double-wetted flat plate are relatively closer to the values of Eq.(\ref{eqhighRe2})\@. Secondly, the $\Delta$ values for the double wetted flat plate tend to converge with the results of Eq.(\ref{eqhighRe2}) at moderate Reynolds numbers ($Re$ = 20 to 100)\@.{ In the range where the $\Delta$ values of the logarithmic and double wetted flat plate equations converge, Eq.(\ref{eqloggeneral}) and Eq.(\ref{eqflat2}) will be identical functions, this is a intriguing  observation because the data that used to derive Eq.(\ref{eqloggeneral}) are coming from a bluff body that we know that generate different flow field than the stream-lined bodies\@. The logarithmic-based equation (Eq.(\ref{eqloggeneral})) will inherit the definitions of $Re$ and $C_D$ from Eq.(\ref{eqflat2}) as our theory describe\@.}  \\ 

To utilize the logarithmic-based equation (Eq.(\ref{eqloggeneral})) for a flat plate, it is necessary to determine the value of the coefficient \( a_1 \). Given that the logarithmic equation (Eq.(\ref{eqloggeneral})) and Eq.(\ref{eqflat2}) share the same derivative within the narrow Reynolds number range of 20 to 100, we can evaluate \( a_1 \) by equating the \( C_D \) value from Eq.(\ref{eqflat2}) at \( Re = 50 \) to Eq.(\ref{eqloggeneral}). Since we are not dealing with the creeping flow regime, the coefficient \( K \) is set to zero. Consequently, the coefficient \( a_1 \) is determined to be 2.795.

The $a_1$ value obtained from the logarithmic equation  is  about $\eval*{-100\frac{2.656-2.795}{2.656}}[2]$\% different from the actual value of  Blasius'\cite{blasius1908grenzschichten} 2.656, indicating a close relationship between the $a_1$ coefficient  and the shear stress at the wall obtained from the  boundary layer theory\@. The value of $a_1$ emerged without using any elements of the boundary layer theory, suggesting that Eq.(\ref{eqloggeneral}) represents a general model for the drag coefficient, at same time it also confirms the existence of the non-slip boundary conditions  \@. \\

\begin{figure}
\begin{center}
\includegraphics [scale=0.8, trim = 0 0 0 0,clip]{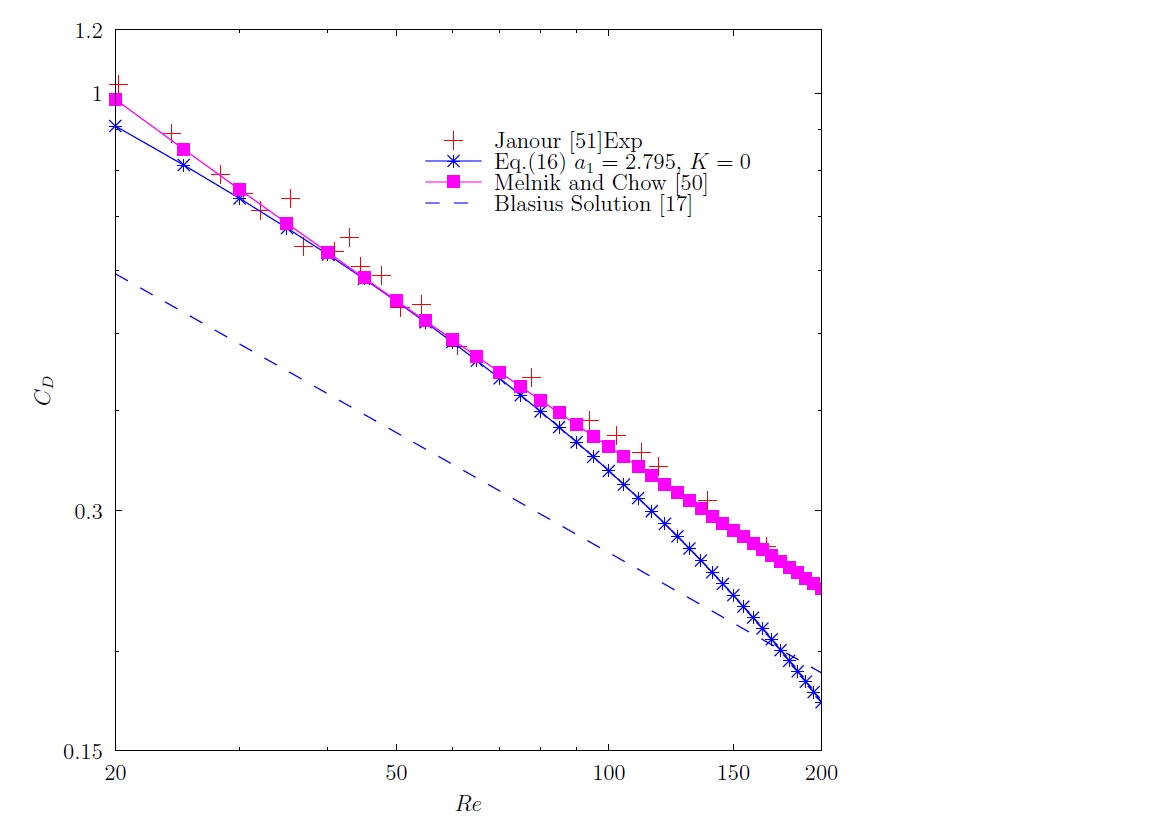}
\end{center}
\caption{Comparison of $C_D$ values predicted by the logarithmic equation, and the boundary layer theory, and experiments for the case of horizontal flat plate\@. }
\label{Fig2plate}
\end{figure}

 The logarithmic equation Eq.(\ref{eqloggeneral}) is compared to the analytical solution from the boundary layer theory and the experimental data of Janour\cite{janour1951resistance} as shown in Figure  \ref{Fig2plate}\@. The comparison shows that the logarithmic equation closely follows the analytical solution in the Reynolds number range of 20 to 100\@. This is expected as the logarithmic equation was derived from the bluff body data of a sphere, where at high Reynolds numbers, form drag dominates over skin friction drag\@.  \\


\subsubsection{Normal flat plate}
The flow over a flat plate normal to its direction is the simplest case of a bluff body. At the edges of the plate, the flow separates. In the middle of the nineteenth century, Kirchhoff and Rayleigh \cite{lisoski1993nominally} calculated the drag force for the normal flat plate. They assumed that the shear layers leave the flat plate at the edges and that the dynamic pressure equals the free stream pressure, and that the total drag force is due to the dynamic pressure. They derived the following formula for the drag coefficient:
\begin{equation}\label{eqnormal_flat_plate}
C_D = \frac{4\pi \sin{\alpha}}{4+\pi \sin{\alpha} }
\end{equation} 

The drag coefficient is based on the frontal area of the plate\@. The angle of attack ($\alpha$) is 90 degrees for a normal flat plate case. Eq. (\ref{eqnormal_flat_plate}) gives a drag coefficient of 0.88, which differs by 
\eval*{$(2.0-0.88)/(2.0)\times 100 $}\% from the  measured value of 2.0 from experiments\@. For the case of the normal flat plate, the drag coefficient data are sparse, and even finding correlations is difficult\@.  We will not use any experimental data or data from correlations to calculate the value of the $a_1$ coefficient,  we will rely on the boundary layer theory instated \@.

The normal flat plate case is a special case within the family of Falkner-Skan boundary layer flows, whose numerical solution is well-known. The general governing equation for the shear stress, derived from the boundary layer equations, is given by the following equation \cite{kundu2015fluid}:
\begin{equation}
\tau = \frac{f^{''}(0)\rho v^2_{\infty}} {\sqrt{Re_x}}
\end{equation}
Where $Re_x$ is the local Reynolds number, the value of $f''(0)$ for the case of a normal flat plate, obtained from the solution of the Falkner-Skan boundary layer equations, is 1.233 \cite{bejan2013convection}. The total frictional drag can be calculated as follows: 
\begin{equation}
F_{D_{f}} = \int_{0}^{L} \tau \,dx = \frac{2.466 \rho v^2_{\infty} L}{\sqrt{Re}}
\end{equation}  
 $L$ is the length of the plate, the frictional drag coefficient is the following: 
\begin{equation}
C_{D{_f}} = \frac{F_{D_{f}}}{\left(1/2\right)\rho v^2_{\infty} L } = \frac{4.932}{\sqrt{Re}}
\end{equation}

following our theory will assign the value of 4.932 for the $a_1$ coefficient of Eq.(\ref{eqloggeneral}) for the case of the normal flat plate\@. 
\begin{table}[h!]
\begin{center}
\begin{tabu} to 1.25\textwidth { | X[c] | X[c] | X[c]| X[c]|}
\hline
&  $Re$&$C_D$&$C_D$ Eq.(\ref{eqloggeneral}) $a_1$ = 4.932, $K$ =0\\
 \hline
 Ref \cite{najjar1998low}(DNS)&250& 2.36& 2.26 ($\eval*{$\frac{CD1-CD2}{CD1}100$}[CD1 = 2.36, CD2 = 2.266][2]$\%)\\
 
Ref\cite{narasimhamurthy2009numerical}(DNS)&$1000$&2.31 & 2.09 ($\eval*{$\frac{CD1-CD2}{CD1}100$}[CD1 = 2.31, CD2 =2.09][2]$\%)\\

Ref \cite{lisoski1993nominally}(Exp)& $5000$&2.0 &2.04 ($\eval*{$-\frac{CD1-CD2}{CD1}100$}[CD1 = 2.0, CD2 =2.04][2]$\%)\\ 

Ref\cite{fage1927flow}(Exp)&$1.5\times 10^{5}$& 2.13 &2.11($\eval*{$\frac{CD1-CD2}{CD1}100$}[CD1 = 2.13, CD2 =2.11][2]$\%)\\  

Ref\cite{tian2014large}(LES)&$1.5\times 10^{5}$& 2.2 &2.11($\eval*{$\frac{CD1-CD2}{CD1}100$}[CD1 = 2.2, CD2 =2.11][2]$\%)\\  
 \hline
 
\end{tabu}
\end{center}
\caption{ Relative error between the predictive $C_D$ coefficient from the 
logarithmic equations, and those form different sources from the literature 
 for different Reynolds numbers for the geometry of normal flat plate\@.}
\label{tabledrag_normal_flat_plate}
\end{table}
We will compare our logarithmic equation with the existing results in the literature, as shown in Table \ref{tabledrag_normal_flat_plate}. Since the drag coefficient for the normal flat plate case is predominantly influenced by form drag (pressure drag), we expect the logarithmic equation to hold at high Reynolds numbers. The results presented in Table \ref{tabledrag_normal_flat_plate} confirm this expectation, with the relative error of the logarithmic equation being below 5\% in the majority of cases tested across a wide range of Reynolds numbers, ranging from $250$ to $10^5$. Moreover, the logarithmic equation 
Eq.(\ref{eqloggeneral}) with $a_1 = 4.932$ and $K = 0$ outperforms the only analytical model available for the normal flat plate case 
Eq.(\ref{eqnormal_flat_plate}). One intriguing aspect of the normal flat plate case is the strong relationship between the $a_1$ coefficient value and the shear stress value derived from boundary layer theory for the same geometry.
\subsubsection{Infinite cylinder} 
The study of fluid flow over infinite cylinders has been a subject of immense  investigation, primarily due to the proliferation of complex, unsteady flow patterns associated with this particular bluff body geometry\@. Simulations using large eddy turbulence models have shown strong dependence of the flow parameters on factors such as grid resolution, time steps, and aspect ratio\@. The system's dimensionality also plays a crucial role, with two-dimensional simulations only reliable up to a Reynolds number of 250 \cite{kravchenko2000numerical}\@
. Beyond that, three-dimensional flow simulations are necessary to adequately capture complex flow structures and obtain accurate results for parameters such as the drag coefficient. Machine learning methods, primarily using different types of neural networks, have been applied to the flow over cylinders and trained on extensive flow field data. However, these studies have yet to accurately predict flow parameters, such as the drag coefficient at high Reynolds numbers \cite{lee2019data, raissi2019deep}
. For instance, despite including $Re$ = 1000 flow data in their training set, Vlachas et al. \cite{vlachas2022multiscale} only achieved a 15\% relative error in their prediction of the drag coefficient at that specific Reynolds number\@. \\

In our current investigation, we will examine the geometry of an infinite cylinder\@. The configuration is similar to the experiments conducted by Wieselsberger \cite{wieselsberger1922further}, where the flow velocity flows perpendicular to the cylinder's axis\@. In the experimental data of Wieselsberger \cite{wieselsberger1922further}, the $Re$ is defined on the bases of the diameter of the cylinder, and the $C_D$ is defined based on the frontal area\@. With this specific geometry in mind, Thom \cite{thom1928boundary} utilized boundary layer theory to calculate the skin friction coefficient for the front part of the cylinder up to $60^{\circ}$ degrees from the forward stagnation point. Through his calculations, he derived the following relation:
\begin{equation}\label{eqcylinederThom1}
C_{D_{f}} = \frac{3.84}{\sqrt{Re}}
\end{equation}
For the rear of the cylinder 
Thom \cite{thom1928boundary} calculated the skin friction coefficient by slightly modifying the coefficient presented in Eq.(\ref{eqcylinederThom1}) based on experimental results\@. As a result, he proposed the following equation as a comprehensive representation of the skin friction for the entire surface of the cylinder:
 \begin{equation}\label{eqcylinederThom2}
C_{D_{f}} = \frac{4.0}{\sqrt{Re}}
\end{equation} 
Takami and Keller \cite{takami1969steady} derived the necessary constants for the Imai \cite{imai1957university} semi-theoretical equation, which is based on the Kirchhoff type of flow, from their numerical findings\@. The Imai \cite{imai1957university} semi-theoretical equation takes the following form:
\begin{equation}
C_D  = \sqrt{0.5+\frac{3.547}{\sqrt{Re}}}
\end{equation}
Munson et al. \cite{gerhart2016munson} utilized data from experiments to construct a semi-empirical equation specifically designed for high Reynolds numbers\@. The resulting equation is as follows:
\begin{equation}
C_D = 1.17+ \frac{5.93}{\sqrt{Re}}
\end{equation}  
To apply the logarithmic-based equation Eq.(\ref{eqloggeneral}) to the cylinder case, it is necessary to determine the value of the coefficient $a_1$\@. We will assign a value of zero to the coefficient of $K$ for the Stokes term, since we our only interested for high $Re$ regime\@. We will calculate $a_1$  utilizing  two methods\@.
The first approach is to calculate the $a_1$ coefficient based on the value of the $C_D$ at $Re$ = 114.58 from the Wieselsberger experiments \cite{wieselsberger1922further}, $C_D$ is equal to 1.37. Using this information, we obtain $a_1$ = 3.865\@.
The second method based on our theory which links the value of the $a_1$ with the frictional 
drag coefficient\@. For this reason we will assign a value of 4.0 to $a_1$ coefficient following the estimations of Thom \cite{thom1928boundary} for frictional drag coefficient for the cylinder Eq.(\ref{eqcylinederThom2})\@. 

\begin{figure}[H]
\begin{center}
\includegraphics [scale=0.8, trim = 0 0 0 0,clip]{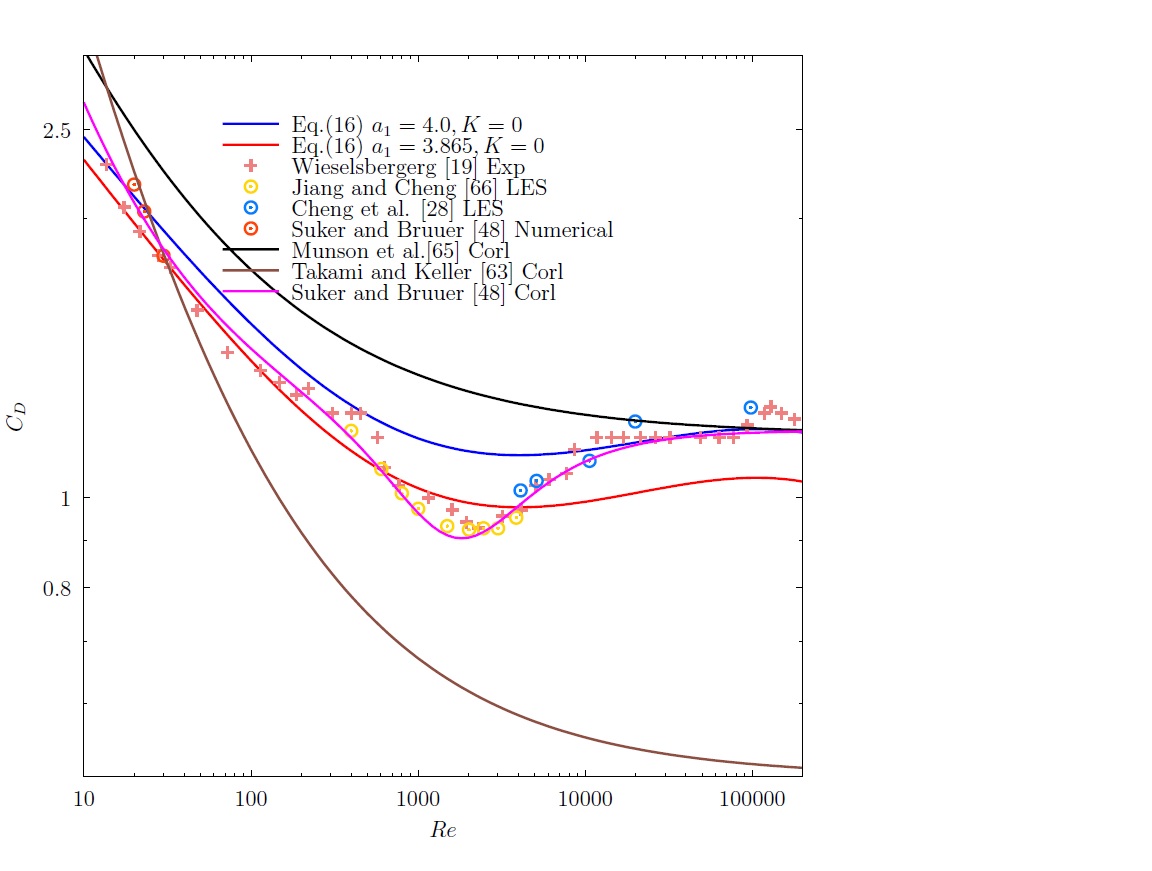}
\end{center}
\caption{Comparison of $C_D$ values predicted by the logarithmic equations, other predictive models, experiments, and numerical simulations for the case of infinite cylinder\@. }
\label{Fig1cylinder}
\end{figure}

The value  of  the $a_1$ coefficient calculated using the  Wieselsberger's\cite{wieselsberger1922further} data  
 differs only by  $\eval*{-100\frac{3.84-3.86}{3.84}}[2]$\%  from the value calculated by Thom \cite{thom1928boundary} for the front part of the cylinder using boundary layer theory, affirming our hypothesis that the $a_1$ coefficient is linked to the cylinder's skin friction\@. The evolution of the drag coefficient with Reynolds number is shown in Figure \ref{Fig1cylinder}\@. The predictive models of Takami and Keller \cite{takami1969steady} and Munson et al. \cite{gerhart2016munson} were unable to produce accurate predictions, but the logarithmic equations were found to be more accurate predicting the  Wieselsberger's\cite{wieselsberger1922further}
 experimental results\@. The logarithmic equations were the first mathematical models of the drag coefficient that were able to accurately predict the results of Wieselsberger's \cite{wieselsberger1922further}
experiments, which are over a century old. Eq.(\ref{eqloggeneral}) with $a_1$ = 3.865, and $K$ = 0 is capable of accurately predicting the experimental results  for Reynolds numbers up to 3000, while Eq.(\ref{eqloggeneral}) with $a_1$ = 4.0, and $K$ = 0 is effective for Reynolds numbers higher than 
$10^{4}$\@. An intriguing observation is that Eq.(\ref{eqloggeneral}) with $a_1$ = 3.865, and $K$ = 0 can predict the minimum point in the drag coefficient curve, which occurs around Reynolds number of 3000, and how close its predictions are to the LES simulations conducted by Jiang and Cheng\cite{jiang2021large}\@. The logarithmic equation Eq.(\ref{eqloggeneral}) with $a_1$ = 3.865, and $K$ = 0 accurately predicts the numerical results of Suker and Bruuer \cite{sucker1975fluiddynamik} for low Reynolds numbers, and Eq.(\ref{eqloggeneral}) with $a_1$ = 4.0, and $K$ = 0 predicts the LES results of Cheng et al. \cite{cheng2017large}\@. The overall performance of the logarithmic equations supports our belief that the rate of change of the drag coefficient with the  Reynolds number is given by Eq.(\ref{eqhighRe2}), which underscores its universality\@. We aim to develop our theory further so that we can arrive at a single equation that can describe the entirety of the cylinder data\@. Table  \ref{table1cylinder} reveals that the accuracy of the logarithmic equation Eq.(\ref{eqloggeneral}) with $a_1$ = 3.865, and $K$ = 0 is satisfactory in the lower and medium Reynolds number regimes\@. Even at a high Reynolds number of $10^5$, the deviation from accuracy is only 14\% which is still acceptable. Conversely, the logarithmic equation Eq.(\ref{eqloggeneral}) with $a_1$ = 4.0, and $K$ = 0 performs exceptionally well in the high Reynolds number range\@.

\begin{figure}[H]
\begin{center}
\includegraphics [scale=0.8, trim = 0 0 0 0,clip]{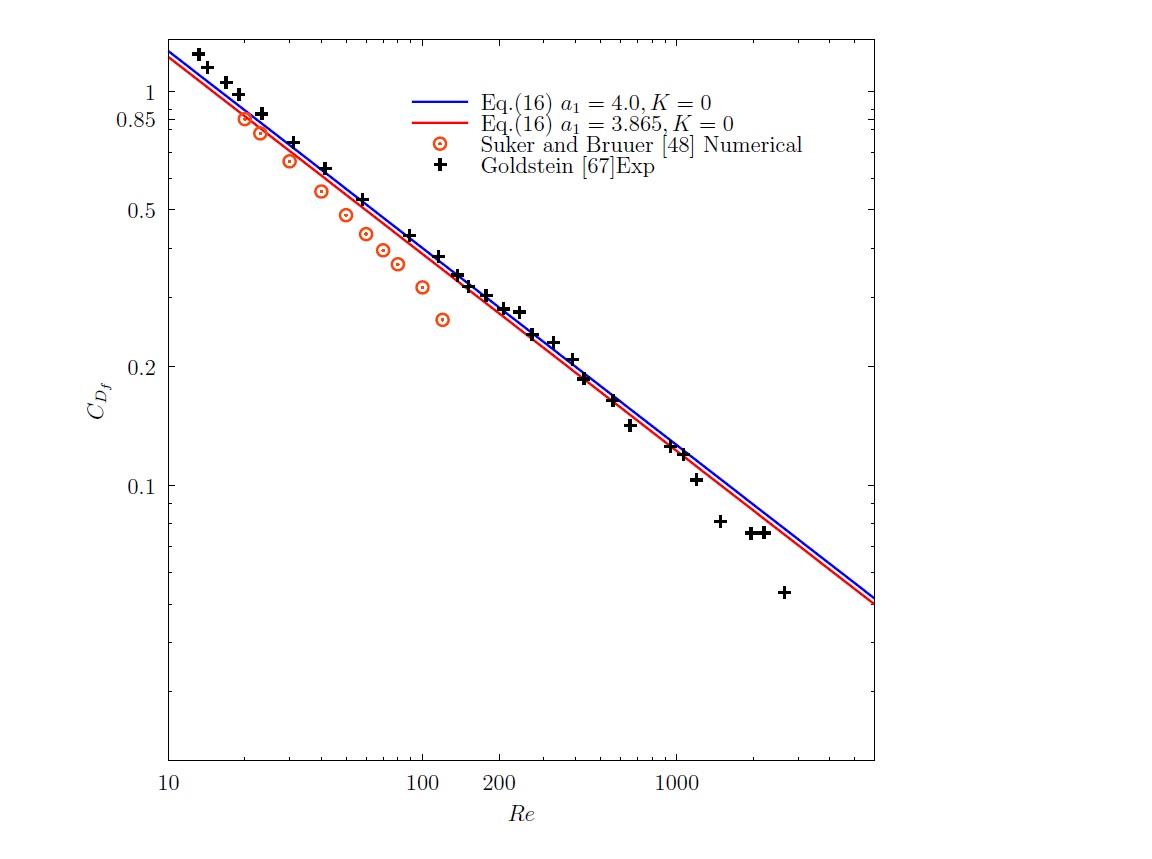}
\end{center}
\caption{Comparison of $C_{D_{f}}$ (skin friction drag coefficient)  values predicted by the logarithmic equations,  experiments, and numerical simulations for the case of infinite cylinder\@. }
\label{Fig2cylinder}
\end{figure}

\begin{table}[H]
\begin{center}
\begin{tabu} to 1.0\textwidth { | X[c] | X[c] | X[c] |X[c] |}
 \hline
  $Re$&Wieselsberger\cite{wieselsberger1922further} & Eq.(\ref{eqloggeneral})$a_1$ = 3.865, $K$ = 0(\%)& Eq.(\ref{eqloggeneral})$a_1$ = 4.0, $K$ = 0(\%) \\ 
 \hline
 47.61
 &1.59 & 1.63($\eval*{$-\frac{CD1-CD2}{CD1}100$}[CD1 = 1.59, CD2 = 1.63][2]$\%) &1.77($\eval*{$-\frac{CD1-CD2}{CD1}100$}[CD1 = 1.59, CD2 = 1.77][2]$\%)\\
  $\eval*{$\frac{573.28 }{10^2}$}[2]$ $\times 10^2$

&1.16
 & 1.07($\eval*{$-\frac{CD1-CD2}{CD1}100$}[CD1 = 1.16, CD2 = 1.07][2]$\%) &1.21($\eval*{$-\frac{CD1-CD2}{CD1}100$}[CD1 =  1.07, CD2 =1.21][2]$\%)\\
 
 $\eval*{$\frac{2302.77 }{10^3}$}[2]$ $\times 10^3$
&0.92

 & 0.98($\eval*{$-\frac{CD1-CD2}{CD1}100$}[CD1 = 0.92, CD2 = 0.98][2]$\%) &1.11($\eval*{$-\frac{CD1-CD2}{CD1}100$}[CD1 = 0.92, CD2 =1.11][2]$\%)\\ 
 
 $\eval*{$\frac{48888 }{10^4}$}[2]$ $\times 10^4$
&1.16
 & 1.04($\eval*{$-\frac{CD1-CD2}{CD1}100$}[CD1 = 1.16, CD2 = 1.04][2]$\%) &1.17($\eval*{$-\frac{CD1-CD2}{CD1}100$}[CD1 = 1.16, CD2 =1.17][2]$\%)\\
  
  $\eval*{$\frac{179207 }{10^5}$}[2]$ $\times 10^5$

&1.21
 & 1.04($\eval*{$-\frac{CD1-CD2}{CD1}100$}[CD1 = 1.21, CD2 = 1.04][2]$\%) &1.17($\eval*{$-\frac{CD1-CD2}{CD1}100$}[CD1 = 1.21, CD2 =1.17][2]$\%)\\ 
\hline
\end{tabu}
\end{center} 
\caption{ Relative error between the predictive $C_D$ coefficient from the logarithmic equations , and the experiments of Wieselsberger\cite{wieselsberger1922further} for different Reynolds numbers  for the geometry of infinite  cylinder\@.  } 

\label{table1cylinder}
\end{table} 
Figure \ref{Fig2cylinder} shows the variation of the skin friction drag coefficient for infinite cylinder\@. The skin friction for logarithmic equations are the following $\dfrac{3.865}{\sqrt{Re}}$, and $\dfrac{4.0}{\sqrt{Re}}$ receptively\@. 
It can be seen from Figure \ref{Fig2cylinder} that both logarithmic equations provide close predictions of the skin friction drag coefficient compared to numerical results obtained by Sucker and Bruuer \cite{sucker1975fluiddynamik} at low Reynolds numbers\@. The predictions also agree with the experimental results by Goldstein \cite{goldstein1938modern} for both low and high Reynolds numbers\@. \\

\begin{figure}[H]
\begin{center}
\includegraphics [scale=0.8, trim = 0 0 0 0,clip]{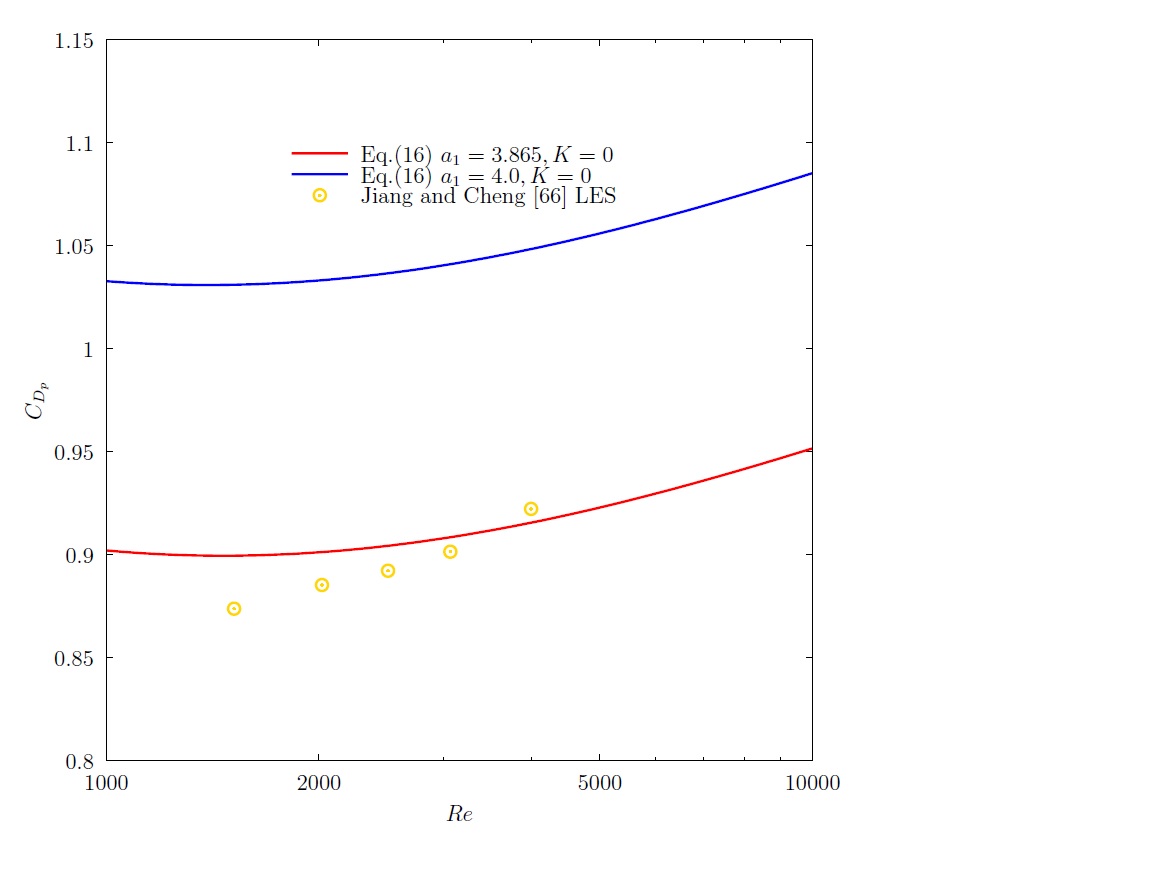}
\end{center}
\caption{Comparison of $C_{D_{p}}$ (form drag coefficient)  values predicted by the logarithmic equations, and numerical simulations for the case of infinite cylinder\@.  }
\label{Fig4cylinder} 
\end{figure}

In comparing the form drag coefficient $C_{D_{p}}$, the results from logarithmic equations as depicted in Figure \ref{Fig4cylinder} are analyzed against the Large-Eddy Simulations (LES) carried out by Jiang and Cheng\cite{jiang2021large}\@. Interestingly both equations capture  the increasing trend of the form drag coefficient with the Reynolds number as described by the results of Jiang and Cheng\cite{jiang2021large}\@. It is noted that the logarithmic equation (Eq.(\ref{eqloggeneral}) with $a_1$ = 3.865, and $K$ = 0) exhibits an excellent agreement with the simulation results of Jiang and Cheng\cite{jiang2021large} across a broad range of Reynolds numbers, with an average relative difference of merely 1.5\% as depicted in Table \ref{table2cylinder}\@. However, a higher deviation is observed in the logarithmic equation (Eq.(\ref{eqloggeneral}) with $a_1$ = 4.0, and $K$ = 0), with an average relative difference of approximately 16\%.  
\begin{table}[H]
\begin{center}
\begin{tabu} to 1.0\textwidth { | X[c] | X[c] | X[c] |X[c] |}
 \hline
  $Re$&Jiang and Cheng\cite{jiang2021large} &  Eq.(\ref{eqloggeneral})$a_1$ = 3.865, $K$ = 0(\%)& Eq.(\ref{eqloggeneral})$a_1$ =4.0, $K$ = 0(\%) \\ 
 \hline
  $\eval*{$\frac{1.516557829568893112 *10^3 }{10^3}$}[2]$ $\times 10^3$

 &0.873 & 0.899($\eval*{$-\frac{CD1-CD2}{CD1}100$}[CD1 = 0.873, CD2 = 0.899][2]$\%) &1.031($\eval*{$-\frac{CD1-CD2}{CD1}100$}[CD1 = 0.873, CD2 = 1.031][2]$\%)\\
 $\eval*{$\frac{2.020521197191781994 *10^3 }{10^3}$}[2]$ $\times 10^3$

&
0.885 & 0.901($\eval*{$-\frac{CD1-CD2}{CD1}100$}[CD1 = 0.885, CD2 = 0.901][2]$\%) &1.033($\eval*{$-\frac{CD1-CD2}{CD1}100$}[CD1 =  0.885, CD2 =1.033][2]$\%)\\
 
 $\eval*{$\frac{2.505542495503466398 *10^3 }{10^3}$}[2]$ $\times 10^3$

&0.892

 & 0.904($\eval*{$-\frac{CD1-CD2}{CD1}100$}[CD1 = 0.892, CD2 = 0.904][2]$\%) &1.036($\eval*{$-\frac{CD1-CD2}{CD1}100$}[CD1 = 0.892, CD2 =1.036][2]$\%)\\ 
 
$\eval*{$\frac{3.994673144433314064 *10^3 }{10^3}$}[2]$ $\times 10^3$
& 0.922
 & 0.915($\eval*{$-\frac{CD1-CD2}{CD1}100$}[CD1 =  0.922, CD2 = 0.915][2]$\%) &1.031($\eval*{$-\frac{CD1-CD2}{CD1}100$}[CD1 =  0.922, CD2 =1.031][2]$\%)\\

\hline
\end{tabu}
\end{center} 
\caption{ Relative error between the predictive $C_D$ coefficient from the logarithmic equations , and the LES simulations of 
Jiang and Cheng\cite{jiang2021large} for different Reynolds numbers for the geometry of infinite  cylinder\@.  } 

\label{table2cylinder}
\end{table} 
 
\subsubsection{Streamline wires, and compression struts}
To better understand the relationship between the $a_1$ coefficient and boundary layer theory, we will compare the predictions of the logarithmic equation with the limited experimental data of  Hoerner \cite{hoerner1965fluid}. Hoerner conducted experiments on two-dimensional streamlined wires and struts with varying thickness-to-chord ratios ($\text{t/c}$) at high Reynolds numbers. Furthermore, Hoerner proposed a correlation that accurately fits the aforementioned data, which is expressed as follows:
\begin{equation}\label{eqfluiddrag}
C_D = \underbrace{\frac{2.66}{\sqrt{Re}}(1+\text{t/c})}_{\text{frictional drag}} +\underbrace{(\text{t/c})^2}_{\text{form drag}}
\end{equation}
Eq.(\ref{eqfluiddrag}) comprises two components. The first component is the frictional drag, that depends  both on the geometry ($\text{t/c}$) and the Reynolds number($Re$). The second component is the form drag, which remains constant and relies solely on the geometry (it is proportional to the quadratic power of $\text{t/c}$). To forecast the experimental results  of Hoerner \cite{hoerner1965fluid} using Eq.(\ref{eqloggeneral}), we will make the following assumption regarding the $a_1$ coefficient:
\begin{equation} \label{eqa1}
a_1 = 2.656+\text{t/c}
\end{equation}
Eq.(\ref{eqa1}) is based on two main assumptions. The first assumption posits that the coefficient $a_1$ is  linked to the shear stress and frictional drag. The second assumption aligns with Hoerner \cite{hoerner1965fluid} correlation, proposing that the frictional drag exhibits a linear dependence on $\text{t/c}$. The constant 2.656 originates from the prefactor in the frictional drag coefficient  for the double-wetted flat plate, derived by Blasius \cite{blasius1908grenzschichten}. By utilizing the values of the $a_1$ coefficient from Eq.(\ref{eqa1}) in order for  Eq.(\ref{eqloggeneral}) to predict the values of $C_D$ of Hoerner \cite{hoerner1965fluid} data, we add extra validity to our assumption that $a_1$ is strongly correlated with shear stress and frictional drag\@.

\begin{figure}[H]
\begin{center}
\includegraphics [scale=0.8, trim = 0 0 0 0,clip]{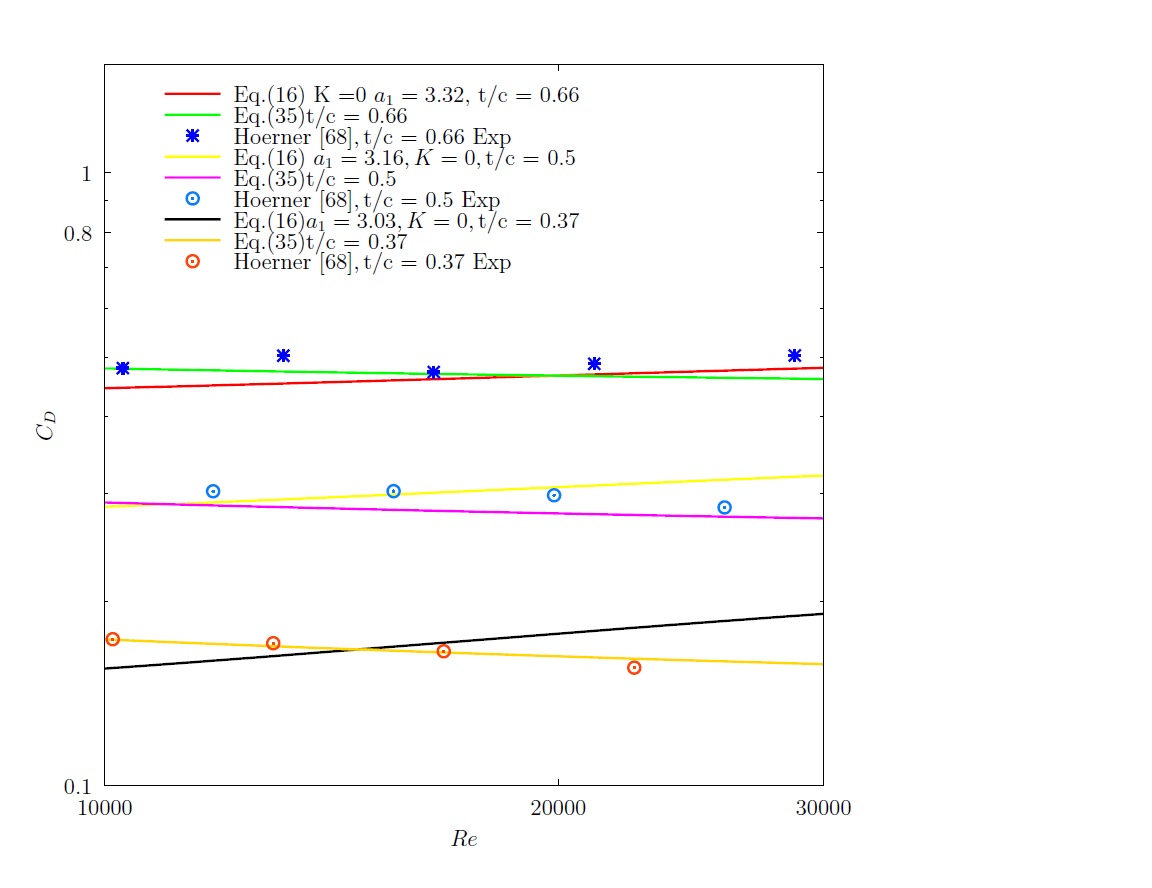}
\end{center}
\caption {Comparison of $C_D$ values predicted by the logarithmic equation, with the correlation and the experimental results of Hoerner \cite{ hoerner1965fluid}. }
\label{Figarofoildrag} 
\end{figure}     

\begin{table}[t]
\centering
\begin{tabular}{|c|*{3}{>{\centering\arraybackslash}p{3.2cm}|}}
\hline
 & $C_D$ Eq.(\ref{eqloggeneral})(\%)& $C_D$ Hoerner \cite{hoerner1965fluid}\\
\hline

$Re = 1\times 10^4$, t/c = 0.66& 0.445 (\eval*{$\frac{CD1-CD2}{CD2}*100$}[CD1 =  0.445,CD2 =0.479][2]\%)  & 0.479 \\

$Re = 1\times 10^4$, t/c = 0.50& 0.285 (\eval*{$\frac{CD1-CD2}{CD2}*100$}[CD1 = 0.285,CD2 = 0.289][2]\%)  & 0.289 \\

$Re = 1\times 10^4$, t/c = 0.37& 0.1553 (\eval*{$\frac{CD1-CD2}{CD2}*100$}[CD1 = 0.1553 ,CD2 =0.173][2]\%)  & 0.173 \\

$Re = 3\times 10^4$, t/c = 0.66& 0.480 (\eval*{$\frac{CD1-CD2}{CD2}*100$}[CD1 = 0.476 ,CD2 =0.461][2]\%)  & 0.461 \\

$Re = 3\times 10^4$, t/c = 0.50& 0.320 (\eval*{$\frac{CD1-CD2}{CD2}*100$}[CD1 = 0.316 ,CD2 = 0.273][2]\%)  & 0.273\\

$Re = 3\times 10^4$, t/c = 0.37& 0.190 (\eval*{$\frac{CD1-CD2}{CD2}*100$}[CD1 =  0.186 ,CD2 = 0.157][2]\%)  & 0.157\\
\hline
\end{tabular}
\caption{Comparison of the predictions of the drag coefficient from the logarithmic equation Eq.(\ref{eqloggeneral}), with the results of Hoerner \cite{hoerner1965fluid}.}
\label{tablefluidragbookdrag}
\end{table}

\begin{table}[H]
\centering
\begin{tabular}{|c|*{3}{>{\centering\arraybackslash}p{3.2cm}|}}
\hline
 & $C_{D_{p}}$ Eq.(\ref{eqloggeneral})(\%)& $C_{D_{p}}$ Hoerner \cite{hoerner1965fluid}\\
\hline
$Re = 1\times 10^4$\, t/c = 0.66& 0.408 (\eval*{$\frac{CD1-CD2}{CD2}*100$}[CD1 =  0.408,CD2 =0.435][2]\%)  &  0.435 \\

$Re = 1\times 10^4$, t/c = 0.50& 0.249 (\eval*{$\frac{CD1-CD2}{CD2}*100$}[CD1 = 0.249,CD2 = 0.25 ][2]\%)  & 0.25 \\

$Re =1\times 10^4$, t/c = 0.37& 0.121 (\eval*{$\frac{CD1-CD2}{CD2}*100$}[CD1 = 0.121 ,CD2 =0.136][2]\%)  & 0.136 \\

$Re = 3\times 10^4$, t/c = 0.66 & 0.457 (\eval*{$\frac{CD1-CD2}{CD2}*100$}[CD1 = 0.457 ,CD2 =0.435][2]\%)  & 0.435 \\
$Re = 3\times 10^4$, t/c = 0.50& 0.298 (\eval*{$\frac{CD1-CD2}{CD2}*100$}[CD1 = 0.298 ,CD2 = 0.25][2]\%)  & 0.25\\
$Re = 3\times 10^4$, t/c = 0.37 & 0.169 (\eval*{$\frac{CD1-CD2}{CD2}*100$}[CD1 =  0.169 ,CD2 = 0.136][2]\%)  & 0.136\\
\hline
\end{tabular}
\caption{Comparison of the predictions of the pressure drag coefficient from the logarithmic equation Eq.(\ref{eqloggeneral}), with the results of Hoerner \cite{hoerner1965fluid}.}
\label{tablefluidragbookpressuredrag}
\end{table}

The comparison between the predictions of Eq.(\ref{eqloggeneral}) and those of Hoerner \cite{hoerner1965fluid} is depicted in Figure \ref{Figarofoildrag}. We observe that the predictions of the drag coefficient from the logarithmic equation Eq.(\ref{eqloggeneral}) increase slightly with the Reynolds number, while the $C_D$ coefficient from Hoerner \cite{hoerner1965fluid} slightly decreases in the Reynolds number regime under investigation. The reason for this discrepancy is that our model, based on data from a full bluff body, while,  the data of Hoerner's  \cite{hoerner1965fluid},  are based on streamlined bodies  where form drag plays a less active role\@. However, as the value of $\text{t/c}$ increases and the body becomes more bluff, both the predictions of the logarithmic equation and the $C_D$ data of Hoerner \cite{hoerner1965fluid} show a constant value of the  drag coefficient for  the range of Reynolds numbers used. \\

The results of the drag coefficient From Figure \ref{Figarofoildrag} solidifies our initial assumption that the value of the $a_1$ coefficient is a linear  function of thickness to the chord ratio of the geometry as given by Eq.(\ref{eqa1})\@ for the current geometries under investigation\@. This is because the predictions of the logarithmic equation (Eq.(\ref{eqloggeneral}) matches with acceptable accuracy the results of Hoerner \cite{hoerner1965fluid}, as shown from the results of Table \ref{tablefluidragbookdrag}\@. The maximum deviation is about 18.47\% for the case of $\text{t/c}$ equal to 0.37, and $Re$ =  $3\times 10^4$\@. Which is a very acceptable deviation, since the logarithmic equation was agnostic about the geometry of such streamed bodies\@. However, if we increase the value of the $\text{t/c}$ to 0.66 the deviation is significantly reduced to only 7.1\% for $Re = 10^4$\@. Now, to show that $a_1$ is related to the shear stress, and the frictional  drag, we will compare our results for the pressure drag, with those of      Hoerner \cite{hoerner1965fluid}, as shown in Table \ref{tablefluidragbookpressuredrag}\@. The pressure drag is given from the following relation: 

\begin{equation}
C_{D_{p}} = C_D - \frac{a_1}{\sqrt{Re}}
\end{equation}
Where $C_D$ is obtained from Eq.(\ref{eqloggeneral})  $K$ = 0.0\@. The relative predictions for the pressure drag are very satisfactory, with an average difference of about 11\%\@. The accurate prediction of the pressure drag by the logarithmic equation indicates a strong correlation between the $a_1$ coefficient and the frictional drag\@. Overall, our results demonstrate that the frictional drag of the body primarily determines the magnitude of the pressure drag it experiences\@. This finding contradicts current theories which suggest that the pressure drag is independent of the frictional drag \cite{batchelor2000introduction}\@. In our previous work \cite{el2022logarithmic}, we showed that the $a_1$ coefficient is also the drag coefficient value that nullifies all dependent terms on $Re$ in the logarithmic-based equation\@. The Reynolds number that corresponds to the value of the $a_1$ is confined in the low Reynolds flow regime in which predominantly governed by frictional forces, for the case of the sphere   is equal to  14.06\@. The description provided above unequivocally illustrates the connection of the $a_1$  coefficient to the frictional flow regime\@.
\subsubsection{Cubes}  
A Cube is another bluff geometry,  {for which} we have experimental data for the drag coefficient at high Reynolds numbers, so we can test further the validity of our logarithmic equation\@. The only correlation that we know  {of} for the drag coefficient of {a} cube is that of Saha \cite{saha2004three}, which is based on numerical simulations\@. It has the following form:
\begin{equation}\label{eqcubepower}
C_D = \frac{24}{Re}(1+0.232Re^{0.628})
\end{equation} 
We will determine the coefficient $a_1$ in Equation of the logarithmic equation (Eq.(\ref{eqloggeneral}))  by utilizing the drag coefficient obtained from the Saha correlation \cite{saha2004three} at a Reynolds number ($Re$) of 100, the value of $K$ = 1.0 following {from} the correlation of Saha\cite{saha2004three}\@. The resulting value of $a_1$ is found to be 3.488\@. \\ 

\begin{figure}[H]
\begin{center}
\includegraphics [scale=0.8, trim = 0 0 0 0,clip]{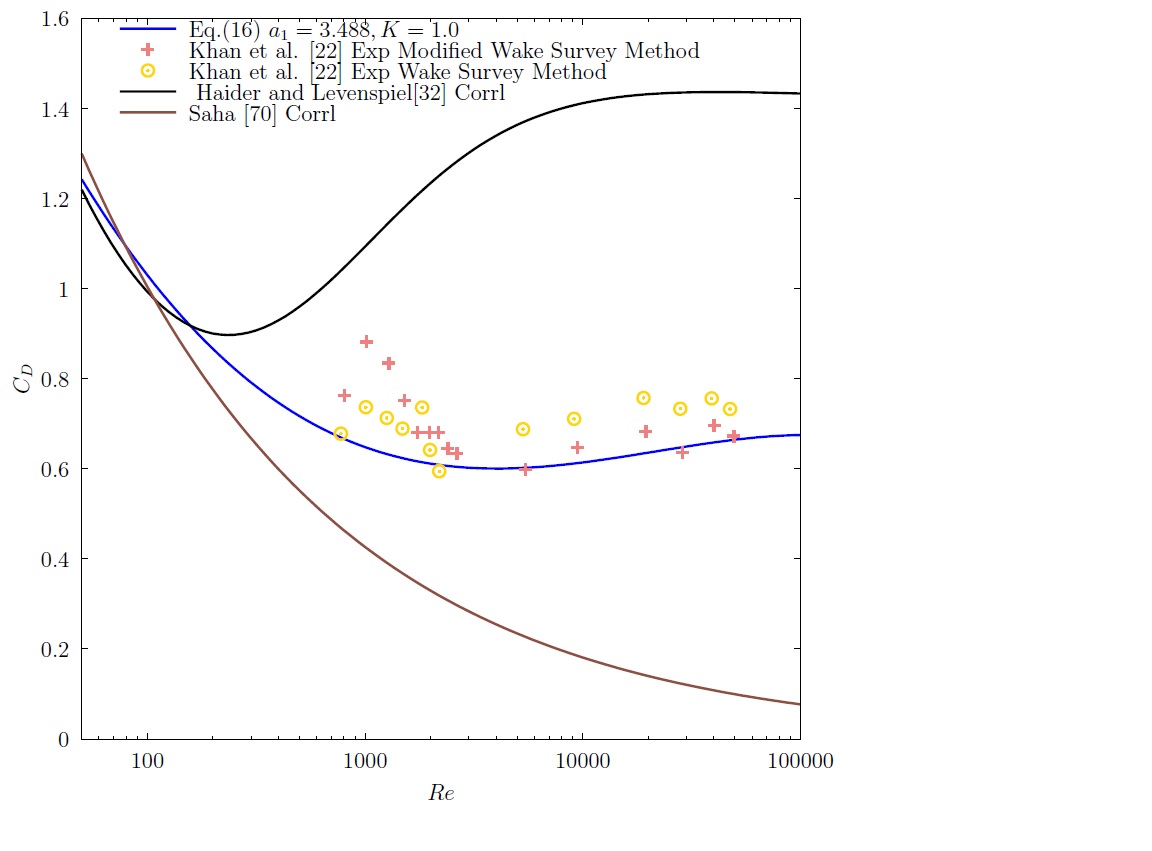}
\end{center}
\caption {Comparison of $C_D$ values predicted by the logarithmic equation, correlations, and  experiments, for the case  of a cube\@.  }
\label{Fig1cube} 
\end{figure}

 We will compare our results to the experimental findings of Khan et al. \cite{khan2018flow} for the case of a cube placed normal to the flow. In addition, we will also include in our comparison the correlations of Saha \cite{saha2004three} and Haider and Levenspiel \cite{happel2012low}, as illustrated in Figure \ref{Fig1cube}\@. Khan et al. \cite{khan2018flow} estimated the drag coefficient from their experimental readings by employing both the Wake survey and the modified Wake survey methods\@.\\

For low Reynolds numbers, the drag coefficient values predicted by the logarithmic equation closely matches those from the correlations of Haider and Levenspiel \cite{happel2012low} and Saha \cite{saha2004three}\@. However, at high Reynolds numbers, the logarithmic equation 
(Eq.(\ref{eqloggeneral}), $a_1$ = 3.488, $K$ = 1.0) predicts the experimental results of Khan et al. \cite{khan2018flow} with acceptable accuracy, capturing the increasing and decreasing trends in their data\@. The logarithmic equation predicts the results of the modified Wake survey method more closely\@. \\

On the other hand, the correlations of Saha \cite{saha2004three} and Haider and Levenspiel \cite{happel2012low} under-predict and over-predict the experimental results of Khan et al. \cite{khan2018flow}, respectively\@.

The drag coefficient for the incoming geometries relies on the concept of a volume-equivalent sphere, while the Reynolds number is determined based on the diameter of this volume-equivalent sphere\@.  
\subsubsection{Oblate spheroids} 
For oblate spheroids, we can compare our theory with a significant number of spheroid geometries, as well as different flow orientations\@.
For the sake of brevity we will focus mainly for case of flow that is parallel to the equatorial diameter of the spheroid ($\alpha = 90^{\circ}$), and for different aspect ratios $p_a$\@. To determine the $a_1$ coefficient in the logarithmic equation, we will utilize the $C_D$ value from  the Ouchene \cite{ouchene2020numerical} correlation and the numerical data provided by Sanjeevi et al. \cite{sanjeevi2018drag}\@. We will be using the values of $C_D$ from the Ouchene correlation \cite{ouchene2020numerical} for aspect ratios of $p_a = 0.8$, $0.5$, and $0.2$ at $Re = 100$ and $\alpha = 90^{\circ}$\@. The corresponding values of $a_1$ that we calculated are $3.361$, $3.365$, and $3.365$, respectively\@. It is interesting to note that the value of the $a_1$ coefficient remains almost constant for all aspect ratio values used\@. Therefore, we have decided to keep the value of the $a_1$ coefficient at $3.365$\@. \cite{bagheri2016drag}

\begin{figure}[h!]
\begin{center}
\includegraphics [scale=0.8, trim = 0 0 0 0,clip]{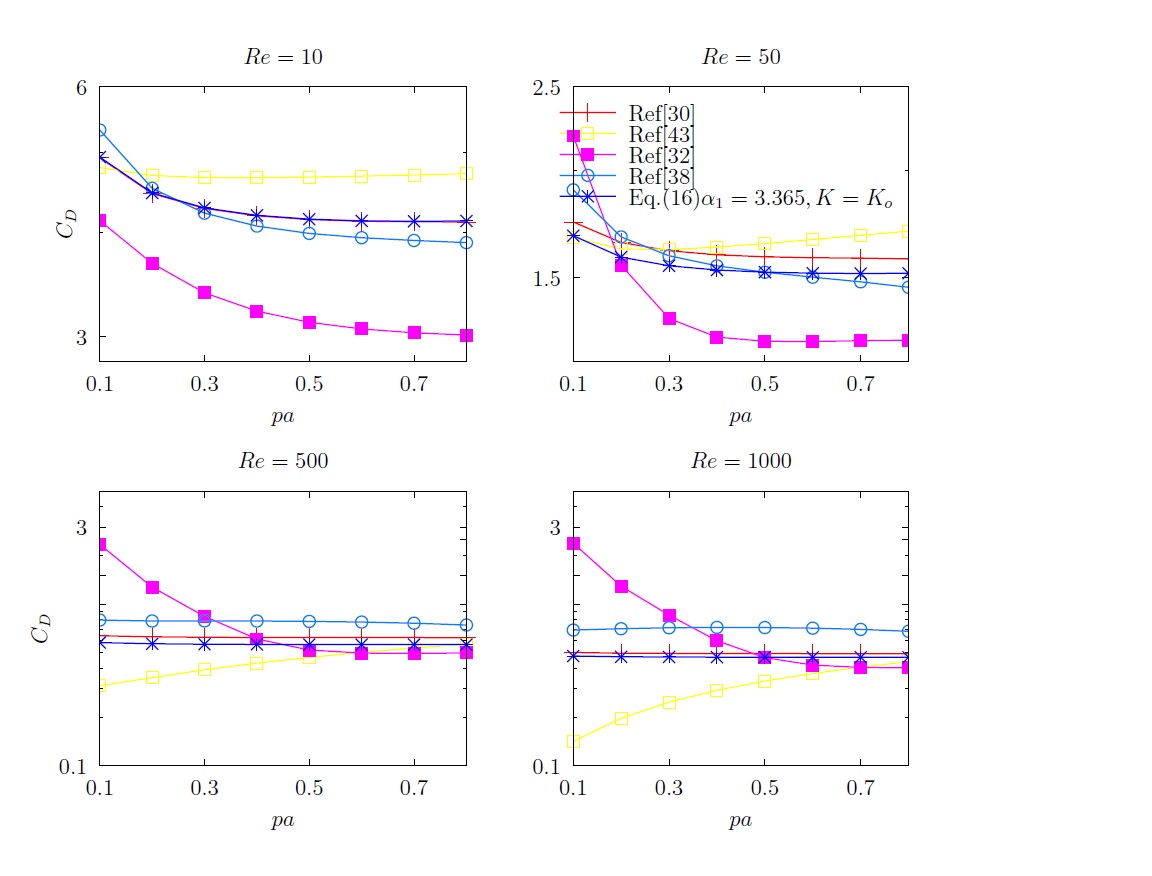}
\end{center}
\caption{Comparison  between the drag coefficient $C_D$ predicted by different predicted models for the case of oblate shaped particle, for different Reynolds numbers, and high aspect ratio values, and $\alpha$ = $90^{\circ}$.}
\label{Fig1oblate}
\end{figure}

We will start by comparing the predictions of our  logarithmic equation  with those of Ouchene \citep{ouchene2020numerical}, and with other predictive models available from the  literature as   shown in Figure \ref{Fig1oblate}\@ for values of $p_a$ $\geq$ 0.1. The drag coefficient predicted by the Ouchene \citep{ouchene2020numerical} correlation follow a pattern  that is independent  of the aspect ratio at high Reynolds numbers, similar to the observation by El Hasadi and Padding \citep{el2019solving}, and Frohlich et al. \cite{frohlich2020correlations} for the prolate spheroids\@. Beyond the Ouchene \citep{ouchene2020numerical} correlation only the logarithmic predictive model of  Eq.(\ref{eqloggeneral}),$a_1 = 3.365$, and $K = K_o$ predicted the aspect ratio independency of the drag  coefficient independently, here what we mean by $K = K_o$ is that the $K$ value in the logarithmic equation is calculated form Eq.(\ref{eq2olog}) for each value of the aspect ratio $p_a$\@. \\

The power based correlations such as Chen et al.\citep{chen2021drag}, Haider and Levenspiel \cite{haider1989drag}, and Holzer and Sommerfeld \cite{holzer2008new} show different degree of accuracy with respect to the Ouchene \cite{ouchene2020numerical} results\@. For the case of the Chen et al. \cite{chen2021drag} correlation  its divergence from the predictions of Ouchene \cite{ouchene2020numerical} is only due to its power based  mathematical structure that they used as we showed in our previous work \cite{el2022logarithmic} in which we showed that power based drag coefficient correlations are poorly extrapolate beyond their training data\@. 
 For the case of the Haider and Levenspiel \cite{haider1989drag} correlation its basic concept that the geometry of the particle (i. e. $\phi$) control the evolution of the drag coefficient in low and high Reynolds number regimes contradicts  the numerical predictions of Ouchene et al.\cite{ouchene2020numerical}\@. The Holzer and Sommerfeld \cite{holzer2008new} is the only correlation beyond the predictive logarithmic equations that approximately predicted the aspect ratio independence of the drag coefficient at high $Re$\@. However, if we zoom further  we will conclude that Holzer and Sommerfeld \cite{holzer2008new} correlation  predicts a weakly non-linear relation between the drag coefficient with the aspect ratio of $p_a \geq$ 0.1, and the Reynolds number\@. \\
\begin{table}[H]
\begin{center}
\begin{tabu} to 0.8\textwidth { | X[c] | X[c] | }
 \hline
  & Average Relative Error (\%) \\ 
 \hline
 Eq.(\ref{eqloggeneral}) $a_1$ = 3.365,$K= K_o$ & 4.36 \%   \\
 
 Chen et al.\citep{chen2021drag} & 19.09\% \\
 Haider and Levenspiel \citep{haider1989drag} & 43.38\%\\
 Holzer and Sommerfeld \citep{holzer2008new} & 18.74\%\\
 
\hline
\end{tabu}
\end{center} 
\caption{ Relative average error of the predictions of  different predictive models with the respect to the results of Ouchene \citep{ouchene2020numerical}. }
\label{table2olog}
\end{table}
The average relative error for different predictive models and correlations with respect to the data of 
Ouchene \cite{ouchene2020numerical}   for the different $Re$ considered in Figure \ref{Fig1oblate}  is given in Table \ref{table2olog}\@. The worst performing correlation is that of Haider and Levenspiel \cite{haider1989drag} with average error of about 43\%, which is not suitable for making accurate  predictions for oblate geometries in general especially for the ones with low  value of the aspect ratio\@. The logarithmic equation Eq.(\ref{eqloggeneral}), out performs all the other predictive models by average error of 4.36\%\@.   One issue in fluid flow investigations, particularly for multiphase problems, is the lack of benchmark cases for validating numerical simulations or experimental results. This makes it difficult to assess the accuracy of Ouchne's \cite{ouchene2020numerical} results. Therefore, further comparisons with available results in the literature must be made\@. Thus, will compare our predictive models with further results from the literature either from numerical simulations or from experiments, for different Reynolds numbers, and different  $p_a$ values\@. Finding relevant experimental results to compare with our predictive models is difficult due to the lack of empirical investigations. For this reason, we selected only one experiential investigation to compare our results with\@. \\

\begin{figure}[H]
\begin{center}
\includegraphics [scale=0.8, trim = 0 0 0 0,clip]{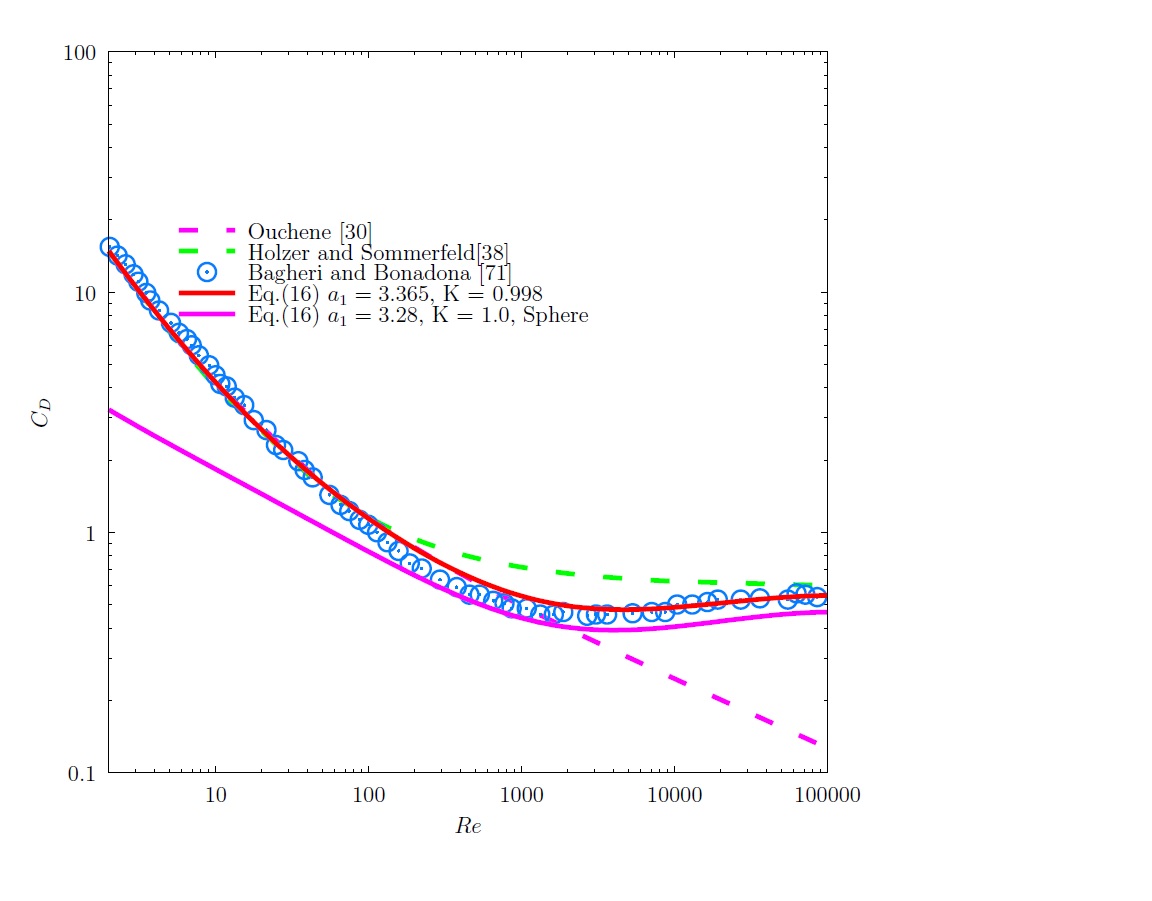}
\end{center}
\caption{Comparison  between the drag coefficient $C_D$ predicted by different predictive  models for the case of oblate shaped particle with the experimental  results of Bagheri et al. \cite{bagheri2016drag} for $p_a$ = 0.5, and for  $\alpha$ = $90^{\circ}$.}
\label{Fig5oblate}
\end{figure}

The only available experimental results for oblate spheroids at high Reynolds numbers are those of 
Bagheri and Bonadona \cite{bagheri2016drag} for the case of an oblate with $p_a = 0.5$, with $\alpha$ = $90^{\circ}$.  The comparison is shown in Figure \ref{Fig5oblate}\@. The logarithmic equation (Eq.(\ref{eqloggeneral}), $a_1 = 3.365$, and $K = 0.998$) predictions are very close to the experimental results of Bagheri and Bonadona \cite{bagheri2016drag} for the entire range of the Reynolds number considered\@. Also, an interesting observation is that the logarithmic-based model captures with great accuracy the local minimum in the drag coefficient curve, of the experimental results of Bagheri and Bonadona \cite{bagheri2016drag}, while all the other correlations failed to do that\@. The accurate predictions of the logarithmic equation solidify our assumption that $a_1$ is constant for aspect ratios in the range $0.2\leq p_a\leq 0.8$\@.
The results of Ouchene \cite{ouchene2020numerical} correlation are close to those of Bagheri and Bonadona \cite{bagheri2016drag} for Reynolds numbers up to 1000. However, for higher Reynolds numbers, its prediction starts to deviate significantly\@. On the other hand, the predictions of the correlation of   Holzer and Sommerfeld \cite{holzer2008new}  over-predict,  the empirical results of Bagheri and Bonadona \cite{bagheri2016drag} especially at the Reynolds range of $10^3$ to $10^4$\@. Also, we made a compression with the experimental results for case of an oblate spheroid of $p_a$ = 0.75, and $\alpha$ = $90^{\circ}$  provided by Cengel and Cimbala \cite{cengel2013ebook} their $C_D$ value is equal to 0.5 for $Re$ = $2\times 10^5$, while our predicted value is 0.541 using the Eq.(\ref{eqloggeneral}) with $a_1$ = 3.365, and $K$ = 0.988\@.

The only numerical results that we have for high Reynolds number flow over an oblate geometry is that of Sanjeevi et al.\cite{sanjeevi2018drag} for the case of $p_a$ = 0.4, for different flow orientations  that includes flow parallel to the equalutral axis ($\alpha = 90^{\circ}$), and polar axis ($\alpha = 0^{\circ}$)\@. In addition, we have calculated the $a_1$ coefficient using the numerical data of Sanjeevi et al. \cite{sanjeevi2018drag} for $Re$ = 100, $\alpha$ = $90^{\circ}$, and $p_a$ = $0.4$. The value we obtained for $a_1$ is 3.226. The difference between the values of the $a_1$ coefficient obtained from the data of Ouchene \cite{ouchene2020numerical} and that of Sanjeevi et al. \cite{sanjeevi2018drag} is due to the different numerical schemes that they used, which produce results with varying levels of accuracy\@. The Reynolds number for both Ouchene \cite{ouchene2020numerical}, and  Sanjeevi et al. \cite{sanjeevi2018drag} data is based on the volume equivalent sphere\@. We need now to calculate the $a_1$ for the case of flow orientation of $\alpha = 0^{\circ}$, and we will do it for both the  Ouchene \cite{ouchene2020numerical} and that of Sanjeevi et al. \cite{sanjeevi2018drag} data\@. For the case of Ouchene \cite{ouchene2020numerical}, we will evaluate the $a_1$ coefficient using the data for the $C_D$ for case of $Re$ = 100, $p_a$ =0.4, and for this case we obtain a value of $a_1$ = 4.242\@. While for the   Sanjeevi et al. \cite{sanjeevi2018drag} data we will use the data of the $C_D$ at $Re$ = 300, for better accuracy, for this case we obtained a value for the $a_1$ equal to 4.186\@. We want to emphasize that Sanjeevi et al. \cite{sanjeevi2018drag} data are discrete points directly obtained from numerical simulations, and not influenced by any correlations\@.     \\

\begin{figure}[H]
\begin{center}
\includegraphics [scale=0.8, trim = 0 0 0 0,clip]{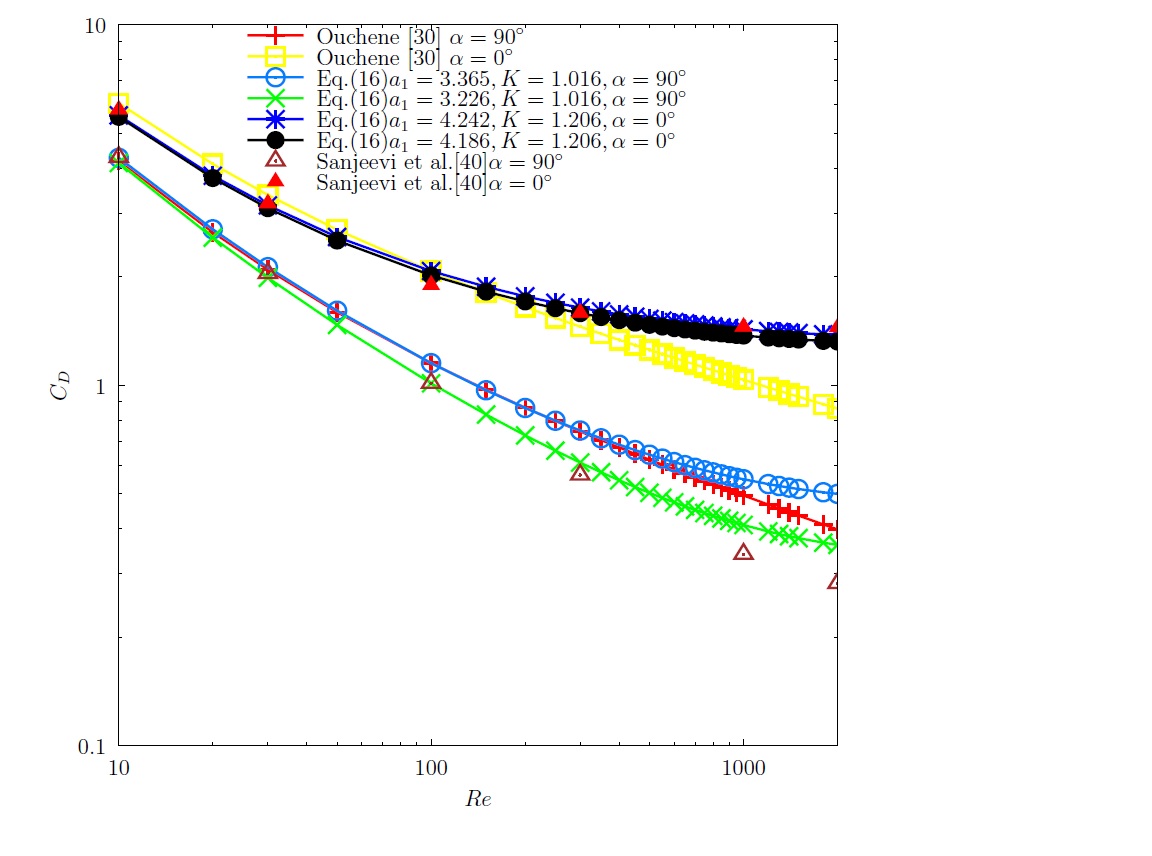}
\end{center}
\caption{Comparison  between the drag coefficient $C_D$ predicted by different predictive  models for the case of oblate shaped particle with the numerical results of Sanjeevi et al.\cite{sanjeevi2018drag} for $p_a$ = 0.4, $\alpha = 0^{\circ}$, and $90^{\circ}$\@.}
\label{Fig3oblate}
\end{figure}

This time we will test our logarithmic equations for two different flow orientations, as shown in Figure \ref{Fig3oblate}\@. For the case of $\alpha = 90^{\circ}$, the logarithmic equation with $a_1 = 3.365$ and $K = 1.016$, which is derived from the Ouchene \cite{ouchene2020numerical} data (Eq.(\ref{eqloggeneral})), follows the Sanjeevi et al. \cite{sanjeevi2018drag} numerical data from low Reynolds numbers up to a value of 100\@. However, for higher Reynolds numbers, there is a slight deviation between the drag coefficient predicted by the logarithmic equation and the numerical data of Sanjeevi et al. \cite{sanjeevi2018drag}\@. On the other hand, the logarithmic equation with $a_1 = 3.226$ and $K = 1.016$, which is derived from the Sanjeevi et al. \cite{sanjeevi2018drag} data (Eq.(\ref{eqloggeneral})), follows the numerical results of Sanjeevi et al. \cite{sanjeevi2018drag} from low to high Reynolds numbers\@.
For the case of $\alpha = 0^{\circ}$, where the oblate particle experiences higher  drag force due to its bluff shape compared to $\alpha = 90^{\circ}$, the logarithmic equations with $a_1 = 4.242$ and $K = 1.206$, and $a_1 = 4.186$ and $K = 1.206$, which are derived from the Ouchene \cite{ouchene2020numerical} and Sanjeevi et al. \cite{sanjeevi2018drag} data (Eq.(\ref{eqloggeneral})), respectively, follow the numerical data of Sanjeevi et al. \cite{sanjeevi2018drag} from low to high Reynolds numbers\@. However, the correlation of Ouchene \cite{ouchene2020numerical} significantly underpredicts the values of the numerical data of Sanjeevi et al. \cite{sanjeevi2018drag}, which shows the limitation of its applicability\@. We have shown that our theory holds for the oblate spheroid with different flow geometries\@.

\subsubsection{Prolate spheroids}
The flow behaviour of prolate spheroids differs from that of oblate spheroids, and it is more sensitive to the geometry aspect ratio. For example, Sanjeevi et al. \cite{sanjeevi2022accurate} found that at low Reynolds numbers ($Re$), the drag coefficient reaches a minimum at $p_a \simeq 1.95$ and then increases as the aspect ratio increases. Therefore, we will separately test our equations in the low [1-2] and high [2-15] aspect ratio ranges\@. We will investigate extensively, the case when the flow is parallel to the major axis of the spheroid  $\alpha$ = $0^{\circ}$\@.  \\

To calculate the coefficient $a_1$, we will use the data from the correlation provided by Sanjeevi et al. \cite{sanjeevi2022accurate} at a Reynolds number of $Re$ = 100, considering different aspect ratios of the prolate spheroid geometry\@.
For the low range of aspect ratios [1-2], we will use the data for the case of $p_a$ = 1.2. From the data, we found that $a_1$ = 3.1416, which is approximately equal to $\pi$.
For the higher range of aspect ratios [2-15], we will calculate the $a_1$ coefficient for three specific aspect ratio values: $p_a$ = 5, 10, and 15\@. The calculated values of $a_1$ are approximately 2.989, 2.994, and 2.997, respectively, which shows that they are equal\@. To simplify, we will approximate the value of $a_1$ equal to 3.00\@.\\

\begin{figure}[H]
\begin{center}
\includegraphics [scale=0.8, trim = 0 0 0 0,clip]{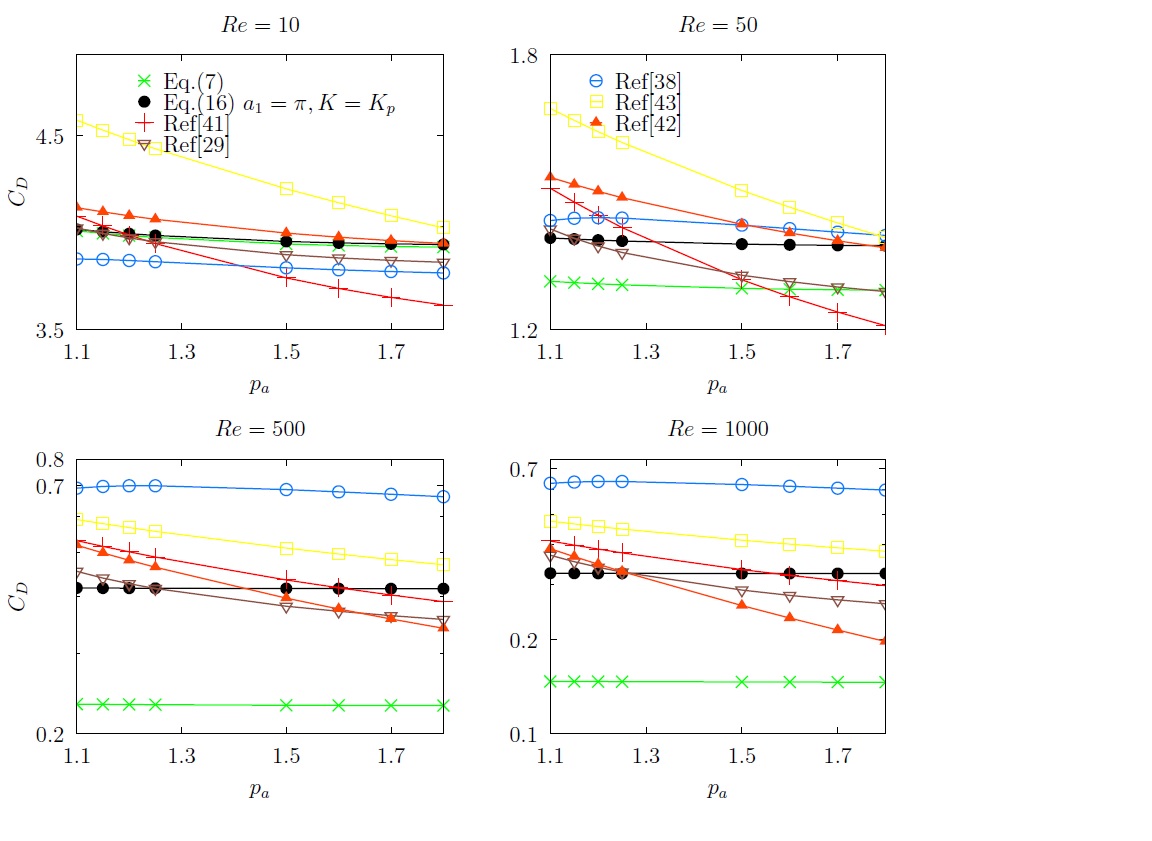}
\end{center}
\caption{Comparison  between the drag coefficient $C_D$ predicted by different predicted models for the case of prolate spheroid, for different Reynolds numbers, and low aspect ratio values, and $\alpha$ = $0^{\circ}$.}
\label{Fig1Porlate}
\end{figure}

For aspect ratios between 1.1 and 1.7, we plot the variation of the drag coefficient with $p_a$ for different $Re$ as shown in Figure \ref{Fig1Porlate}\@. We have minimal data points for the drag coefficient in this aspect ratio range, even from CFD investigations\@. Therefore, the correlations we will use to compare our results with, usually only contain a single point in this aspect ratio range in their fitting data\@. For the case of $Re$  = 10, all the logarithmic-based equations produce results that are close to the results of the majority of the correlations that are based on CFD simulations, while the correlations of Chen et al.\cite{chen2021drag}, and Holzer and  Sommerfeld \cite{holzer2008new}, they over, and under predicts respectively the results of the CFD based correlations\@. Increasing the Reynolds further to values  50 and 500 in the medium Reynolds number regime, we observe that the values of logarithmic equation Eq.(\ref{eq1plog}) start to deviate and under predicts the results of the correlations of the CFD simulations\@.  We mean by CFD-based correlations the correlations of Sanjeevi et al. \cite{sanjeevi2022accurate}, Ouchene et al.\cite{ouchene2016new}, and  Frohlich et al. \cite{frohlich2020correlations}\@. Interestingly,the logarithmic equation (Eq.(\ref{eqloggeneral}),$a_1$ = $\pi$, and $K=K_p$) predictions are close to the those of Sanjeevi et al. \cite{sanjeevi2022accurate}, and  Frohlich et al. \cite{frohlich2020correlations}\@. If we increase further the Reynold number to 1000, the logarithmic-based equations keep similar predictive behaviour as in the medium Reynolds number flow regime, with the predictions of the logarithmic equation (Eq.(\ref{eqloggeneral}), $a_1$ = $\pi$, and $K=K_p$) getting closer to the predictions of Sanjeevi et al. \cite{sanjeevi2022accurate}, and Ouchene et al.\cite{ouchene2016new}, it is expected that the deviation will be increased from the predictions of  Frohlich et al. \cite{frohlich2020correlations} since their correlations is valid only up to Reynolds number 100\@.  Still, there needs to be more information on how to drag coefficient varies in this specific aspect ratio range to have a better conclusion of functional dependence of the drag coefficient on the geometry \@.      \\

\begin{table}[H]
\begin{center}
\begin{tabu} to 0.8\textwidth { | X[c] | X[c] | }
 \hline
  & Average Relative Error (\%) \\ 
 \hline
 Eq.(\ref{eq1plog}) &23.58 \% \\
 Eq.(\ref{eqloggeneral})$a_1$ = $\pi$, $K$ =$K_p$  & 8.18 \%   \\
 Chen et al.\cite{ouchene2016new} &23.68\% \\
 Ouchene et al. \cite{ouchene2020numerical} & 8.69\%\\
 Holzer and Sommerfeld \citep{holzer2008new} & 48.27\%\\
 Frohlich et al.\cite{frohlich2020correlations} & 8.046\%\\
\hline
\end{tabu}
\end{center} 
\caption{ Relative average error of the predictions of  different predictive models with the respect to the results of Sanjeevi et al. \cite{sanjeevi2022accurate} for prolate spheroids with low aspect ratios and  $\alpha$ = $0^{\circ}$. }
\label{table1p}
\end{table} 
Table \ref{table1p} shows the relative error of the logarithmic-based models, and the available correlations with respect to the Sanjeevi et al. \cite{sanjeevi2022accurate} correlation\@. The logarithmic-based equation Eq.(\ref{eq1plog}), and Chen et al. \cite{chen2021drag} correlation show a substantial difference from the results of Sanjeevi et al. \cite{sanjeevi2022accurate}\@. While the logarithmic-based equation (Eq.(\ref{eqloggeneral})$a_1$ = $\pi$, $K$ =$K_p$ ) shows a relative error of 8.18\%, which is similar to the relative errors of Ouchene \cite{ouchene2016new} and Frohlich et al. \cite{frohlich2020correlations}\@.  The predictions of Holzer and Sommerfeld \cite{holzer2008new} have the worst relative error of 48.27\%\@. This over prediction may be attributed to the divergence of the $\Delta$ for the Holzer and Sommerfeld \cite{holzer2008new} correlation from those of Eq.(\ref{eqhighRe2}) as shown in Figure \ref{Fig1shpherdiv}\@. \\

\begin{figure}[H]
\begin{center}
\includegraphics [scale=0.8, trim = 0 0 0 0,clip]{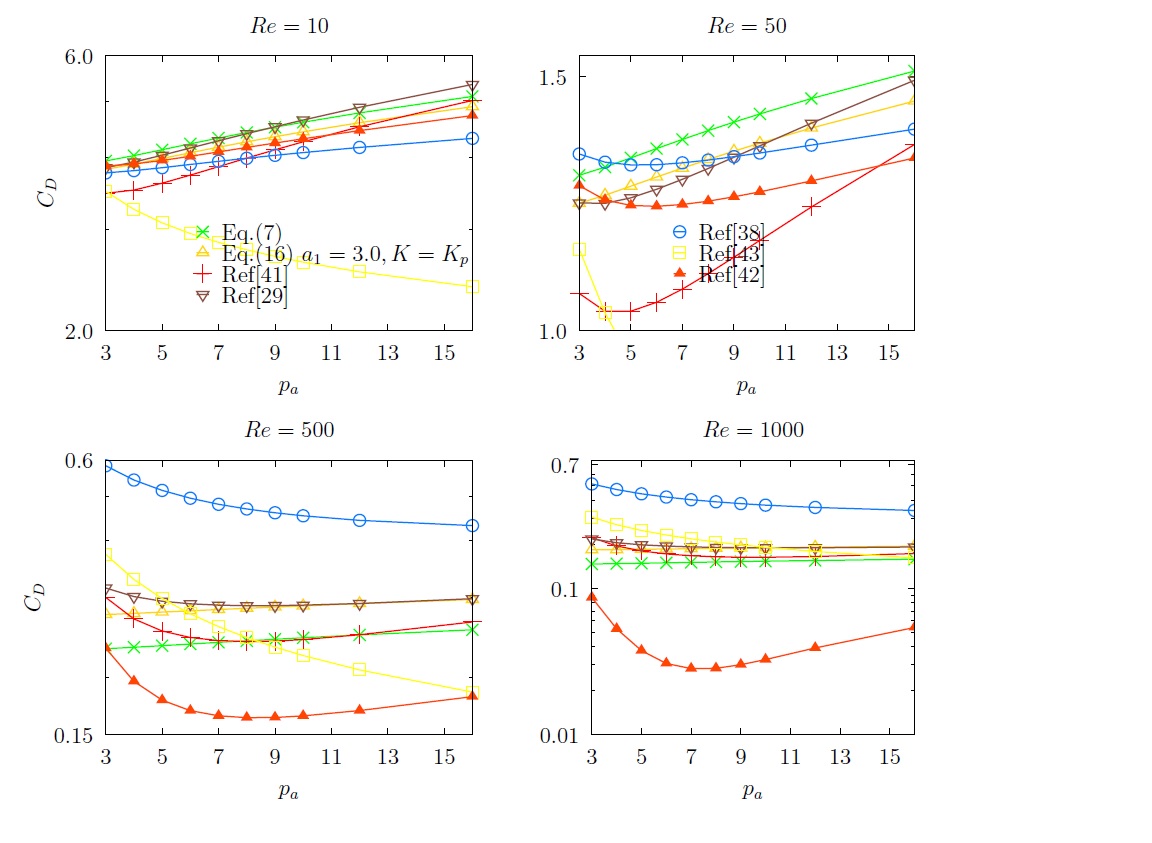}
\end{center}
\caption{Comparison  between the drag coefficient $C_D$ predicted by different predicted models for the case of prolate  shaped particle, for different Reynolds numbers, and high aspect ratio values, and $\alpha$ = $0^{\circ}$.}
\label{Fig2Porlate}
\end{figure}

For values of $p_a$ greater than or equal to 2, the drag coefficient increases as the aspect ratio increases for low Reynolds numbers (10 and 50), as shown in Figure \ref{Fig2Porlate}\@. Most predictive models follow this trend, except for the Chen et al.\cite{chen2021drag} correlation, which exhibits non-physical behavior\@. This is because the Chen et al.\cite{chen2021drag} correlation does not accurately account for the Stokes term\@. At higher Reynolds numbers,the logarithmic-based equation Eq.(\ref{eq1plog}) and the logarithmic equation 
(Eq.(\ref{eqloggeneral})$a_1 = 3.0,$ and $K = K_p$)  predictions are close to those of the Sanjeevi et al.\cite{sanjeevi2022accurate} correlation\@. The predictions of the Sanjeevi et al.\cite{sanjeevi2022accurate} and Ouchene\cite{ouchene2016new} correlations show that the drag coefficient is independent of the aspect ratio for high Reynolds numbers, which is consistent with our  predictions  in our previous work \cite{el2019solving} for prolate geometry\@. The Holzer and Sommerfeld \cite{holzer2008new} correlation fails to predict the interdependence of the drag coefficient on the aspect ratio at high Reynolds numbers because two of its terms, $\frac{3}{\sqrt{Re}}\frac{1}{\phi^{\frac{3}{4}}}$ and $0.42\times 10^{0.4\left(-\log(\phi)\right)^{0.2}}\frac{1}{\psi}$, show  aspect ratio dependency at high Reynolds numbers, indicating that the theoretical framework they use in their predictive model does not correctly reflect the physics of the problem\@. \\

Table \ref{table2p} show average relative error for  the different models, and correlations for the case   of Figure \ref{Fig2Porlate}\@.  The logarithmic based Eq.(\ref{eq1plog}) that used the training data of a single aspect ratio performs significantly better then the Frohlich\cite{frohlich2020correlations}, and Holzer and Sommerfeld\cite{holzer2008new}, which they use multiple  aspect ratio prolates to build their correlations\@. The logarithmic equation (Eq.(\ref{eqloggeneral}), $a_1$ = 3.0, $K$ = $K_p$) performs better than the other correlations\@. Also, we noticed from the different geometries we used so far in the paper that the more streamlined the body is, the smaller the value of the $a_1$ is  \@.

\begin{table}[H]
\begin{center}
\begin{tabu} to 0.8\textwidth { | X[c] | X[c] | }
 \hline
  & Average Relative Error (\%) \\ 
 \hline
 Eq.(\ref{eq1plog}) &11.94 \% \\
 Eq.(\ref{eqloggeneral}) $a_1$ = 3.0, $K$ = $K_p$  & 3.15 \%   \\
 Chen et al.\cite{ouchene2016new} &25.07\% \\
 Ouchene et al. \cite{ouchene2020numerical} & 11.92\%\\
 Holzer and Sommerfeld \citep{holzer2008new} & 46.71\%\\
 Frohlich et al.\cite{frohlich2020correlations} & 32.07\%\\

\hline
\end{tabu}
\end{center} 
\caption{ Relative average error of the predictions of  different predictive models with the respect to the results of Sanjeevi et al. \cite{sanjeevi2022accurate} for prolate spheroids with high aspect ratios, and $\alpha$ = $0^{\circ}$ . } 
\label{table2p}
\end{table}

\begin{figure}[H]
\begin{center}
\includegraphics [scale=0.8, trim = 0 0 0 0,clip]{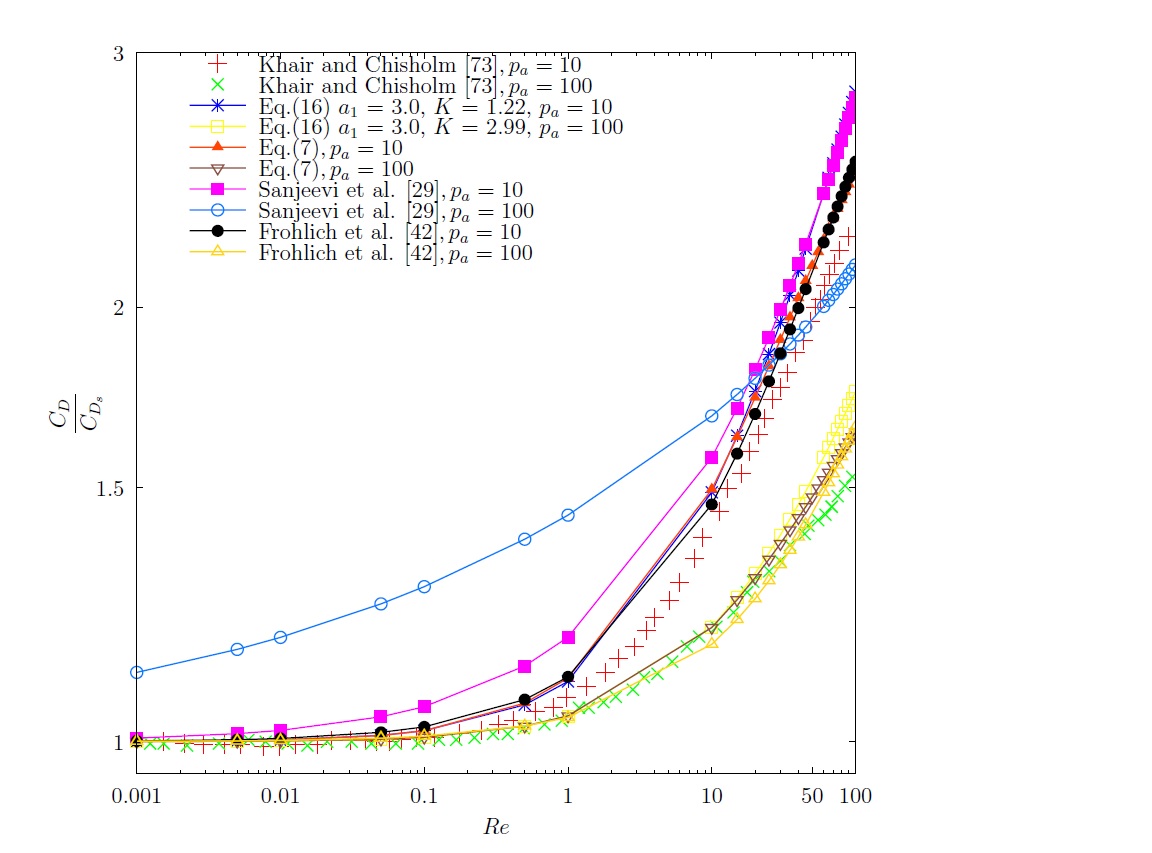}
\end{center}
\caption{Comparison of $\frac{C_D}{C_{D_s}}$ values from Khair and Chisholm's \cite{khair2018higher} numerical results and predictive models for prolate spheroids with aspect ratios $p_a$ = 10 and 100 at various Reynolds numbers for $\alpha$ = $0^{\circ}$.}
\label{Fig4Porlate}
\end{figure}

 To gain a more comprehensive understanding of the accuracy of our predictions, we will compare them to the numerical results of Khair and Chisholm \cite{khair2018higher} for two aspect ratios of $p_a$ = 10, 100, over a range of Reynolds numbers up to 100\@. We will examine the variation of $\frac{C_D}{C_{D_s}}$ with Reynolds number, as shown in Figure \ref{Fig4Porlate}\@. Both logarithmic equation Eq.(\ref{eq1plog}), and the logarithmic equation (Eq.(\ref{eqloggeneral}),$a_1$ = 3.0, $K$ = 1.22, 2.99), and the Frohlich correlation \cite{frohlich2020correlations} closely match the numerical results of Khair and Chisholm for both aspect ratios\@. We did not compare our drag coefficient results to the analytical solution provided by Khair and Chisholm \cite{khair2018higher} as their solution is only valid for Reynolds numbers up to 20\@. \\

\begin{figure}[H]
\begin{center}
\includegraphics [scale=0.8, trim = 0 0 0 0,clip]{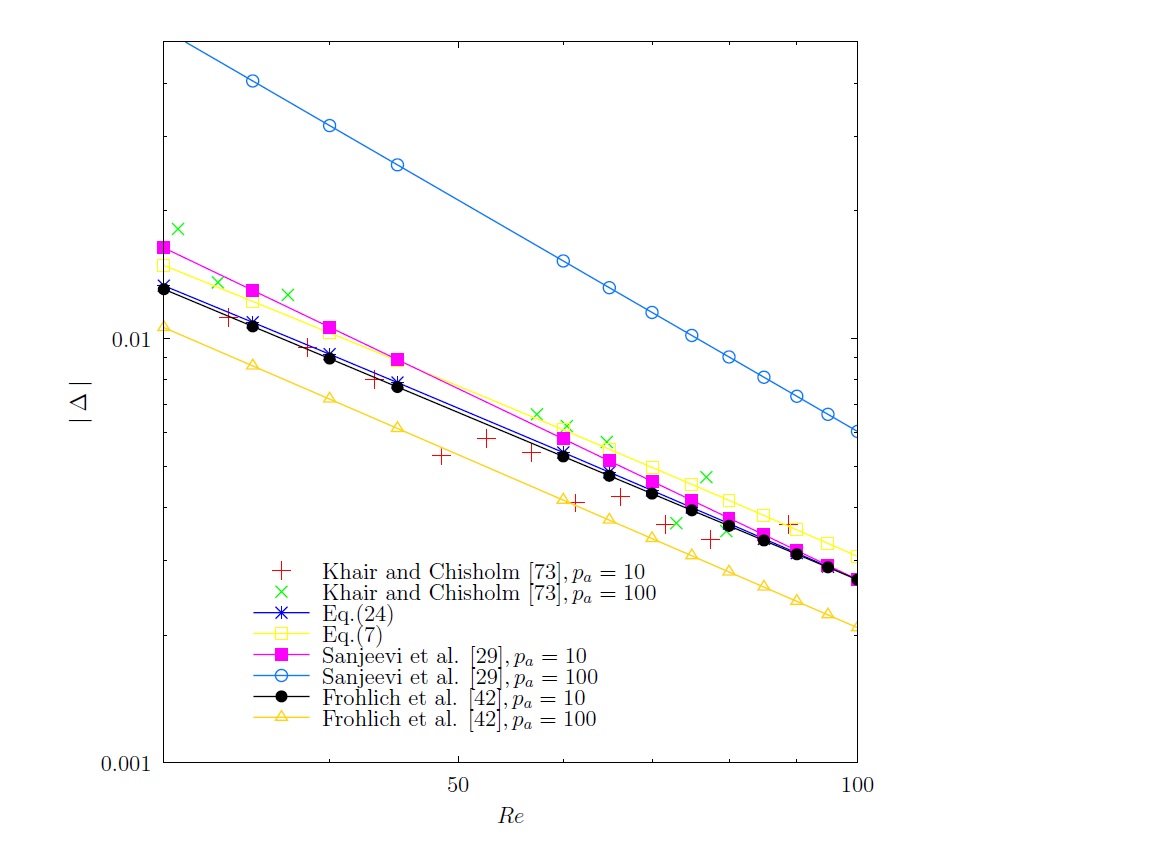}
\end{center}
\caption{Comparison of $\Delta$ values from Khair and Chisholm's \cite{khair2018higher} numerical results and from predictive models for prolate spheroids with aspect ratios $p_a$ = 10 and 100 at various Reynolds numbers, and for $\alpha$ = $0^{\circ}$.}
\label{Fig5Porlate}
\end{figure}

Figure \ref{Fig5Porlate} compares $\Delta$ values obtained from the numerical results of Khair and Chisholm \cite{khair2018higher}\@. We calculated the $\Delta$ for the results of Khair and Chisholm \cite{khair2018higher} by digitally extracting discrete points from Figure 4 of their paper and then calculating the gradient\@. The $\Delta$ values from Khair and Chisholm's \cite{khair2018higher} numerical investigation are close to our results from the logarithmic equations\@. This indicates that the $\Delta$ values from the logarithmic equations Eq.(\ref{eqhighRe2}) and Eq.(\ref{eq1plog}) accurately reflect the behaviour of Navier-Stokes equations solutions and are not limited to any particular drag coefficient correlation\@. The fluctuations in the  $\Delta$  results of Khair and Chisholm \cite{khair2018higher} is due to the fact we calculated the gradient from discrete points which were not uniform, adding to that the values of $\Delta$ are small of the order less than $O(10^{-2})$\@.
The $\Delta$ values for the case of $p_a$ = 100, from Sanjeevi et al. \cite{sanjeevi2022accurate} correlation  are significantly deviating from those of  Khair, and Chisholm \cite{khair2018higher}, and this is the main reason that they over-predicted the values of drag coefficient as shown in Figure \ref{Fig4Porlate} previsuly\@. Frohlich \cite{frohlich2020correlations} correlation results show a slight dependency on the aspect ratio. However, this is inconsistent with the results of the direct numerical simulations of Khair and Chisholm \cite{khair2018higher}\@. \\

We will test our theory for a flow perpendicular to the major axis $\alpha$ = $90^{\circ}$ for two prolate spheroids with $p_a$ values of 5 and 10\@. For this purpose, we will use the drag coefficient from the correlation of Sanjeevi et al. \cite{sanjeevi2022accurate} at $Re$ = 300\@. The obtained values for $a_1$ are as follows: $a_1$ = 4.063 for $p_a$ = 5.0 and $a_1$ = 4.7 for $p_a$ = 15.0\@. We used a higher Reynolds number to evaluate the $a_1$ coefficient because Sanjeevi et al. \cite{sanjeevi2022accurate} reported that their predictions overestimate the drag coefficient values in the Stokes regime compared to the analytical solutions\@.\\

\begin{figure}[H]
\begin{center}
\includegraphics [scale=0.8, trim = 0 0 0 0,clip]{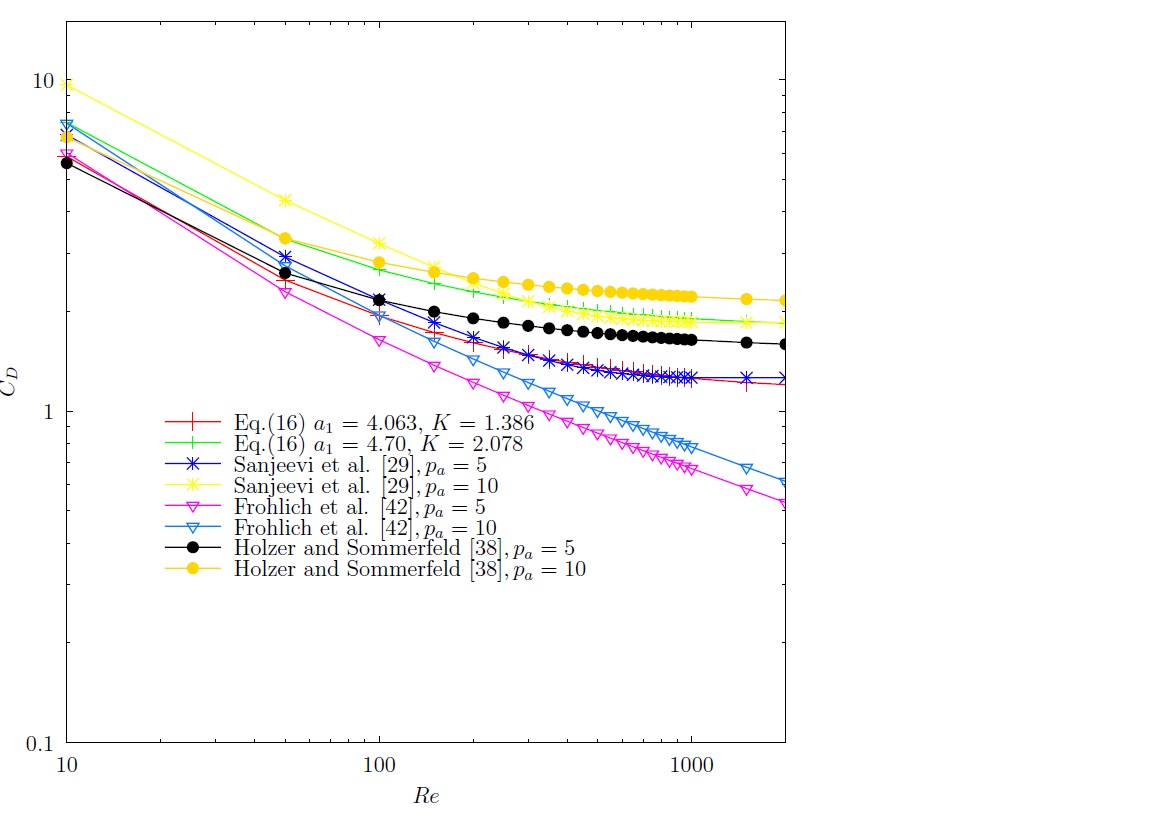}
\end{center}
\caption {Comparison of $C_D$ values predicted by the logarithmic equations, and different correlations for the case of prolate spheroids with $p_a$ = 5, 10 and for $\alpha$ = $90^{\circ}$\@.}
\label{Fig1verticalporlate} 
\end{figure}

The comparison of the drag coefficients for two selected aspect ratios is illustrated in Figure \ref{Fig1verticalporlate}\@. In the low Reynolds number regime, the logarithmic equations are in close agreement with the correlation of  Frohlich et al. \cite{frohlich2020correlations} that is based on numerical simulations and the general correlation of Holzer and Sommerfeld \cite{holzer2008new}\@. At high Reynolds numbers, the drag coefficients obtained from the logarithmic equations match perfectly with the results of Sanjeevi et al. \cite{sanjeevi2022accurate}, suggesting that the rate of change of the drag coefficient follows Eq.(\ref{eqhighRe2}) even under extreme flow conditions as at $\alpha = 90^{\circ}$\@. For the case of $\alpha = 0^{\circ}$, the $a_1$ coefficient was found to be constant, independent of the aspect ratio\@. However, for $\alpha = 90^{\circ}$, the $a_1$ coefficient strongly depends on the aspect ratio. In addition, increasing the aspect ratio increased the skin drag for $\alpha = 90^{\circ}$\@.\\

We will compare the results obtained from the logarithmic-based equation with the high-fidelity CFD simulations for a prolate spheroid with an aspect ratio of $p_a = 6.0$ and an orientation angle of $\alpha = 45^{\circ}$ with respect to the flow velocity, as described in the work of Jiang et al. \cite{jiang2014laminar}\@. To perform this comparison, we first calculated the parameter $a_1$ using the predicted value of the drag coefficient, $C_D$, obtained from the correlation developed by Sanjeevi et al. \cite{sanjeevi2022accurate} at a Reynolds number of 300\@. Our calculations revealed that $a_1$ = 3.607\@. 

We have presented our comparison results in Table \ref{table1porlate45}\@. Our findings further confirm our theory, with only a slightly significant difference observed at the lowest Reynolds number\@. The advantage of using our approach is that it significantly reduces the number of simulations required for each geometry and orientation\@. The reduction in the number of simulations is because we only need a single result for the drag coefficient at a moderate Reynolds number to predict the evolution of $C_D$ for the specific geometry and orientation\@. \\

\begin{table}[h!]

\begin{center}
\begin{tabu} to 1.2\textwidth { | X[c] | X[c] | X[c] |X[c] |}
 \hline
  $Re$& Jiang et al.\cite{jiang2014laminar}& Sanjeevi et al.\cite{sanjeevi2022accurate}  (\%)& Eq.(\ref{eqloggeneral}), $a_1$ = 3.607,$K$ = 1.276(\%) \\ 
 \hline
 91

 &1.691 & 1.662($\eval*{$-\frac{CD1-CD2}{CD1}100$}[CD1 = 1.691, CD2 = 1.662][2]$\%) &1.514($\eval*{$-\frac{CD1-CD2}{CD1}100$}[CD1 = 1.691, CD2 =1.514][2]$\%)\\
363

&
1.040 &0.954($\eval*{$-\frac{CD1-CD2}{CD1}100$}[CD1 = 1.040, CD2 = 0.954][2]$\%) &0.964($\eval*{$-\frac{CD1-CD2}{CD1}100$}[CD1 = 1.040, CD2 =0.964][2]$\%)\\
 
 1817

&0.805

 & 0.770($\eval*{$-\frac{CD1-CD2}{CD1}100$}[CD1 = 0.805, CD2 =0.770][2]$\%) &0.750($\eval*{$-\frac{CD1-CD2}{CD1}100$}[CD1 = 0.805, CD2 =0.750][2]$\%)\\ 

\hline
\end{tabu}
\end{center} 
\caption{ Relative error between the predictive $C_D$ coefficient from the logarithmic equations , and other correlations with the numerical results of 
Jiang  et al.\cite{jiang2021large} for different Reynolds numbers for the geometry of prolate with $p_a$ = 6.0, and $\alpha$ = $45^{\circ}$\@.  }
\label{table1porlate45} 
\end{table}
For the subcritical flow regime, we have only a few experimental points to compare our results with for the case of the prolate geometry at two different flow orientations, $\alpha = 0^\circ$ and $\alpha = 90^\circ$. For the case of $\alpha = 0^\circ$, we use Eq. (\ref{eqloggeneral}) with $a_1 = \pi$ and 3.0, which we previously tested\@. For the case of $\alpha = 90^\circ$, we evaluate $a_1$ using the drag coefficient ($C_D$) value obtained from the correlation of Holzer and Sommerfeld \citep{holzer2008new} at $Re = 50$ and $p_a = 1.33$, where $C_D$ is equal to 1.6. The results of the comparison are shown in Table \ref{table1porlatecomparsion}.\\

For the case of prolate spheroid geometry with aspect ratios of $p_a = 2$ and 4.0, and $\alpha = 0^\circ$, the logarithmic equation (Eq. \ref{eqloggeneral}) with $a_1 = \pi$ shows excellent agreement with the experiments. Meanwhile, for the case of the prolate spheroid with $p_a = 8.0$ and $\alpha = 0^\circ$, the logarithmic equation (Eq. \ref{eqloggeneral}) with $a_1 = 3.0$ predicts the experimental results closely. The current comparison for the case of $\alpha = 0^\circ$ solidifies our assumption that, for this particular orientation, the value of the $a_1$ coefficient fluctuates between $\pi$ and 3.0. However, the exact value of $p_a$ at which the transition in the value of $a_1$ occurs is not known.
\begin{table}[H]

\begin{center}
\begin{tabu} to 1.2\textwidth { | X[c] | X[c] | X[c] |X[c] |}
 \hline
 & Eq.(\ref{eqloggeneral}), $a_1$ = $\pi$ (\%)&  Eq.(\ref{eqloggeneral}), $a_1$ = 3.0 (\%)& Eq.(\ref{eqloggeneral}), $a_1$ = 3.349(\%) \\ 
 \hline
 Ref\cite{cengel2013ebook}, $p_a$ = 2.0,$Re$ = $2\times 10^5$, $C_D$ = 0.3, $\alpha$ = $0^{\circ}$

 &0.317($\eval*{$-\frac{CD1-CD2}{CD1}100$}[CD1 = 0.3, CD2 = 0.317][2]$\%),$K$ = 0.955 & 0.176($\eval*{$-\frac{CD1-CD2}{CD1}100$}[CD1 = 0.3, CD2 = 0.176][2]$\%),$K$ = 0.955 &- \\
 \hline
Ref\cite{cengel2013ebook}, $p_a$ = 4.0,$Re$ = $2\times 10^5$, $C_D$ = 0.3, $\alpha$ = $0^{\circ}$& 0.317($\eval*{$-\frac{CD1-CD2}{CD1}100$}[CD1 = 0.3, CD2 =  0.317][2]$\%),$K$ =1.006 & 0.176($\eval*{$-\frac{CD1-CD2}{CD1}100$}[CD1 = 0.3, CD2 = 0.176][2]$\%),$K$ = 1.006&-\\
\hline
Ref\cite{cengel2013ebook}, $p_a$ = 8.0,$Re$ = $2\times 10^5$, $C_D$ = 0.2, $\alpha$ = $0^{\circ}$&0.317($\eval*{$-\frac{CD1-CD2}{CD1}100$}[CD1 = 0.2, CD2 =  0.317][2]$\%),$K$ =1.157& 0.186($\eval*{$-\frac{CD1-CD2}{CD1}100$}[CD1 = 0.2, CD2 =0.186][2]$\%),$K$ = 1.157 &-\\
\hline
Ref\cite{knudsen1959fluid}, $p_a$ = 1.33, $Re$ = $1\times 10^5$, $C_D$ = 0.6, $\alpha$ = $90^{\circ}$&-&-&0.538($\eval*{$-\frac{CD1-CD2}{CD1}100$}[CD1 = 0.6, CD2 = 0.538][2]$\%),$K$ = 1.026\\

\hline
\end{tabu}
\end{center} 
\caption{ Relative error between the predictive $C_D$ coefficient from the logarithmic equations and the experimental results of Cengel and Cimbala \cite{cengel2013ebook}, and Knudsen and Katz \cite{knudsen1959fluid}, for different prolate spheroids geometers, and flow orientations at high Reynolds numbr regime\@.}
\label{table1porlatecomparsion} 
\end{table}
\subsubsection{Spherocylinders}
\begin{figure}[H]
\begin{center}
\includegraphics [scale=0.8, trim = 0 0 0 0,clip]{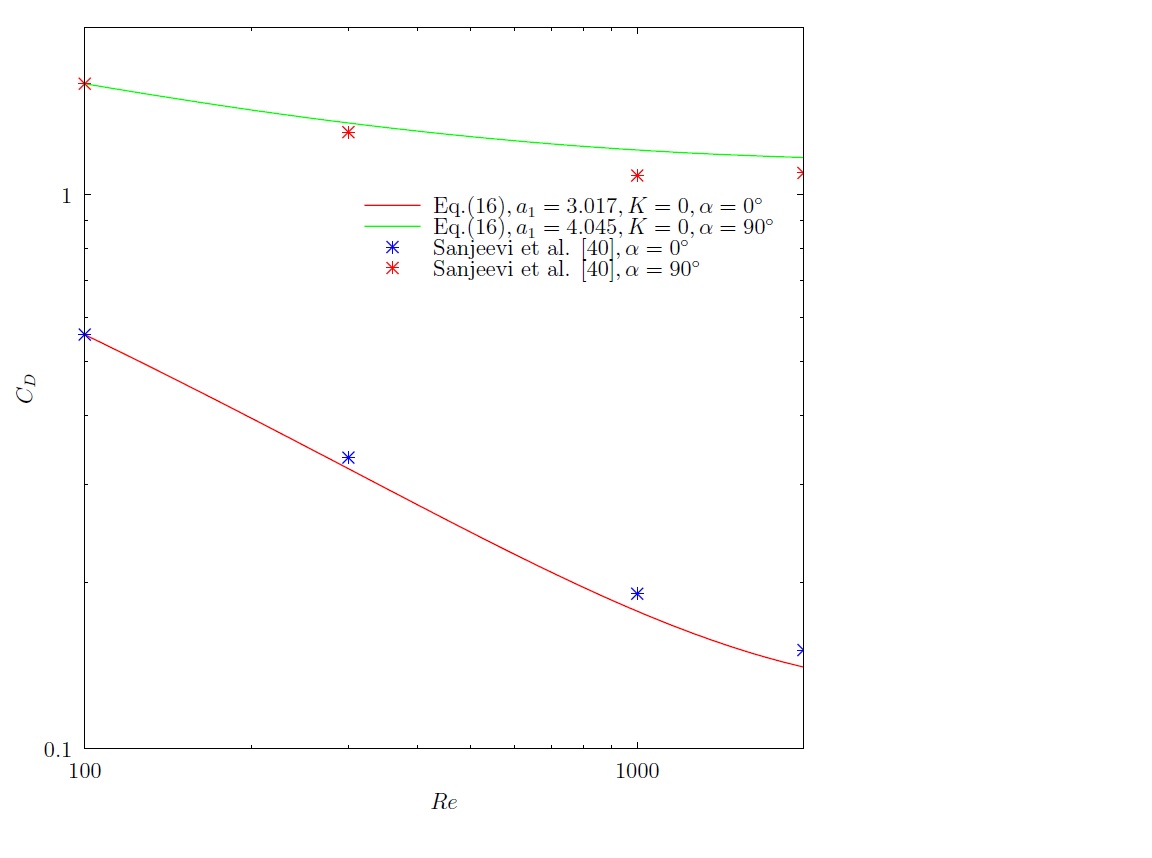}
\end{center}
\caption {Comparison of $C_D$ values predicted by the logarithmic equations, with the numerical results of Sanjeevi et al.\cite{sanjeevi2018drag} for the case of a spherocylinder with aspect ratio of 4.0 for $\alpha$ = $0^{\circ}$, and  $90^{\circ}$\@. }

\label{Fig1sphrocy} 
\end{figure}

For the case of spherocylinders, we will test our theory against the data of Sanjeevi et al. \cite{sanjeevi2018drag} for a spherocylinder with an aspect ratio of 4.0 at two angles of attack, $\alpha = 0^\circ$ and $\alpha = 90^\circ$\@. We will also compare our results with the correlation of Feng and Michaelides \cite{feng4435415general} for $pa = 2.0$ at the same angles of attack as used by Sanjeevi et al. \cite {sanjeevi2018drag}\@.
The discrete data provided by Sanjeevi et al. \cite{sanjeevi2018drag} allows us to directly compare our theory with the approximate solution of the Navier-Stokes equations\@. To evaluate the coefficient $a_1$ for the two orientations, we will use the drag coefficient at $Re = 100$  from Sanjeevi et al. \cite{sanjeevi2018drag} data, and assign a zero value to the coefficient $K$. The obtained values for $a_1$ are 3.017 for $\alpha = 0^\circ$ and 4.045 for $\alpha = 90^\circ$\@.
The comparison between the logarithmic equations (Eq. (\ref{eqloggeneral}), $a_1 = 3.017$, $K = 0$) and (Eq. (\ref{eqloggeneral}), $a_1 = 4.045$, $K = 0$) and the numerical results of Sanjeevi et al. \cite{sanjeevi2018drag} is depicted in Figure \ref{Fig1sphrocy}\@. The logarithmic equations accurately predict the numerical results of Sanjeevi et al. \cite{sanjeevi2018drag} for both angles of attack. The value of $a_1$ indicates that when $\alpha = 90^\circ$, the spherocylinder particle behaves approximately like an infinite cylinder.

Recently Feng and Michaelides\cite{feng4435415general} provided a correlation for the drag coefficient for different flow orientations, from their numerical simulations\@. The correlation is complicated function in the Reynolds number, and the aspect ratio\@. We will evaluate $a_1$, at $Re$ = 100 from Feng and Michaelides\cite{feng4435415general} correlation\@. The comparison   between the logarithmic equations and the drag coefficients from the  Feng and Michaelides\cite{feng4435415general}\@ are shown in Table \ref{tableFengSphero}. Our results compare well for the two orientation with the results Feng and Michaelides\cite{feng4435415general}, only for the case of $Re$ = 300, where there is a slight deviation\@. 

\begin{table} [H]
\begin{center}
\begin{tabu} to 1.2\textwidth { | X[c] | X[c] | X[c] |X[c] |X[c]|}
 \hline
  $Re$& Feng and Michaelides\cite{feng4435415general} $\alpha$ = $0^{\circ}$& Eq.(\ref{eqloggeneral}), $a_1$ = 3.142, $K$ =0, $\alpha$ = $0^{\circ}$ \ (\%)&Feng and Michaelides\cite{feng4435415general} $\alpha$ = $90^{\circ}$&Eq.(\ref{eqloggeneral}), $a_1$ = 3.566,$K$ = 0,$\alpha$ = $90^{\circ}$ (\%) \\ 
 \hline
 50
 &0.907 & 0.896($\eval*{$-\frac{CD1-CD2}{CD1}100$}[CD1 = 0.907, CD2 = 0.896][2]$\%) &1.342&1.320($\eval*{$-\frac{CD1-CD2}{CD1}100$}[CD1 = 1.342, CD2 =1.320][2]$\%)\\
 100
 &0.683 & 0.683($\eval*{$-\frac{CD1-CD2}{CD1}100$}[CD1 = 0.683, CD2 = 0.683][2]$\%) &1.108&
 1.108($\eval*{$-\frac{CD1-CD2}{CD1}100$}[CD1 = 1.108, CD2 =1.108][2]$\%)\\
 200
 &0.504 & 0.520($\eval*{$-\frac{CD1-CD2}{CD1}100$}[CD1 = 0.504, CD2 = 0.520][2]$\%) &0.906&
 0.944($\eval*{$-\frac{CD1-CD2}{CD1}100$}[CD1 = 0.944, CD2 =0.906][2]$\%)\\
 300
 &0.445 & 0.418($\eval*{$-\frac{CD1-CD2}{CD1}100$}[CD1 = 0.445, CD2 = 0.418][2]$\%) &0.796&
 0.869($\eval*{$-\frac{CD1-CD2}{CD1}100$}[CD1 = 0.796, CD2 =0.869][2]$\%)\\
\hline
\end{tabu}
\end{center} 
\caption{Relative error between the predictive $C_D$ coefficient from the logarithmic equations , with the correlation of Feng and Michaelides\cite{feng4435415general}
 for different Reynolds numbers for the geometry of spherocylinder with $p_a$ = 2.0, and $\alpha$ = $0^{\circ}$, $90^{\circ}$\@.}
\label{tableFengSphero} 
\end{table}

\subsubsection{Irregular non-spherical particles}

Finally, we will implement our theory for irregular non-spherical particles\@. These types of particles are complicated to characterize geometrically\@. Recently, Lain et al. \cite{lain4317628sphericity} developed a correlation for the drag coefficient of irregular non-spherical particles using sphericity as the main geometrical characterization parameter\@. They performed thousands of direct numerical simulations for random orientations of particles, considering Reynolds numbers up to 200 and a narrow range of sphericity between 0.7 and 0.95\@. They derived a correlation similar to Haider and Levenspiel \cite{haider1989drag}, as shown in Eq. (\ref{eqpowg1}), only the coefficients in Eq.(\ref{eqpowg2}) are different \@.
The correlation proposed by Lain et al. \cite{lain4317628sphericity} provides an opportunity to test the universality of the $\Delta$ parameter beyond regular-shaped particles\@. We will evaluate the coefficient $a_1$ at $Re = 100$ for two values of sphericity, $\phi = 0.7$ and $\phi = 0.8$, with $K = 1.0$\@. The results of the logarithmic equations are listed in Table \ref{tableiraqular}, and our predictions for the drag coefficient align closely with the results of Lain et al. \cite{lain4317628sphericity}, confirming the universality of our theory\@.\\

The Reynolds number used in all correlations presented from oblate subsections until now is based on the volume-equivalent sphere diameter\@.\\

\begin{figure}[H]
\begin{center}
\includegraphics [scale=0.8, trim = 0 0 0 0,clip]{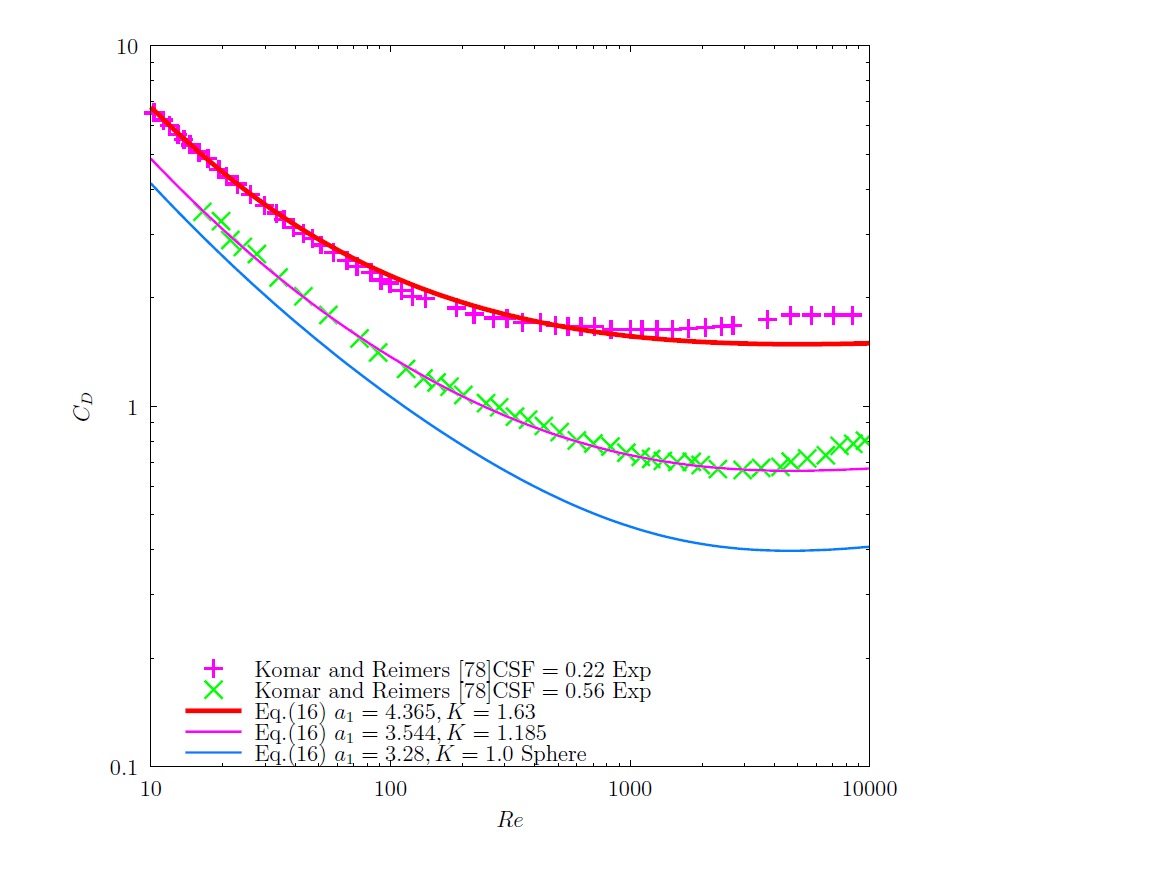}
\end{center}
\caption {Comparison of $C_D$ values predicted by the logarithmic equation, and the experimental  results of  Komar and Reimeres\cite{komar1978grain} for the case of different irregular non-spherical particle geometries\@. }
\label{Figgrian} 
\end{figure}

We will compare the predictions of the drag coefficient from our logarithmic-based model with two different sources of experimental results for irregular non-spherical particles. The first set of experimental results is from Komar and Reimers \cite{komar1978grain}, for various pebbles with different Corey shape factors (CSF). Their drag coefficient data are mainly for Reynolds numbers in the range of 0.07 to 1.5. However, they used data from the experiments of Corey \cite{corey1949influence} for high Reynolds numbers up to $10^5$, creating fitting curves that connect their data with those of Corey \cite{corey1949influence}. They mentioned that the accuracy of the fitting deteriorates for Reynolds numbers higher than $10^4$ due to uncertainty in the Corey \cite{corey1949influence} data in that specific range. 
For the case of $\text{CSF} = 0.22$, we calculated the value of $K$ by using the value of $C_D = 96.15$ at $Re = 0.407$ from the data of Komar and Reimers \cite{komar1978grain}, and found that $K = 1.63$. For calculating the value of $a_1$, we used the value of $C_D = 5.068$ at $Re = 15.89$. For the case of $\text{CSF} = 0.56$, using the same procedure, we evaluated $K$ using the value of $C_D = 94.8$ at $Re = 0.3$, and found that $K = 1.18$. We found the value of $a_1$ using the value of $C_D = 3.46$ at $Re = 16.41$, determining that $a_1 = 3.54$. 
Our predictions using the logarithmic equation (Eq. \ref{eqloggeneral}) show excellent agreement with the results of Komar and Reimers \cite{komar1978grain}, as illustrated in Figure \ref{Figgrian}. \\

\begin{figure}[H]
\begin{center}
\includegraphics [scale=0.8, trim = 0 0 0 0,clip]{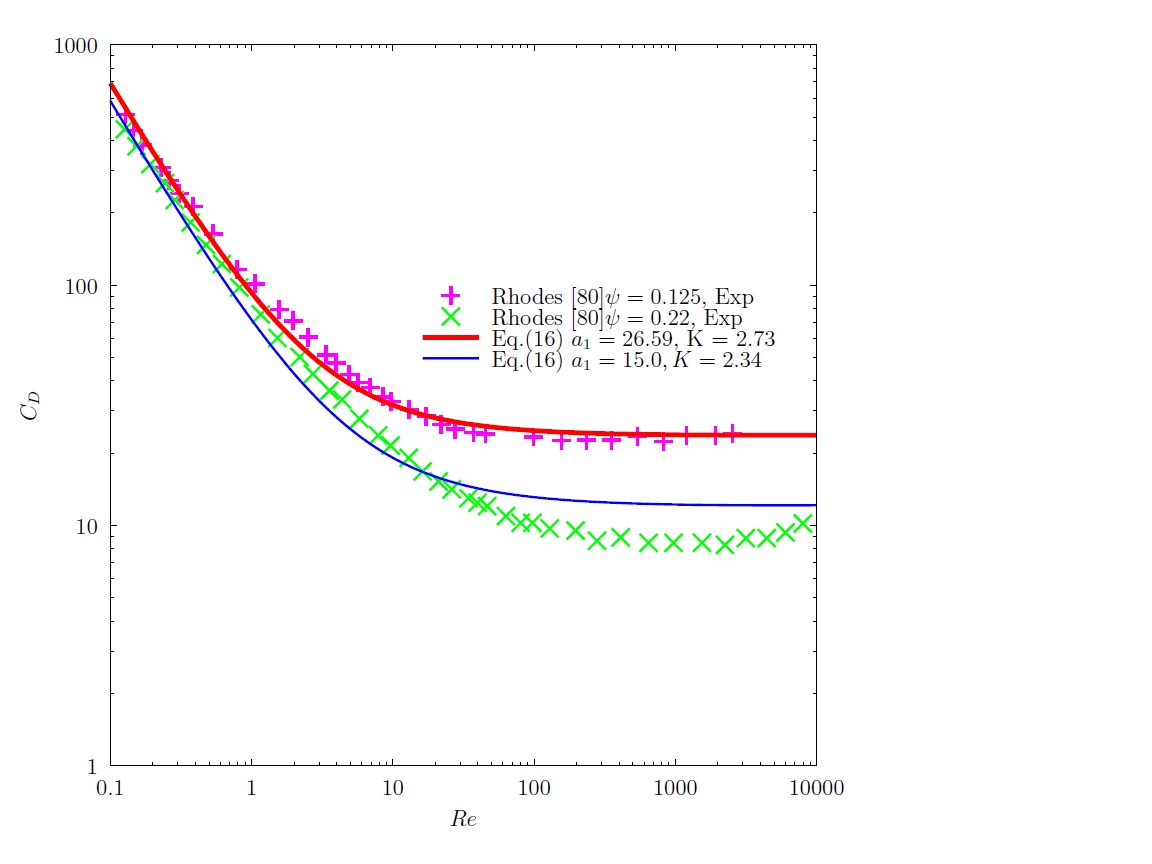}
\end{center}
\caption {Comparison of $C_D$ values predicted by the logarithmic equation, and the experimental  results of Rhodes\cite{rhodes2024introduction} for the case of different irregular non-spherical particle geometries\@. }
\label{FigRhodesnonspherical} 
\end{figure}

Finally, we will compare our results with those of Rhodes \cite{rhodes2024introduction} for non-spherical irregular particles with different values of sphericity ($\psi$). The comparison is shown in Figure \ref{FigRhodesnonspherical}. We used two points from the curve of Rhodes \cite{rhodes2024introduction} to calculate the coefficients $K$ and $a_1$ for the logarithmic equation (Eq. (\ref{eqloggeneral})). Specifically, for the case of $\psi = 0.125$, we calculated the value of $K$ using the value of $C_D$ at $Re = 0.127$, finding $K = 2.734$. To evaluate the value of $a_1$, we used the value of $C_D$ at $Re = 17.847$, finding $a_1 = 26.596$. For the case of $\psi = 0.22$, we used the value of $C_D$ at $Re = 0.126$, finding $K = 2.34$, and we used the value of $C_D$ at $Re = 17.33$, finding $a_1 = 15.0$. The main conclusion from Figure \ref{FigRhodesnonspherical} is that our theory applies to irregular particle shapes, and that the asymptotic value of the drag coefficient can be predicted using values of the drag coefficient at low Reynolds numbers. Irrespective of the shape of the bluff object, they all follow a general law described by Eq. (\ref{eqhighRe2}).

\begin{table} [H]  
\begin{center}
\begin{tabu} to 1.2\textwidth { | X[c] | X[c] | X[c] |X[c] |X[c]|}
 \hline
  $Re$& Lain et al.\cite{lain4317628sphericity},$\phi$ = 0.8 & Eq.(\ref{eqloggeneral}), $a_1$ = 3.630, $K$ =1.0, $\phi = 0.8$ \ (\%)&Lain et al.\cite{lain4317628sphericity},$\phi$ = 0.7&Eq.(\ref{eqloggeneral}), $a_1$ = 3.870,$K$ = 1,$\phi$ =0.7 (\%) \\ 
 \hline
 50
 &1.893 & 1.867($\eval*{$-\frac{CD1-CD2}{CD1}100$}[CD1 = 1.893, CD2 = 1.867][2]$\%) &2.181&2.108($\eval*{$-\frac{CD1-CD2}{CD1}100$}[CD1 = 2.181, CD2 =2.108][2]$\%)\\
 100
 &1.414 & 1.413($\eval*{$-\frac{CD1-CD2}{CD1}100$}[CD1 = 1.414, CD2 = 1.413][2]$\%) &1.654&
1.653($\eval*{$-\frac{CD1-CD2}{CD1}100$}[CD1 = 1.654, CD2 =1.653][2]$\%)\\
 150
 &1.236 & 1.231($\eval*{$-\frac{CD1-CD2}{CD1}100$}[CD1 = 1.236, CD2 = 1.231][2]$\%) &1.458&
 1.471($\eval*{$-\frac{CD1-CD2}{CD1}100$}[CD1 = 1.458, CD2 =1.471][2]$\%)\\
 200
 &1.140& 1.128($\eval*{$-\frac{CD1-CD2}{CD1}100$}[CD1 = 1.140, CD2 =1.128][2]$\%) &1.354&
 1.369($\eval*{$-\frac{CD1-CD2}{CD1}100$}[CD1 = 1.354, CD2 =1.369][2]$\%)\\
\hline
\end{tabu}
\end{center} 
\caption{Relative error between the predictive $C_D$ coefficient from the logarithmic equations , with the correlation of Lain et al.\cite{lain4317628sphericity}
 for different Reynolds numbers for different irregular particles with $\phi$ = 0.8, and 0.7\@.}
\label{tableiraqular} 
\end{table}
\subsubsection{Simple power law equation}
We want to further test our theory, which states that the derivative of the drag coefficient with respect to the Reynolds number is not dependent on the geometry of the bluff body beyond the Stokes flow regime. This time, we will use a power-based equation similar to the structure proposed by Abraham \cite{abraham1970functional}, and Takami and Keller \cite{takami1969steady} for the case of cavity flow. The equation is as follows:
\begin{equation}\label{eqpowersimple}
C_D = a_{p_{1}}+\frac{a_{p_{2}}}{Re}+\frac{a_{p_{3}}}{\sqrt{Re}}
\end{equation}  
 
In our previous work \cite{el2022logarithmic}, the symbolic regression algorithm found a similar equation to that of Eq.~(\ref{eqpowersimple}) for the case of the sphere. For Eq.~(\ref{eqpowersimple}), the coefficient $a_{p_2}$ is the Stokes resistance and can be determined from the shape geometry of the bluff body, while the term $\frac{a_{p_3}}{\sqrt{Re}}$ is the only one that plays a role in the derivative of the $C_D$ with respect to $Re$. According to our theory, this term must be universal and does not depend on the shape of the object. Thus, we can obtain its value from the case of the sphere, which we found that it is 4.119 (Ref \cite{el2022logarithmic}, Eq.~(9)). It will remain only to determine the value of $a_{p_1}$, which will be done by the same method as the one we used to determine the coefficient $a_1$ for the logarithmic-based equation, which was to use a single value of the $C_D$ at a certain $Re$ for the investigated geometry. We will shortly show that $a_{p_1}$ is related to the asymptotic value of the form drag. The derivative of the drag coefficient with respect to the Reynolds number for the simple power equation is the following:

\begin{equation}\label{eqderivative_power}
\Delta_p = -\frac{2.059}{Re^{\frac{3}{2}}}
\end{equation}

\begin{figure}[H]
\begin{center}
\includegraphics [scale=0.8, trim = 0 0 0 0,clip]{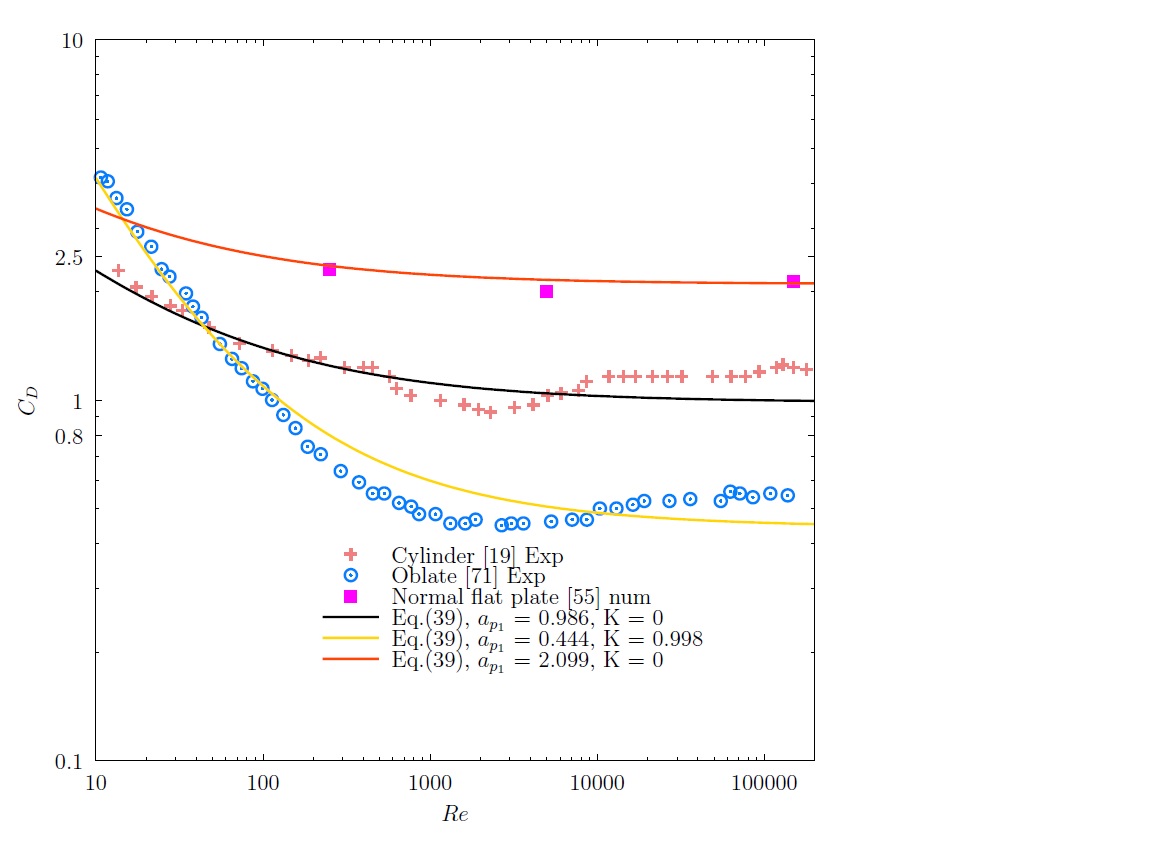}
\end{center}
\caption {Comparison of $C_D$ values predicted by the power based  equation \ref{eqpowersimple}, for different geometries of bluff bodies\@.  }
\label{Figpowereqdifferntgeometries} 
\end{figure}

To evaluate the behaviour of Eq. (\ref{eqpowersimple}), we will test its predictions for different bluff body geometries as shown in Figure \ref{Figpowereqdifferntgeometries}. The values of $C_D$ used to determine the coefficient $a_{p_3}$ are listed in Table \ref{tablepowerap} \@. Also, if we compare the $a_{p_{1}}$ values listed in table \ref{tablepowerap} to the asymptotic value of the form drag for sub-critical flow regime from the Figure \ref{Figpowereqdifferntgeometries} we see that thier values are very close \@. Eq. (\ref{eqpowersimple}) predicts the values of $C_D$ for different geometries with acceptable accuracy, which further confirms our theory that the derivative of the drag coefficient with respect to the Reynolds number remains consistent across various bluff body geometries. This behaviour is similar to that described by the logarithmic equation (Eq. (\ref{eqloggeneral}))

It was previously thought that the significant differences in the wake structure between two-dimensional and three-dimensional bodies would lead to different drag coefficient responses to changes in flow, hoowever, this is not the case. The existence of free constants $a_1$ and $a_{p_1}$ in Eq. (\ref{eqloggeneral}) and Eq. (\ref{eqpowersimple}) can be explained by the general property of drag coefficient predictive models for different geometries: which are a family of functions that share the  same derivative with respect to the Reynolds number\@. According to the mean value theorem, functions with this property contain a free constant, implying that the asymptotic form drag coefficient ($a_{p_1}$) constant existence  is a result of this theorem, the same implies to the $a_1$ coefficient for the logarithmic equation\@.
Eq. (\ref{eqpowersimple}) also demonstrates that the boundary layer thickness around the body plays a crucial role in the drag coefficient, with a global value represented by $a_{p_3}$ that remains invariant across different body geometries.
Beyond the differences in mathematical structure between the power-based equation (Eq. (\ref{eqpowersimple})) and the logarithmic equation (Eq. (\ref{eqloggeneral})), the value of the free constant in each equation represents a different flow regime. For example, in the former, it represents the form drag, while in the latter, it represents the frictional drag. In general, the logarithmic equation provides better predictions than the power-based equation.

\begin{figure}[H]
\begin{center}
\includegraphics [scale=0.8, trim = 0 0 0 0,clip]{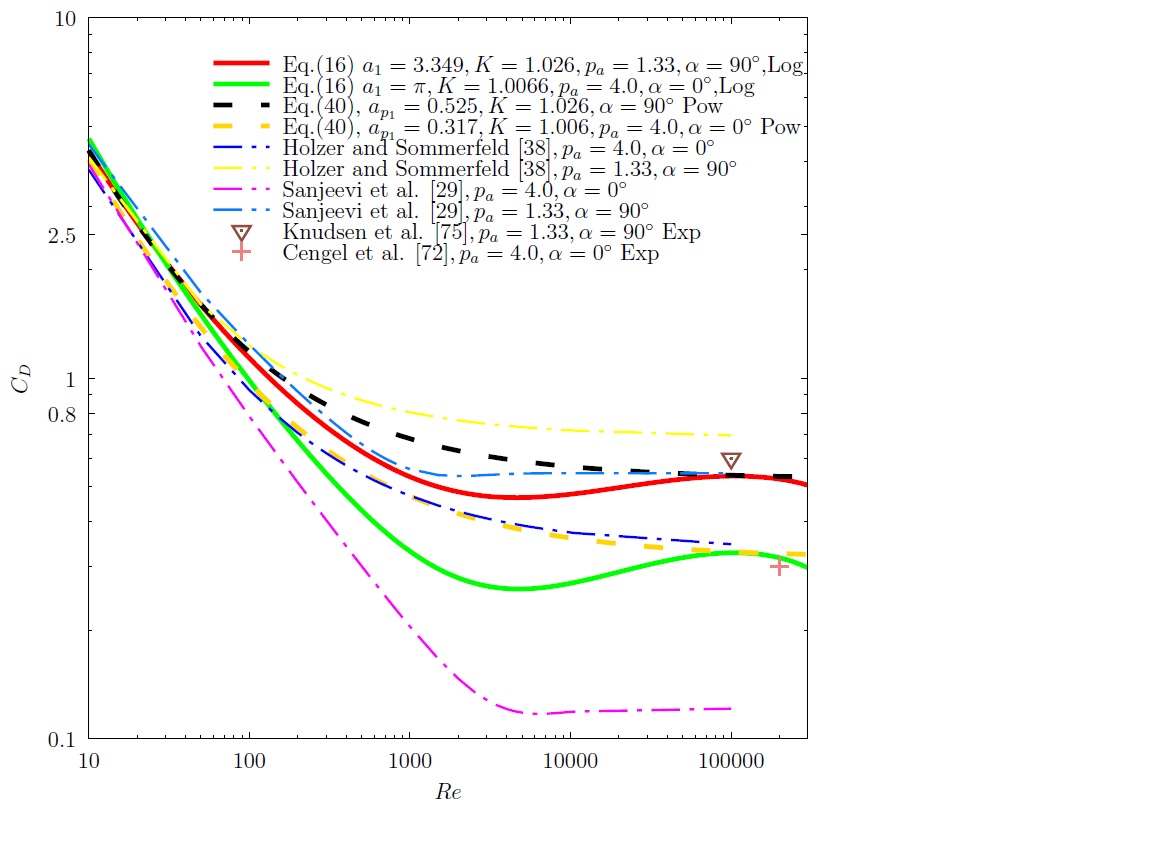}
\end{center}
\caption { compression of the $C_D$ predictions of power based equation (Eq.(\ref{eqpowersimple}), and the logarithmic based equation  Eq.(\ref{eqloggeneral}) with those from the literature  for different prolate spheroid geometries\@.    }
\label{Figpowerspheriod} 
\end{figure}

We will compare the predictions of the power-based general equation with the logarithmic equation, available correlations from the literature, and a few experimental data points, as shown in Figure \ref{Figpowerspheriod}. From Figure \ref{Figpowerspheriod}, we observe that the predictions of Eq. (\ref{eqpowersimple}) are consistent with the available data in the literature for specific spheroid geometries. The predictions of the power equation and the logarithmic equation align, especially in the very high and low Reynolds number regimes, for both geometries included in the comparison. It is noteworthy that both $a_1$ and $a_{p_1}$ are computed using the same data points.
The results in Figure \ref{Figpowerspheriod} demonstrate that the derivatives $\Delta$ and $\Delta_p$ are linked, even though $\Delta$ is more accurate than $\Delta_p$. The power equation clearly shows that the asymptotic values of the drag coefficient in the subcritical regime can be determined from flow regimes at low Reynolds numbers where frictional forces play an essential role. For example, we obtained the values of $a_{p_1}$ for the two different spheroids with $p_a = 4.0$ and $1.33$ respectively  using the values of $C_D$ at $Re = 10$ and $50$\@. The $a_{p_1}$ values  are consistent with experimental values reported by \cite{cengel2013ebook,knudsen1959fluid}, at the  far edge of the subcritical regime. \\

The current predictions challenge the prevailing perspective that asymptotic form drag values can only be predicted from  data at high Reynolds number flow regimes only\@. The correlations available in the literature often incorporate the asymptotic value of the form drag artificially into their mathematical structure. Further investigation is necessary to fully understand why different bluff body geometries exhibit the same response to changes in the Reynolds number. This phenomenon could be related to the observations of Roshko \cite{roshko1955wake}, who found that the wake structure is similar for different cylinder geometries\@.

\begin{table} [H]  
\begin{center}
\begin{tabu} to 1.2\textwidth { | X[c] | X[c] | X[c] |X[c] |}
 \hline
  Geometry& $Re_u$ & $C_{D_u}$ &$a_{p_1}$, $K$ Eq.(\ref{eqpowersimple} \\ 
 \hline
 Oblate $p_a$ =0.5, $\alpha$ = $90^{\circ}$ & 55.28& 1.432 \cite{bagheri2016drag}& 0.444,0.998 \\
 \hline
 Cylinder & 114.58 & 1.376 \cite{wieselsberger1922further}& 0.986,0\\
 \hline 
 Normal flat plate&250& 2.36 \cite{najjar1998low}& 2.099, 0\\
 \hline 
 Prolate spheroid $p_a$ = 4.0, $\alpha$ = $0^{\circ}$&10& 4.032 Eq.(\ref{eqloggeneral})& 0.318, 1.006\\
 \hline 
 Prolate spheroid $p_a$ = 1.33, $\alpha$ = $90^{\circ}$&50& 1.6 \cite{holzer2008new}& 0.52, 1.0269\\
 \hline
 \end{tabu}
\end{center} 
\caption{  Listing of the values of $Re$ and $C_D$ used to determine the $a_{p_1}$ coefficient for the power-based equation (Eq. (\ref{eqpowersimple})).   }
\label{tablepowerap} 
\end{table}

\subsection{Conclusions}

We showed that after a rigorous comparison with available results in the literature, the drag coefficient can be evaluated with high accuracy by Eq.(\ref{eqloggeneral})  {which was } initially derived for the case of a sphere but can be applied to any bluff body geometry by linearly adding a constant ($a_1$ coefficient)  which  depends on the shape and   orientation of the  {body}\@. The logarithmic Eq.(\ref{eqloggeneral}) is applicable for the case of incompressible uniform flow\@. We draw the following conclusions:

\begin{itemize}
\item One of the key contributions of this investigation is the novel finding that, {in} the inertial flow regime, the rate of change of the drag coefficient with respect to the Reynolds number is independent of the geometry or orientation of the object\@. This property has profound implications for drag coefficient prediction of different shapes, and we believe it is directly linked to the vorticity generated in the fluid by the solid boundary\@.

\item We have discovered a strong correlation between the $a_1$ coefficient of the logarithmic equation and the frictional drag predicted by the boundary layer theory. Moreover, we have found that only using logarithmic equations we can get values of the $a_1$ coefficient that is linked to  boundary layer theory\@. The  link between {the} $a_1$ coefficient  and  boundary layer theory may be a result of a yet undiscovered theory in fluid mechanics\@.

\item By utilizing Eq.(\ref{eqloggeneral}), we can significantly reduce the computational time required to determine the drag coefficient for a single {body} at a specific flow orientation over a wide range of Reynolds numbers\@. This efficiency arises from the fact that we only need to compute the drag coefficient at a single Reynolds number, and predict the rest by implementing the logarithmic equation Eq.(\ref{eqloggeneral}) \@.
\item We showed that by  using the physics informed symbolic regression algorithm \cite{el2019solving, el2022logarithmic} there is a possibility to obtain accurate solutions that can be compared well with  high fidelity numerical solutions by using  empirical data, and previous knowledge {of} analytical solutions\@.  \\

Also, we found that the power-based equation (Eq.(\ref{eqpowersimple})) can predict the drag coefficient of the different bluff geometries, and we showed that we can predict the asymptotic pressure drag value at the subcritical flow regime by using Eq.(\ref{eqpowersimple}), from $C_D$ data at low Reynolds regime\@.  

\end{itemize} 
 \section*{Acknowledgements}
The first author  thanks Dimitra Damianidou, and Mohamed Fakhri El Hasadi for the enlightening discussions about the subject.  Finally, the authors thank the European Research Council for its financial support under its consolidator grant scheme, contract no. 615096 (NonSphereFlow). 
\bibliography{22} 

\begin{thebibliography}{10}

\bibitem{vogel2020life}
S.~Vogel, {\em Life in moving fluids: the physical biology of flow-revised and
  expanded second edition}.
\newblock Princeton university press, 2020.

\bibitem{cohen2019fluid}
I.~Cohen, S.~C. Whitehead, and T.~Beatus, ``Fluid dynamics and control of
  insect flight,'' {\em Nature Reviews Physics}, vol.~1, no.~11, pp.~638--639,
  2019.

\bibitem{el2016simulating}
Y.~M. El~Hasadi and M.~Crapper, ``Simulating the hydrodynamics of
  self-propelled colloidal clusters using stokesian dynamics,'' {\em
  Biomicrofluidics}, vol.~10, no.~6, 2016.

\bibitem{el2017self}
Y.~M. El~Hasadi and M.~Crapper, ``Self-propelled nanofluids a path to a highly
  effective coolant,'' {\em Applied Thermal Engineering}, vol.~127,
  pp.~857--869, 2017.

\bibitem{el2020self}
Y.~M. El~Hasadi and M.~Crapper, ``Self-propelled nanofluids a coolant inspired
  from nature with enhanced thermal transport properties,'' {\em Journal of
  Molecular Liquids}, vol.~313, p.~113548, 2020.

\bibitem{hucho1993aerodynamics}
W.~Hucho and G.~Sovran, ``Aerodynamics of road vehicles,'' {\em Annual review
  of fluid mechanics}, vol.~25, no.~1, pp.~485--537, 1993.

\bibitem{anderson2011ebook}
J.~Anderson, {\em EBOOK: Fundamentals of Aerodynamics (SI units)}.
\newblock McGraw hill, 2011.

\bibitem{eckert2019ludwig}
M.~Eckert, {\em Ludwig Prandtl: A Life for Fluid Mechanics and Aeronautical
  Research}.
\newblock Springer, 2019.

\bibitem{winslow2018basic}
J.~Winslow, H.~Otsuka, B.~Govindarajan, and I.~Chopra, ``Basic understanding of
  airfoil characteristics at low reynolds numbers (10 4--10 5),'' {\em Journal
  of Aircraft}, vol.~55, no.~3, pp.~1050--1061, 2018.

\bibitem{hansen2015aerodynamics}
M.~O. Hansen, {\em Aerodynamics of wind turbines}.
\newblock Routledge, 2015.

\bibitem{mahajan2019fluid}
V.~V. Mahajan, J.~Mehmood, Y.~M. El~Hasadi, and J.~T. Padding, ``Fluid medium
  effect on stresses in suspensions of high-inertia rod-like particles,'' {\em
  Chemical Engineering Science: X}, vol.~3, p.~100030, 2019.

\bibitem{darrigol2005worlds}
O.~Darrigol, {\em Worlds of flow: A history of hydrodynamics from the
  Bernoullis to Prandtl}.
\newblock Oxford University Press, 2005.

\bibitem{stokes1851effect}
G.~G. Stokes, {\em On the effect of the internal friction of fluids on the
  motion of pendulums}, vol.~9.
\newblock Pitt Press Cambridge, Cambridge, UK, 1851.

\bibitem{stewartson1981d}
K.~Stewartson, ``D’alembert’s paradox,'' {\em SIAM review}, vol.~23, no.~3,
  pp.~308--343, 1981.

\bibitem{Prandtl1904}
L.~Prandtl, ``Über flussigkeitsbewegung bei sehr kleiner reibung,'' in {\em
  Verhandlg. III. Intern. Math. Kongr. Heidelberg}, pp.~484--491, Math
  Congress, 1904.

\bibitem{day1990no}
M.~A. Day, ``The no-slip condition of fluid dynamics,'' {\em Erkenntnis},
  vol.~33, no.~3, pp.~285--296, 1990.

\bibitem{blasius1908grenzschichten}
H.~Blasius, {\em Grenzschichten in Fl{\"u}ssigkeiten mit kleiner Reibung}.
\newblock University of Gottingen, Germany, 1908.

\bibitem{birkhoff1960study}
G.~Birkhoff and A.~Hydrodynamics, {\em study in Logic, Fact and Similitude}.
\newblock Princeton Univ. Press, 1960.

\bibitem{wieselsberger1922further}
C.~Wieselsberger, ``Further information on the laws of fluid resistance,''
  tech. rep., 1922.

\bibitem{churchill2013viscous}
H.~Brenner, {\em Viscous flows: the practical use of theory}.
\newblock Butterworth-Heinemann, 2013.

\bibitem{achenbach1972experiments}
E.~Achenbach, ``Experiments on the flow past spheres at very high reynolds
  numbers,'' {\em Journal of Fluid Mechanics}, vol.~54, no.~3, pp.~565--575,
  1972.

\bibitem{khan2018flow}
M.~H. Khan, P.~Sooraj, A.~Sharma, and A.~Agrawal, ``Flow around a cube for
  reynolds numbers between 500 and 55,000,'' {\em Experimental Thermal and
  Fluid Science}, vol.~93, pp.~257--271, 2018.

\bibitem{proudman1957expansions}
I.~Proudman and J.~Pearson, ``Expansions at small reynolds numbers for the flow
  past a sphere and a circular cylinder,'' {\em Journal of Fluid Mechanics},
  vol.~2, no.~3, pp.~237--262, 1957.

\bibitem{kaplun1957asymptotic}
S.~Kaplun and P.~Lagerstrom, ``Asymptotic expansions of navier-stokes solutions
  for small reynolds numbers,'' {\em Journal of Mathematics and Mechanics},
  vol.~6, no.~5, pp.~585--593, 1957.

\bibitem{thom1928boundary}
A.~Thom, ``The boundary layer of the front portion of a cylinder,'' tech. rep.,
  HM Stationery Office, 1928.

\bibitem{frossling1958evaporation}
N.~Fr{\"o}ssling, ``Evaporation, heat transfer, and velocity distribution in
  two-dimensional and rotationally symmetrical laminar boundary-layer flow,''
  {\em Fysiografiska Saellskapets Handlingar}, vol.~51, no.~NACA-TM-1432, 1958.

\bibitem{dennis1971calculation}
S.~Dennis and J.~Walker, ``Calculation of the steady flow past a sphere at low
  and moderate reynolds numbers,'' {\em Journal of Fluid Mechanics}, vol.~48,
  no.~4, pp.~771--789, 1971.

\bibitem{cheng2017large}
W.~Cheng, D.~Pullin, R.~Samtaney, W.~Zhang, and W.~Gao, ``Large-eddy simulation
  of flow over a cylinder with: a skin-friction perspective,'' {\em Journal of
  Fluid Mechanics}, vol.~820, pp.~121--158, 2017.

\bibitem{sanjeevi2022accurate}
S.~K. Sanjeevi, J.~F. Dietiker, and J.~T. Padding, ``Accurate hydrodynamic
  force and torque correlations for prolate spheroids from stokes regime to
  high reynolds numbers,'' {\em Chemical Engineering Journal}, vol.~444,
  p.~136325, 2022.

\bibitem{ouchene2020numerical}
R.~Ouchene, ``Numerical simulation and modeling of the hydrodynamic forces and
  torque acting on individual oblate spheroids,'' {\em Physics of Fluids},
  vol.~32, no.~7, p.~073303, 2020.

\bibitem{proudman1960example}
I.~Proudman, ``An example of steady laminar flow at large reynolds number,''
  {\em Journal of Fluid Mechanics}, vol.~9, no.~4, pp.~593--602, 1960.

\bibitem{haider1989drag}
A.~Haider and O.~Levenspiel, ``Drag coefficient and terminal velocity of
  spherical and nonspherical particles,'' {\em Powder technology}, vol.~58,
  no.~1, pp.~63--70, 1989.

\bibitem{clift1978bubbles}
R.~Clift, ``Bubbles, drops and particles,'' {\em Drops and Particles},
  vol.~117, 1978.

\bibitem{el2019solving}
Y.~M. El~Hasadi and J.~T. Padding, ``Solving fluid flow problems using
  semi-supervised symbolic regression on sparse data,'' {\em AIP Advances},
  vol.~9, no.~11, p.~115218, 2019.

\bibitem{brown2003sphere}
P.~P. Brown and D.~F. Lawler, ``Sphere drag and settling velocity revisited,''
  {\em Journal of environmental engineering}, vol.~129, no.~3, pp.~222--231,
  2003.

\bibitem{acrivos1965asymptotic}
A.~Acrivos and J.~Goddard, ``Asymptotic expansions for laminar
  forced-convection heat and mass transfer,'' {\em Journal of Fluid Mechanics},
  vol.~23, no.~2, pp.~273--291, 1965.

\bibitem{el2022logarithmic}
Y.~M. El~Hasadi and J.~T. Padding, ``Do logarithmic terms exist in the drag
  coefficient of a single sphere at high reynolds numbers?,'' {\em Chemical
  Engineering Science}, vol.~265, p.~118195, 2023.

\bibitem{holzer2008new}
A.~H{\"o}lzer and M.~Sommerfeld, ``New simple correlation formula for the drag
  coefficient of non-spherical particles,'' {\em Powder Technology}, vol.~184,
  no.~3, pp.~361--365, 2008.

\bibitem{happel2012low}
J.~Happel and H.~Brenner, {\em Low Reynolds number hydrodynamics: with special
  applications to particulate media}, vol.~1.
\newblock Springer Science \& Business Media, 2012.

\bibitem{sanjeevi2018drag}
S.~K. Sanjeevi, J.~Kuipers, and J.~T. Padding, ``Drag, lift and torque
  correlations for non-spherical particles from stokes limit to high reynolds
  numbers,'' {\em International Journal of Multiphase Flow}, 2018.

\bibitem{ouchene2016new}
R.~Ouchene, M.~Khalij, B.~Arcen, and A.~Tani{\`e}re, ``A new set of
  correlations of drag, lift and torque coefficients for non-spherical
  particles and large reynolds numbers,'' {\em Powder Technology}, vol.~303,
  pp.~33--43, 2016.

\bibitem{frohlich2020correlations}
K.~Fr{\"o}hlich, M.~Meinke, and W.~Schr{\"o}der, ``Correlations for inclined
  prolates based on highly resolved simulations,'' {\em Journal of Fluid
  Mechanics}, vol.~901, 2020.

\bibitem{chen2021drag}
Y.~Chen, P.~Jiang, T.~Xiong, W.~Wei, Z.~Fang, and B.~Wang, ``Drag and heat
  transfer coefficients for axisymmetric nonspherical particles: A lbm study,''
  {\em Chemical Engineering Journal}, vol.~424, p.~130391, 2021.

\bibitem{flett19582742}
T.~M. Flett, ``2742. a mean value theorem,'' {\em The Mathematical Gazette},
  vol.~42, no.~339, pp.~38--39, 1958.

\bibitem{schlichting2016boundary}
H.~Schlichting and K.~Gersten, {\em Boundary-layer theory}.
\newblock Springer, Berlin Heidelberg, Germany, 2016.

\bibitem{abraham1970functional}
F.~F. Abraham, ``Functional dependence of drag coefficient of a sphere on
  reynolds number,'' {\em The Physics of Fluids}, vol.~13, no.~8,
  pp.~2194--2195, 1970.

\bibitem{abadi2016tensorflow}
M.~Abadi, A.~Agarwal, P.~Barham, E.~Brevdo, Z.~Chen, C.~Citro, G.~S. Corrado,
  A.~Davis, J.~Dean, M.~Devin, {\em et~al.}, ``Tensorflow: Large-scale machine
  learning on heterogeneous distributed systems,'' {\em arXiv preprint
  arXiv:1603.04467}, 2016.

\bibitem{sucker1975fluiddynamik}
D.~Sucker and H.~Brauer, ``Fluiddynamik bei quer angestr{\"o}mten zylindern,''
  {\em W{\"a}rme-und Stoff{\"u}bertragung}, vol.~8, no.~3, pp.~149--158, 1975.

\bibitem{blasius1907boundarylayers}
H.~Blasius, {\em Boundary layers in liquids with low friction}.
\newblock printed by BG Teubner, 1907.

\bibitem{melnik1975asymptotic}
R.~Melnik and R.~Chow, ``Asymptotic theory of two-dimensional trailing-edge
  flows,'' {\em NASA. Langley Res. Center Aerodynamic Analysis Requiring
  Advanced Computers, Pt. 1}, 1975.

\bibitem{janour1951resistance}
Z.~Janour, ``Resistance of a plate in parallel flow at low reynolds numbers,''
  tech. rep., 1951.

\bibitem{lisoski1993nominally}
D.~Lisoski, ``Nominally 2-dimensional flow about a normal flat plate (ph. d.
  thesis),'' {\em California Institute of Technology}, 1993.

\bibitem{kundu2015fluid}
P.~K. Kundu, I.~M. Cohen, and D.~R. Dowling, {\em Fluid mechanics}.
\newblock Academic press, 2015.

\bibitem{bejan2013convection}
A.~Bejan, {\em Convection heat transfer}.
\newblock John wiley \& sons, 2013.

\bibitem{najjar1998low}
F.~M. Najjar and S.~Balachandar, ``Low-frequency unsteadiness in the wake of a
  normal flat plate,'' {\em Journal of Fluid Mechanics}, vol.~370,
  pp.~101--147, 1998.

\bibitem{narasimhamurthy2009numerical}
V.~D. Narasimhamurthy and H.~I. Andersson, ``Numerical simulation of the
  turbulent wake behind a normal flat plate,'' {\em International Journal of
  Heat and Fluid Flow}, vol.~30, no.~6, pp.~1037--1043, 2009.

\bibitem{fage1927flow}
A.~Fage and F.~Johansen, ``On the flow of air behind an inclined flat plate of
  infinite span,'' {\em Proceedings of the Royal Society of London. Series A,
  Containing Papers of a Mathematical and Physical Character}, vol.~116,
  no.~773, pp.~170--197, 1927.

\bibitem{tian2014large}
X.~Tian, M.~C. Ong, J.~Yang, and D.~Myrhaug, ``Large-eddy simulation of the
  flow normal to a flat plate including corner effects at a high reynolds
  number,'' {\em Journal of Fluids and Structures}, vol.~49, pp.~149--169,
  2014.

\bibitem{kravchenko2000numerical}
A.~G. Kravchenko and P.~Moin, ``Numerical studies of flow over a circular
  cylinder at re d= 3900,'' {\em Physics of fluids}, vol.~12, no.~2,
  pp.~403--417, 2000.

\bibitem{lee2019data}
S.~Lee and D.~You, ``Data-driven prediction of unsteady flow over a circular
  cylinder using deep learning,'' {\em Journal of Fluid Mechanics}, vol.~879,
  pp.~217--254, 2019.

\bibitem{raissi2019deep}
M.~Raissi, Z.~Wang, M.~S. Triantafyllou, and G.~E. Karniadakis, ``Deep learning
  of vortex-induced vibrations,'' {\em Journal of Fluid Mechanics}, vol.~861,
  pp.~119--137, 2019.

\bibitem{vlachas2022multiscale}
P.~R. Vlachas, G.~Arampatzis, C.~Uhler, and P.~Koumoutsakos, ``Multiscale
  simulations of complex systems by learning their effective dynamics,'' {\em
  Nature Machine Intelligence}, vol.~4, no.~4, pp.~359--366, 2022.

\bibitem{takami1969steady}
H.~Takami and H.~B. Keller, ``Steady two-dimensional viscous flow of an
  incompressible fluid past a circular cylinder,'' {\em The Physics of Fluids},
  vol.~12, no.~12, pp.~II--51, 1969.

\bibitem{imai1957university}
I.~Imai, ``University of maryland tech,'' {\em Note, no. BN-104}, 1957.

\bibitem{gerhart2016munson}
P.~M. Gerhart, A.~L. Gerhart, and J.~I. Hochstein, {\em Munson, Young and
  Okiishi's fundamentals of fluid mechanics}.
\newblock John Wiley \& Sons, 2016.

\bibitem{jiang2021large}
H.~Jiang and L.~Cheng, ``Large-eddy simulation of flow past a circular cylinder
  for reynolds numbers 400 to 3900,'' {\em Physics of Fluids}, vol.~33, no.~3,
  p.~034119, 2021.

\bibitem{goldstein1938modern}
S.~Goldstein, {\em Modern developments in fluid dynamics: an account of theory
  and experiment relating to boundary layers, turbulent motion and wakes},
  vol.~2.
\newblock Clarendon Press, 1938.

\bibitem{hoerner1965fluid}
S.~F. Hoerner, ``Fluid dynamic drag, published by the author,'' {\em Midland
  Park, NJ}, pp.~16--35, 1965.

\bibitem{batchelor2000introduction}
G.~Batchelor, {\em An introduction to fluid dynamics}.
\newblock Cambridge university press, Cambridge, UK, 2000.

\bibitem{saha2004three}
A.~K. Saha, ``Three-dimensional numerical simulations of the transition of flow
  past a cube,'' {\em Physics of Fluids}, vol.~16, no.~5, pp.~1630--1646, 2004.

\bibitem{bagheri2016drag}
G.~Bagheri and C.~Bonadonna, ``On the drag of freely falling non-spherical
  particles,'' {\em Powder Technology}, vol.~301, pp.~526--544, 2016.

\bibitem{cengel2013ebook}
Y.~Cengel and J.~Cimbala, {\em Ebook: Fluid mechanics fundamentals and
  applications (si units)}.
\newblock McGraw Hill, 2013.

\bibitem{khair2018higher}
A.~S. Khair and N.~G. Chisholm, ``A higher-order slender-body theory for
  axisymmetric flow past a particle at moderate reynolds number,'' {\em Journal
  of Fluid Mechanics}, vol.~855, pp.~421--444, 2018.

\bibitem{jiang2014laminar}
F.~Jiang, J.~P. Gallardo, and H.~I. Andersson, ``The laminar wake behind a 6: 1
  prolate spheroid at 45 incidence angle,'' {\em Physics of Fluids}, vol.~26,
  no.~11, p.~113602, 2014.

\bibitem{knudsen1959fluid}
J.~G. Knudsen, D.~L. Katz, and R.~E. Street, ``Fluid dynamics and heat
  transfer,'' {\em Physics Today}, vol.~12, no.~3, pp.~40--44, 1959.

\bibitem{feng4435415general}
Z.~Feng and E.~Michaelides, ``A general and accurate correlation for the drag
  on spherocylinders,'' {\em Available at SSRN 4435415}.

\bibitem{lain4317628sphericity}
S.~Lain, C.~Castang, D.~Garc{\'\i}a, and M.~Sommerfeld, ``Sphericity based
  correlations for flow resistance coefficients of non-spherical particles of
  irregular shape beyond the stokes regime,'' {\em Available at SSRN 4317628}.

\bibitem{komar1978grain}
P.~D. Komar and C.~Reimers, ``Grain shape effects on settling rates,'' {\em The
  Journal of Geology}, vol.~86, no.~2, pp.~193--209, 1978.

\bibitem{corey1949influence}
A.~T. Corey, ``Influence of shape on the fall velocity of sand grains,'' 1949.

\bibitem{rhodes2024introduction}
M.~J. Rhodes and J.~Seville, {\em Introduction to particle technology}.
\newblock John Wiley \& Sons, 2008.

\bibitem{roshko1955wake}
A.~Roshko, ``On the wake and drag of bluff bodies,'' {\em Journal of the
  aeronautical sciences}, vol.~22, no.~2, pp.~124--132, 1955.

\end{thebibliography}
\bibliographystyle{ieeetr}

\end{document}